\documentclass[12pt,a4paper,twoside]{thesis}
\usepackage{amsmath,bm,graphicx,psfrag,layout}
\usepackage{txfonts}
\usepackage[bookmarks=true,bookmarksnumbered=true,%
pdftitle={Unitary Fermi gas in the epsilon expansion},%
pdfauthor={Yusuke Nishida (University of Tokyo)}]{hyperref}

\newcommand\eps{\epsilon}
\renewcommand\d{\partial}
\newcommand\grad{\bm{\nabla}}
\newcommand\p{{\bm{p}}}
\newcommand\q{{\bm{q}}}
\renewcommand\k{{\bm{k}}}
\newcommand\ep{\varepsilon_\p}
\newcommand\eq{\varepsilon_\q}
\newcommand\ek{\varepsilon_\k}
\newcommand\Ep{E_\p}
\newcommand\Eq{E_\q}
\newcommand\Ek{E_\k}
\newcommand\+{\dagger}
\newcommand\<{\langle}
\renewcommand\>{\rangle}
\newcommand\eF{\varepsilon_{\mathrm{F}}}
\newcommand\kF{k_{\mathrm{F}}}
\newcommand\Tr{\mathop{\mathrm{Tr}}}
\newcommand\eb{\varepsilon_\mathrm{b}}
\newcommand\Veff{V_\mathrm{eff}}
\newcommand\muF{{\mu_\mathrm{F}}}
\newcommand\muB{{\mu_\mathrm{B}}}
\newcommand\vs{\bm v_\mathrm s}
\newcommand\ev{\varepsilon_{m\vs}}
\newcommand\Hc{H_\mathrm c}
\newcommand\wph{\omega_\mathrm{ph}}
\newcommand\fermi{f_\mathrm{F}}
\newcommand\bose{f_\mathrm{B}}
\newcommand\Tc{{T_\mathrm{c}}}
\newcommand\mr{m_\mathrm{r}}

\def\Jvol<#1,#2,#3>{#1}
\def\Jpage<#1,#2,#3>{#2}
\def\Jyear<#1,#2,#3>{#3}
\newcommand\journal[1]{\textbf{\Jvol<#1>}, \Jpage<#1> (\Jyear<#1>)}
\newcommand\PRL[1]{Phys.\ Rev.\ Lett.\ \journal{#1}}
\newcommand\PRA[1]{Phys.\ Rev.\ A \journal{#1}}
\newcommand\PRB[1]{Phys.\ Rev.\ B \journal{#1}}
\newcommand\PRC[1]{Phys.\ Rev.\ C \journal{#1}}
\newcommand\PRD[1]{Phys.\ Rev.\ D \journal{#1}}
\newcommand\RMP[1]{Rev.\ Mod.\ Phys.\ \journal{#1}}
\newcommand\PLA[1]{Phys.\ Lett.\ A \journal{#1}}
\newcommand\PLB[1]{Phys.\ Lett.\ B \journal{#1}}
\newcommand\NPA[1]
{Nucl.\ Phys.\ \textbf{A\Jvol<#1>}, \Jpage<#1> (\Jyear<#1>)}

\newcommand\PR[1]{Phys.\ Rep.\ \journal{#1}}

\newcommand\NJP[1]{New J.\ Phys.\ \journal{#1}}
\newcommand\JLTP[1]{J.\ Low Temp.\ Phys.\ \journal{#1}}

\begin{document}

\pagestyle{empty}\pagenumbering{roman}\setcounter{page}{1}
\addcontentsline{toc}{section}{Title}

\begin{center}
\vspace*{3cm}
{\huge\textbf{Unitary Fermi gas in the $\epsilon$ expansion}}\\

\vfill\vspace*{1cm}
{\Large\textbf{Yusuke Nishida}}\\
\vspace*{1cm}
{\large\textbf{\textit{Department of Physics, University of Tokyo}}}\\
\vspace*{1cm}
{\large\textbf{December 2006}}

\vfill
{\large\textbf{\textsf{PhD Thesis}}}\\
\vspace*{1cm}
\includegraphics[width=5cm,clip]{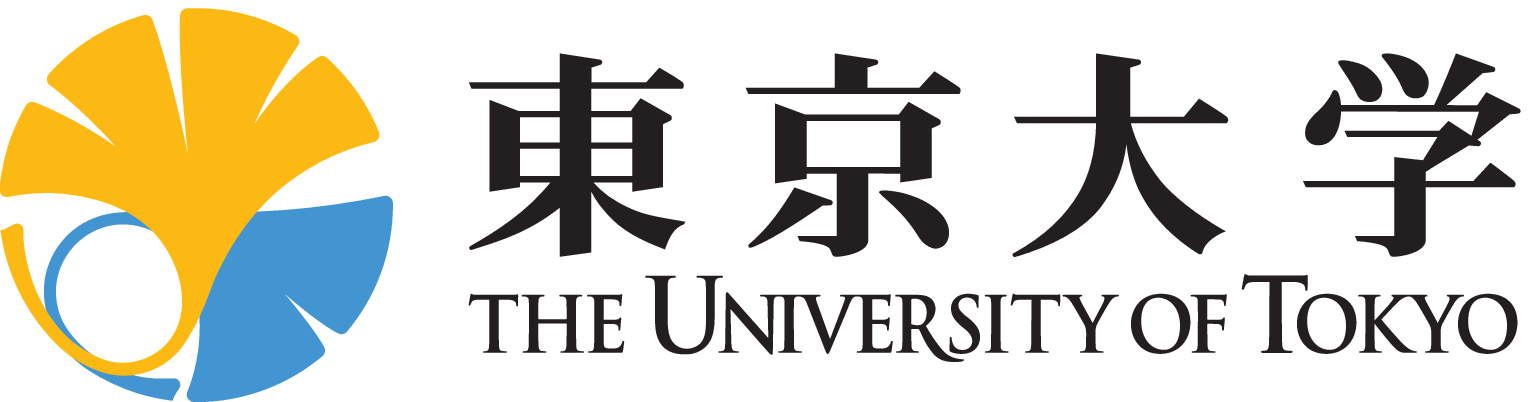}\\
\vspace*{2cm}
\end{center}

\chapter*{Abstract \label{sec:abstract}}
\pagestyle{plain}\addcontentsline{toc}{section}{Abstract}

We construct systematic expansions around four and two spatial
dimensions for a Fermi gas near the unitarity limit.  Near four spatial
dimensions such a Fermi gas can be understood as a weakly-interacting
system of fermionic and bosonic degrees of freedom.  To the leading and
next-to-leading orders in the expansion over $\eps=4-d$, with $d$ being
the dimensionality of space, we calculate the thermodynamic functions
and the fermion quasiparticle spectrum as functions of the binding
energy of the two-body state.  We also show that the unitary Fermi gas
near two spatial dimensions reduces to a weakly-interacting Fermi gas
and calculate the thermodynamic functions and the fermion quasiparticle
spectrum in the expansion over $\bar\eps=d-2$.  

Then the phase structure of the polarized Fermi gas with equal and
unequal masses in the unitary regime is studied using the $\eps$
expansion.  We find that at unitarity in the equal mass limit, there is
a first-order phase transition from the unpolarized superfluid state to
a fully polarized normal state.  On the BEC side of the unitarity point,
in a certain range of the two-body binding energy and the mass
difference, we find a gapless superfluid phase and a superfluid phase
with spatially varying condensate.  These phases occupy a region in the
phase diagram between the gapped superfluid phase and the polarized
normal phase. 

Thermodynamics of the unitary Fermi gas at finite temperature is also 
investigated from the perspective of the expansion over $\eps$.  We
show that the thermodynamics is dominated by bosonic excitations in the
low temperature region $T\ll\Tc$.  Analytic formulas for the
thermodynamic functions as functions of the temperature are derived to
the lowest order in $\eps$ in this region.  In the high temperature
region where $T\sim\Tc$, bosonic and fermionic quasiparticles are
excited and we determine the critical temperature $\Tc$ and the
thermodynamic functions around $\Tc$ to the leading and next-to-leading
orders in $\eps$ and $\bar\eps$.  

Finally we discuss the matching of the two systematic expansions around
four and two spatial dimensions in order to extract physical observables
at $d=3$.  We find good agreement of the results with those from recent
Monte Carlo simulations.

\tableofcontents

\chapter{Introduction \label{sec:intro}}
\pagestyle{headings}\pagenumbering{arabic}

\subsection*{BCS-BEC crossover and the unitarity limit}
\addcontentsline{toc}{section}{BCS-BEC crossover and the unitarity limit}

Interacting fermionic systems appear in various subfields of physics.
The Bardeen--Cooper--Schrieffer (BCS) mechanism shows that if the
interaction is attractive, the Fermi surface is unstable toward the
formation of Cooper pairs and the ground state of the system universally 
exhibits the superfluidity or superconductivity~\cite{Cooper,BCS}.  Such
phenomena have been observed in the metallic
superconductor~\cite{Onnes}, the superfluid $^3$He~\cite{Osheroff}, the
high-$\Tc$ superconductor~\cite{Bednorz}, and recently, in the ultracold
atomic gases such as $^{40}$K~\cite{Regal04} or
$^6$Li~\cite{Zwierlein04} under optical traps.  Possibilities of the
superfluid nuclear matter~\cite{Bailin,Dean}, the color
superconductivity~\cite{Rajagopal,Alford}, and the neutrino
superfluidity~\cite{Kapusta} are also discussed in literatures, which
will be important to astrophysics such as the physics of neutron stars.
Among others, experiments on the ultracold atomic gases have the
remarkable feature where the strength of attraction between atoms is
arbitrarily tunable through the magnetic-field induced Feshbach
resonances~\cite{Stwalley,Tiesinga}.  Their interaction predominantly
arises from binary $s$-wave collisions whose strength can be
characterized by the $s$-wave scattering length $a$.  Across the
Feshbach resonance, $a$ can in principal be tuned to take any value from
$-0$ to $-\infty$ and from $+\infty$ to $+0$~\cite{s_length1,s_length2}.
Therefore, the experiments of the ultracold atomic gases provide an
ideal field for studying interacting fermionic systems and testing
various many-body techniques developed for related problems. 

\begin{figure}[tp]
 \begin{center}
  \includegraphics[width=0.95\textwidth,clip]{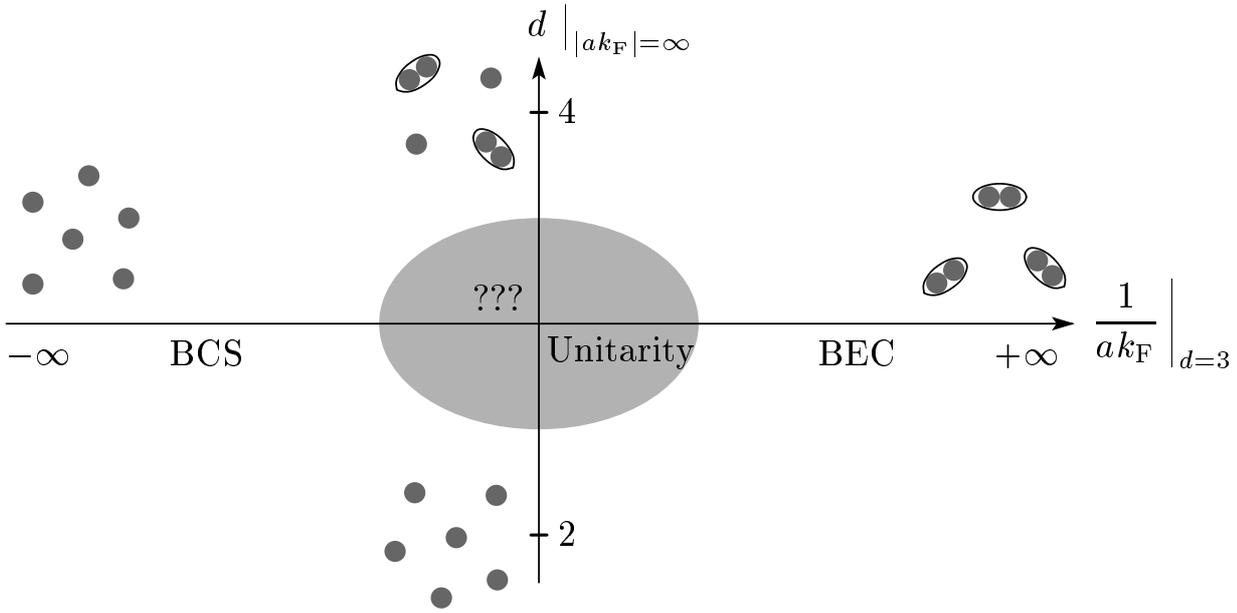}
  \caption{The BCS-BEC crossover problem in the plane of the inverse
  scattering length $1/a\kF$ and the spatial dimension $d$.  There are
  four limits where the system becomes non-interacting;
  $a\kF\to\pm0$, $d\to4$, and $d\to2$.  \label{fig:BCS-BEC}} 
 \end{center}
\end{figure}

With the use of the experimental technique of the magnetic-field tunable 
Feshbach resonance, the longstanding idea of the crossover from the BCS
state to the Bose--Einstein condensation
(BEC)~\cite{Eagles,Leggett,Nozieres,Randeria-review} has been recently
realized in laboratories.  The schematic phase diagram of the BCS-BEC
crossover problem is shown in Fig.~\ref{fig:BCS-BEC} as a function of
the dimensionless variable $1/(a\kF)$ with $\kF$ being the Fermi
momentum (horizontal axis).  Here the effective range of the interaction
$r_0$ is assumed to be small $r_0\kF\to0$.  As long as the attractive
interaction between fermions is weak (BCS regime where $a\kF<0$), the
system exhibits the BCS ground state characterized by the condensation
of the loosely bound Cooper pairs.  On the other hand, if the attractive
interaction is strong enough (BEC regime where $0<a\kF$), two fermions
form a bound molecule and the ground state of the system is described by
the BEC of the tightly bound molecules. These two apparently different
situations are considered to be smoothly connected without the phase
transition, which means the ground state in the whole regime of the
$1/(a\kF)$ axis is the superfluidity/superconductivity.

The both ends of the phase diagram correspond to weakly-interacting
systems; a weakly-interacting Fermi gas in the BCS limit $a\kF\to-0$
and a weakly-interacting Bose gas in the BEC limit $a\kF\to+0$.  In
these two limits, the usual Green's function techniques for the
many-body problem are applicable and we can understand the properties of
the system in terms of the perturbative expansion over
$|a\kF|\ll1$~\cite{Fetter-Walecka}.  An exception is the superfluidity
in the BCS limit where the paring gap $\Delta\sim e^{\pi/(2a\kF)}$ is a
non-perturbative effect and never appears to any finite orders in the
perturbative $a\kF$ expansion.  On the other hand, there exists a
strongly-interacting regime in the middle of the phase diagram where
$|a\kF|\gg1$.  The point of the infinite scattering length
$|a\kF|\to\infty$ corresponds to the threshold where the zero energy
bound molecule is formed and provides the conceptual boundary between
the BCS regime and the BEC regime.  The perturbative expansion over
$|a\kF|$ obviously breaks down in this regime because the expansion
parameter is no longer small.  Such a regime is frequently referred to
as the unitary regime because the infinite scattering length
$|a\kF|\to\infty$ is the ``unitarity limit'' where the $s$-wave cross
section reaches the maximal value allowed by the unitarity constraint of
the scattering matrix.

The Fermi gas in the unitarity limit, which we refer to as the unitary
Fermi gas, has attracted intense attention across many subfields of
physics.  As we mentioned, the system can be experimentally realized in
atomic traps using the Feshbach resonance and has been extensively 
studied~\cite{OHara02,Dieckmann02,Bourdel03,Regal03,Strecker03,%
Cubizolles03,Jochim03-PRL,Greiner03,Zwierlein03,Jochim03,Regal04,%
Regal04-lifetime,Bartenstein04,Zwierlein04,Kinast04,%
Bartenstein04-collective,Bourdel04,Chin04,Kinast04-breakdown,Kinast05,%
Partridge05,Zwierlein05-vortices,Zwierlein05,Partridge06,Zwierlein06,%
Shin06,Partridge06-2}.
Furthermore, since the fermion density is the only dimensionful scale of 
the unitary Fermi gas, its properties are expected to be
\textit{universal}, i.e., independent of details of the interparticle
interaction.  The unitary Fermi gas is thus an idealization of dilute
nuclear matter~\cite{Bertsch}, where the neutron-neutron $s$-wave
scattering length $a_{nn}\simeq-18.5\,\mathrm{fm}$ is much larger than
the effective range $r_0\simeq2.7\,\mathrm{fm}$~\cite{Gardestig}, and
may be relevant to the physics of neutron stars.  It has been also
suggested that its understanding may be important for the high-$\Tc$
superconductivity~\cite{highTc}.  The Fermi gas in the unitary regime
provides a new interesting regime other than the conventional
weak-coupling BCS and BEC regimes and is worthwhile to study
experimentally and theoretically.

\subsection*{Experimental achievements}
\addcontentsline{toc}{section}{Experimental achievements}

Before going to details of the theoretical treatment of this thesis, we
briefly summarize related experimental achievements in the trapped
fermionic atoms over the past decade.  Typically $10^5\sim10^6$ atoms
are confined in the optical potential and can be cooled down to the
temperature $\sim0.1\,E_\mathrm{F}$ using the laser cooling and
evaporative cooling~\cite{Phillips}, where $E_\mathrm{F}$ is the Fermi
energy of order micro-Kelvin.  The degenerate Fermi gas of atoms was
first reported in~\cite{DeMarco}, where the non-classical momentum
distribution and total energy of the gas were observed.  The
magnetic-field induced Feshbach resonance make it possible to realize
the degenerate Fermi gas in the strongly-interacting regime where
$|a|\kF\gg1$.  Expansion dynamics of such a strongly-interacting
degenerate Fermi gas was studied and an anisotropic expansion was
observed when the gas is released from the optical trap~\cite{OHara02}.
The hydrodynamics of expanding strongly-interacting fermionic systems
has also attracted special interest of nuclear physics community in
connection with the strongly-interacting matter created in the
experiment of Relativistic Heavy Ion
Collisions~\cite{BRAHMS,PHOBOS,STAR,PHENIX,BNL,Hatsuda}. 

The next goal after creating the degenerate Fermi gas of atoms in the
strongly-interacting regime was to establish the superfluidity.  First a
long-lived molecular gas was created from the atomic Fermi gas using the
Feshbach
resonance~\cite{Regal03,Strecker03,Cubizolles03,Jochim03-PRL,Regal04-lifetime}
and the condensation of the weakly bound molecules was observed on the
BEC side
$a\kF>0$~\cite{Greiner03,Zwierlein03,Jochim03,Bartenstein04,Bourdel04,Partridge05}. 
The BEC of the bound molecules formed by fermionic atoms is in complete
analogy with that of bosonic atoms achieved earlier using $^{87}$Rb or
$^{23}$Na~\cite{Anderson95,Davis95}.  The sudden onset of a peak in the
momentum distribution of the molecules was reported indicating the phase
transition to the BEC state.  Soon after the realization of the
molecular condensates on the BEC side, the condensation of fermionic
atom pairs was observed on the BCS side
$a\kF<0$~\cite{Regal04,Zwierlein04}.  Here a pairwise projection of the
fermionic atoms onto molecules by a rapid sweep from negative $a$ to
positive $a$ was used to measure the momentum distribution of the
fermionic atom pairs.  As an evidence of the superfluidity, 
collective excitations of the trapped Fermi gas were studied, where the
measured frequencies and damping rates are plausibly explained assuming
the
superfluidity~\cite{Kinast04,Bartenstein04-collective,Kinast04-breakdown}. 
Another evidence of the superfluidity was an appearance of the pairing
gap measured using the radio frequency spectroscopy~\cite{Chin04}.  The
critical temperature as an onset of the superfluidity was presented from
the measurement of the heat capacity~\cite{Kinast05}.  Finally the
definitive evidence of the superfluidity was provided by the observation
of vortex lattices in a rotating Fermi gas~\cite{Zwierlein05-vortices}. 

The striking properties of these trapped atomic systems lay not only on
the tunability of the interaction strength between atoms but also on
their high designability of various interesting systems.  For example,
Fermi gases with a population imbalance between the two fermion
components (finite polarization) have been already realized and
studied~\cite{Zwierlein05,Partridge06,Zwierlein06,Shin06,Partridge06-2}. 
The phase separation in the trapping potential was observed where the
superfluid core with equal densities is surrounded by the normal gas
with unequal densities.  Also the realization of Feshbach resonances is
available between two different atomic species~\cite{Stan04,Inoue04}.
While such Feshbach resonances are currently achieved only between
fermionic and bosonic atoms, it may be possible in future to realize the
Feshbach resonances between two different species of fermionic atoms. 
It will bring us a new ground to investigate a Fermi gas with finite
mass difference between different fermion species.  Such asymmetric
systems of fermions with density and mass imbalances will be also
interesting as a prototype of high density quark matter in the core of
neutron stars, where the density and mass asymmetries exist among
different quark
flavors~\cite{Alford99,Shovkovy,Huang,Gubankova03,Alford03}.

\subsection*{Theoretical treatment and $\eps$ expansion}
\addcontentsline{toc}{section}{Theoretical treatment and $\eps$ expansion}

The breathtaking experimental progresses with high tunability and
designability have stimulated various theoretical studies in the new
regime of the unitarity limit.  The austere simplicity of the unitary
Fermi gas, however, implies great difficulties for theoretical
treatment, because there seems to be no parameter for a perturbation
theory.  The usual Green's function techniques for the many-body problem
is completely unreliable here since the expansion parameter $a\kF$
becomes infinite in the unitarity limit.  Although the mean-field type 
approximations (with or without fluctuations) are often adopted to give
a qualitative picture of the BCS-BEC crossover problem, they are not
controlled approximations in the strong-coupling unitary
regime~\cite{Eagles,Leggett,Nozieres,Randeria-review,Melo,Engelbrecht,%
Haussmann,Holland,Timmermans,Ohashi,Milstein,Perali,Liu,Nishida-Abuki,%
Abuki,Zwerger}.  A challenging problem for many-body theorists is to
establish a systematic approach to investigate the Fermi gas in the
unitary regime.

At very high temperatures, a systematic expansion in terms of the
fugacity $z=e^{\mu/T}$ (virial expansion) is applicable where $\mu$
denotes the chemical potential and $T$ denotes the
temperature~\cite{Ho,Horowitz,Rupak_finite-T}.  However, the virial
expansion is restricted to the normal state and can not describe the
phase transition to the superfluid state.  Considerable progress on the
study of the unitary Fermi gas has been recently made by Monte Carlo
simulations both at zero
temperature~\cite{Carlson2003,Chang2004,Chen:2003vy,Astrakharchik2004,%
Carlson:2005kg,Lee} and finite
temperature~\cite{Wingate-Tc,bulgac-Tc,Lee-Schafer,burovski-Tc,Akkineni-Tc},
but these simulations also have various limitations.  For example, Monte
Carlo simulations can not treat systems with a population or mass
imbalance between two fermion species due to fermion sign problem.  Also
they can not answer questions related to dynamics like the dynamical
response functions and the kinetic coefficients.  Analytic treatments of
the problem at any temperature, if exist, will be extremely useful and
give insights that are not obvious from numerics.

Recently we have proposed a new analytic approach for the unitary Fermi
gas based on the systematic expansion in terms of the dimensionality of
space~\cite{Nishida-Son1,Nishida-Son2,Nishida_finite-T}, utilizing the
specialty of \textit{four} or \textit{two} spatial dimensions in the
unitarity limit~\cite{nussinov04}.  In this approach, one would extend
the problem to arbitrary spatial dimensions $d$ with keeping the
scattering length to be infinite $|a\kF|\to\infty$ (vertical axis in
Fig.~\ref{fig:BCS-BEC}).  Then we find two non-interacting limits on the
$d$ axis, which are $d=4$ and $d=2$.  Accordingly, slightly below four
or slightly above two spatial dimensions, the unitary Fermi gas becomes
weakly-interacting, where a ``perturbative expansion'' is available.
Actually the unitary Fermi gas near four spatial dimensions can be
understood as a weakly-interacting system of fermionic and bosonic
quasiparticles, while near two spatial dimensions it reduces to a
weakly-interacting Fermi gas.  A small parameter of the perturbative
expansion there is $\eps=4-d$ near four dimensions or $\bar\eps=d-2$
near two dimensions.  The weakness of the interaction near four and two
spatial dimensions can be also seen from the viewpoint of the
renormalization group fixed point; the fixed point describing the
unitarity limit approaches to the non-interacting Gaussian fixed point
near $d=4$ and $d=2$~\cite{Sauli,Sachdev}.  After performing all
calculations treating $\eps$ or $\bar\eps$ as a small expansion
parameter, results for the physical case of three spatial dimensions are
obtained by extrapolating the series expansions to $\eps\,(\bar\eps)=1$,
or more appropriately, by matching the two series expansions.

The $\eps$ expansion around four spatial dimensions has been developed
to calculate thermodynamic functions and the fermion quasiparticle
spectrum at zero temperature near the unitarity limit to the leading and
next-to-leading orders in $\eps$.  The results were found to be quite
consistent with those obtained by the Monte Carlo simulations and the
experiments~\cite{Nishida-Son1,Nishida-Son2}.  Then the $\eps$ expansion
was extended to investigate the thermodynamics of the unitary Fermi gas
at finite temperature $T$, below and above the critical temperature
$T=\Tc$~\cite{Nishida_finite-T}.  It should be noted that the critical
dimension of a superfluid-normal phase transition is also four, which
makes weak-coupling calculations reliable at any temperature for the
small $\eps$.  The $\eps$ expansion was also successfully applied to
study atom-dimer and dimer-dimer scatterings in vacuum and found to give
results quite close to non-perturbative numerical solutions at
$d=3$~\cite{Rupak-dimer}.  Thus there are compelling reasons to hope
that the limit $d\to4$ is not only theoretically interesting but also
practically useful, despite the fact that the expansion parameter $\eps$
is one at $d=3$.  We recall that the $\eps$ expansion has been extremely
fruitful in the theory of the second order phase
transition~\cite{WilsonKogut}.  In addition to the above-mentioned
works, the phase structure of the polarized Fermi gas near the unitarity
limit has been investigated based on the $\eps$
expansion~\cite{Nishida-Son2,Rupak-polarized}.  The
next-to-next-to-leading order correction to the thermodynamic functions
at zero temperature was computed in~\cite{Arnold} and the BCS-BEC
crossover was studied by the $\eps$ expansion~\cite{Chen}.  Very
recently, another systematic approach to the Fermi gas in the unitary
regime has been proposed; the two-component Fermi gas is generalized to
a $2N$-component Fermi gas with $\mathrm{Sp}(2N)$ invariant
interaction.  Then the inverse of the component number $1/N$ is used for
the small expansion parameter~\cite{Sachdev,Veillette}.

In this thesis, we give a comprehensive study of the unitary Fermi gas
from the perspective of the $\eps$ expansion.  We start with the study
of two-body scattering in vacuum in the unitarity limit for arbitrary
spatial dimensions $2<d<4$, which will clarify why and how systematic
expansions around four and two spatial dimensions are possible for the
unitary Fermi gas (Chap.~\ref{sec:vacuum}).  In Chap.~\ref{sec:4d}, we
give a detailed account of the $\eps$ expansion for the unitary Fermi
gas and show the results on the thermodynamic functions and the fermion
quasiparticle spectrum in the unitary regime to the leading and
next-to-leading orders in $\eps$.  Then the $\eps$ expansion is applied
to the unitary Fermi gas with unequal densities of two components, which
we call  ``polarized'' Fermi gas.  The phase structure of the polarized
Fermi gas as a function of the two-body binding energy is investigated
with equal fermion masses in Chap.~\ref{sec:polarization} and with
unequal fermion masses in Chap~\ref{sec:unequal}.  We also show in
Chap.~\ref{sec:2d} that there exists a systematic expansion for the
Fermi gas in the unitarity limit around two spatial dimensions.  In
Chap.~\ref{sec:matching}, we make an exploratory discussion to connect
the two systematic expansions around four and two spatial dimensions in
order to extract results at $d=3$.  Chaps.~\ref{sec:below-Tc} and
\ref{sec:above-Tc} are devoted to the thermodynamics of the unitary
Fermi gas at finite temperature.  In Chap.~\ref{sec:below-Tc}, analytic
formulas for the thermodynamic functions in the low temperature region
$T\ll\Tc$ are shown as functions of the temperature to the lowest order
in $\eps$.  The behavior of the thermodynamic functions above $\Tc$ to
the leading and next-to-leading orders in $\eps$ is discussed in
Chap.~\ref{sec:above-Tc}.  In particular, we put emphasis on the
determination of the critical temperature $\Tc$ and the thermodynamic
functions at $\Tc$ by matching the $\eps$ expansion with the expansion
around two spatial dimensions.  All the results thus obtained are
compared to those from the recent Monte Carlo simulations.  Finally,
summary and concluding remarks are given in Chap.~\ref{sec:summary}. 

The chapters \ref{sec:vacuum}--\ref{sec:matching} in this thesis are
based on the author's works in collaboration with
D.~T.~Son~\cite{Nishida-Son1,Nishida-Son2,NS_unequal-mass}, while the
chapters \ref{sec:below-Tc} and \ref{sec:above-Tc} are based on the
author's paper~\cite{Nishida_finite-T}.

\chapter{Two-body scattering in vacuum \label{sec:vacuum}} 
The special role of four and two spatial dimensions in the unitarity
limit has been recognized by Nussinov and Nussinov~\cite{nussinov04}.
They noticed that at infinite scattering length, the
wavefunction of two fermions with opposite spin 
behaves like $R(r)=1/r^{d-2}$ at small $r$, where $r$ is the separation 
between two fermions.  Therefore, the normalization integral of
the wavefunction takes the form of
\begin{equation}\label{eq:normalization}
 \int\!d\bm{r}R(r)^2=\int_{r_0}dr\frac{1}{r^{d-3}},
\end{equation}
which has a singularity at $r_0\to0$ in high dimensions $d\geq4$.
From this observation, it is concluded that the unitary Fermi gas at
$d\to4$ become a non-interacting Bose gas.  In particular, at fixed Fermi 
momentum the energy per particle goes to zero as $d\to 4$ from below.
On the other hand, in low dimensions $d\leq2$, the attractive potential
with any strength causes zero energy bound states, and hence, the
threshold of the appearance of the first two-body bound state
corresponds to the zero coupling.  It follows that the Fermi gas in the
unitary limit corresponds to a non-interacting Fermi gas at $d\to2$ and
the energy per particle approaches to that of the free Fermi gas in this
limit.  This singular character of $d=2$ was also recognized in the
earlier work~\cite{randeria-2d}.

In this Chapter, we shall give precise meaning for their intuitive
arguments at $d=4$ and $2$ from the perspective of diagrammatic
approach.  These two dimensions will provide foundations to construct
systematic expansions for the unitary Fermi gas. 
Here we concentrate on the two-body problem.

\section{Near four spatial dimensions}

\begin{figure}[tp]
 \begin{center}
  \includegraphics[width=\textwidth,clip]{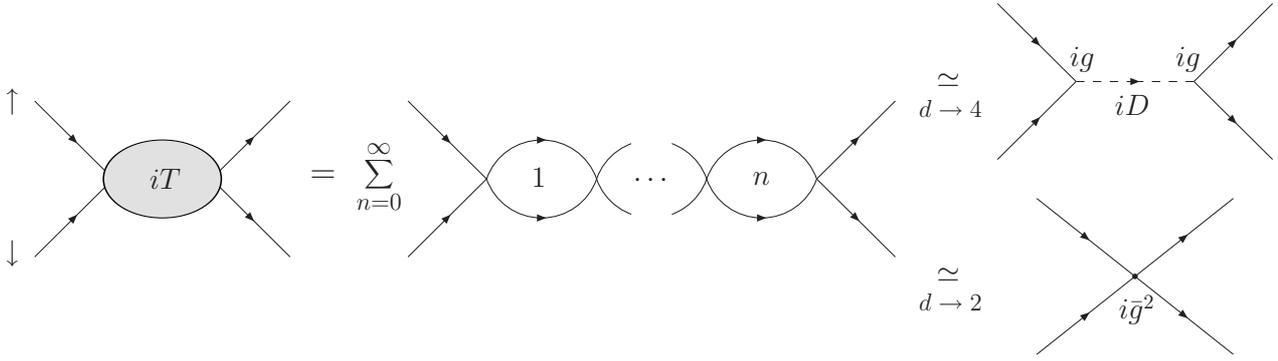}
  \caption{Two-fermion scattering in vacuum in the unitarity limit. The
  $T$-matrix near four spatial dimensions is expressed by the propagation
  of boson with the small effective coupling $g$, while it reduces to a
  contact interaction with the small effective coupling $\bar g^2$ near
  two spatial dimensions. \label{fig:scattering}}  
 \end{center}
\end{figure}

The system under consideration is described by a Lagrangian density with
a local four-Fermi interaction (here and below $\hbar=1$):
\begin{equation}\label{eq:L_vacuum}
 \mathcal{L} = \sum_{\sigma=\uparrow,\downarrow} \psi_\sigma^\dagger
  \left(i\d_t+\frac{\grad^2}{2m_\sigma}\right)\psi_\sigma
  +c_0\,\psi^\dagger_\uparrow\psi^\dagger_\downarrow
  \psi_\downarrow\psi_\uparrow,
\end{equation}
where $m_\sigma\ (=m_\uparrow,m_\downarrow)$ is the fermion mass and
$c_0$ is the bare attractive coupling between two fermions.  Here we
consider the fermions with equal masses $m_\uparrow=m_\downarrow=m$ and
generalize our discussion to the unequal mass case in
Chap.~\ref{sec:unequal}.

The $T$-matrix of the two-body scattering is given by the geometric 
series of bubble diagrams depicted in Fig.~\ref{fig:scattering}. 
As a result of the summation, its inverse can be written as
\begin{equation}\label{eq:c_0}
 \begin{split}
  T(p_0,\p)^{-1}&=\frac1{c_0}+i\int\!\frac{dk_0d\k}{(2\pi)^{d+1}}\,
  \frac1{\frac{p_0}2-k_0-\varepsilon_{\frac\p2-\k}+i\delta} 
  \,\frac1{\frac{p_0}2+k_0-\varepsilon_{\frac\p2+\k}+i\delta} \\
  &=\frac1{c_0}-\int\!\frac{d\k}{(2\pi)^d}\,
  \frac1{2\ek-p_0+\frac\ep2-i\delta},
 \end{split}
\end{equation}
where $\varepsilon_\p=p^2/2m$ is the kinetic energy of non-relativistic
particles.  The integral in Eq.~(\ref{eq:c_0}) is divergent and needs to be
regularized.  In this thesis we shall work in dimensional regularization:
integrals are evaluated for those values of $d$ where they converge 
[for the integral in Eq.~(\ref{eq:c_0}) this corresponds to $d<2$] and
then analytically continued to other values of $d$.  In this
regularization scheme, the integral in Eq.~(\ref{eq:c_0}) vanishes for
$p_0=\p=0$ (scatterings at threshold).  Therefore the limit of infinite
scattering length where $T(0,\bm0)=\infty$ corresponds to $c_0=\infty$.
  
The integration over $\k$ in Eq.~(\ref{eq:c_0}) can be evaluated explicitly
\begin{equation}\label{eq:T^-1}
 T(p_0,\p)^{-1}=-\Gamma\!\left(1-\frac d2\right)
  \left(\frac{m}{4\pi}\right)^{\frac d2}
 \left(-p_0+\frac\ep2-i\delta\right)^{\frac d2-1}.
\end{equation}
This expression has a pole at $d=4$ originating from the ultraviolet
divergence of the $\k$ integration. 
Substituting $d=4-\eps$ and expanding in terms of $\eps$, the $T$-matrix
near four spatial dimensions becomes
\begin{equation}\label{eq:}
 iT(p_0,\p)=-\frac{8\pi^2\eps}{m^2}\frac{i}{p_0-\frac\ep2+i\delta}+O(\eps^2).
\end{equation}
If we define
\begin{equation}\label{eq:g}
 g^2=\frac{8\pi^2\eps}{m^2}
\end{equation}
and
\begin{equation}\label{eq:D_vacuum}
  D(p_0,\p) = \left(p_0 
  - \frac{\ep}2 + i\delta\right)^{-1}.
\end{equation}
the $T$-matrix, to leading order in $\eps$, can be written as
\begin{equation}\label{eq:T}
 iT(p_0,\p)\simeq(ig)^2 iD(p_0,\p).
\end{equation}

The function $D(p_0,\p)$ is the propagator of a particle with mass
$2m$.  It is natural to interpret this particle a bound state of two
fermions at threshold.  We will refer to this particle simply as 
the \textit{boson}.  
Eq.~(\ref{eq:T}) states that the two-fermion scattering near
four spatial dimensions can be thought of as a process that occurs through 
the propagation of an intermediate boson, 
as depicted in Fig.~\ref{fig:scattering}. The effective coupling
of two fermions into the boson is given by $g$. An  important point is
that the effective coupling $g\sim\eps^{1/2}$ is small near four
dimensions. This fact indicates the possibility to construct a
perturbative expansion for the unitary Fermi gas near four spatial
dimensions in terms of the small parameter $\eps$.

\section{Near two spatial dimensions}
Similarly, the perturbative expansion around two spatial dimensions is
possible.  This is because the inverse of $T$-matrix in 
Eq.~(\ref{eq:T^-1}) has another pole at $d=2$.
Substituting $d=2+\bar\eps$ and expanding in terms of
$\bar\eps$, the $T$-matrix near two spatial dimensions becomes 
\begin{equation}
 iT(p_0,\p)=i\frac{2\pi}m\bar\eps+O(\bar\eps^2).
\end{equation}
If we define the effective coupling at $2+\bar\eps$ dimensions as 
\begin{equation}\label{eq:g-bar}
 \bar g^2=\frac{2\pi}m\bar\eps, 
\end{equation}
the $T$-matrix to the leading order in $\bar\eps$ can be written as
\begin{equation}
 iT(p_0,\p)\simeq i\bar g^2.
\end{equation}
We see that the $T$-matrix near two spatial dimensions reduces to that of
a contact interaction with the small effective coupling 
$\bar g^2\sim\bar\eps$ as depicted in Fig.~\ref{fig:scattering}. 
In this case, the boson propagator $D(p)$ in Eq.~(\ref{eq:T})
corresponds to just a constant $-1$.  We note that the same effective
couplings near four and two spatial dimensions in Eqs.~(\ref{eq:g}) and
(\ref{eq:g-bar}) can be obtained by the fixed point of the
renormalization group flow describing the unitarity
limit~\cite{Sauli,Sachdev}.  We defer our discussion of the expansion
over $\bar\eps=d-2$ to Chap.~\ref{sec:2d} and concentrate on the
expansion over $\eps=4-d$.

\section{Binding energy of two-body state}
We shall be interested not only in the physics right at the unitarity
point, but also in the vicinity of it.  In other words, we shall assume
that $1/c_0$ can be nonzero in dimensional regularization.  The case of
$c_0<0$ corresponds to the BEC side of the unitarity point, and $c_0>0$
corresponds to the BCS side.  To facilitate a comparison with the
physics in three dimensions, we shall find the relation between the
coupling $c_0$, assuming that we are on the BEC side, $c_0<0$, and
the binding energy of the two-body state at arbitrary spatial dimensions
$2<d<4$. 

The binding energy $\eb$, defined to be positive $\eb>0$, 
is obtained as the location of the pole of the
$T$-matrix at zero external momentum: $T(-\eb,\bm0)^{-1}=0$.  From 
Eqs.~(\ref{eq:c_0}) and (\ref{eq:T^-1}), we see $c_0$ is related with
$\eb$ via the following equation 
\begin{equation}\label{eq:eb}
 \frac1{c_0}=\Gamma\!\left(1-\frac d2\right)
  \left(\frac{m}{4\pi}\right)^{\frac d2}\eb^{\,\frac d2-1}.
\end{equation}
Near four spatial dimensions, the relationship between $\eb$ and $c_0$
to the next-to-leading order in $\eps=4-d$ is given by
\begin{equation}\label{eq:eb-4d}
 \frac1{c_0}\simeq-\frac{\eb}{2\eps}\left(\frac{m}{2\pi}\right)^2
  \left[1+\frac\eps2\left(1-\gamma\right)\right]
  \left(\frac{m\eb}{4\pi}\right)^{-\frac\eps2}.
\end{equation}
In three spatial dimensions, the relationship between $\eb$ and the
scattering length $a=-mc_0/4\pi$ becomes $\eb=(ma^2)^{-1}$.  If we
recall that the Fermi energy is related to $\kF$ through
$\eF=\kF^{\,2}/(2m)$, one finds the following relationship holds at
$d=3$: 
\begin{equation}\label{eq:ebeF}
  \frac\eb\eF = \frac2{(a\kF)^{2}}.  
\end{equation}
We will use these relations to compare results in the expansion over
$\eps=4-d$ with the physics at three spatial dimensions.

\chapter{Unitary Fermi gas around four spatial dimensions
\label{sec:4d}}

\section{Lagrangian and Feynman rules}
Employing the idea of the previous Chapter, we now construct the $\eps$
expansion for the Fermi gas near the unitarity limit. 
We start with the Lagrangian density in Eq.~(\ref{eq:L_vacuum}) with 
chemical potentials $\mu_\uparrow$ and $\mu_\downarrow$ introduced to
two different components: 
\begin{equation}
 \mathcal{L} = \sum_{\sigma=\uparrow,\downarrow} \psi_\sigma^\dagger
  \left(i\d_t+\frac{\grad^2}{2m}+\mu_\sigma\right)\psi_\sigma
  +c_0\,\psi^\dagger_\uparrow\psi^\dagger_\downarrow
  \psi_\downarrow\psi_\uparrow.
\end{equation}
After making the Hubbard--Stratonovich
transformation~\cite{Stratonovich,Hubbard}, the Lagrangian 
density in the Nambu--Gor'kov formalism can be written as 
\begin{equation}\label{eq:L}
  \mathcal{L} = \Psi^\+\left(i\d_t + \frac{\sigma_3\grad^2}{2m}
  + \mu\sigma_3 + H\right)\Psi - \frac1{c_0} \phi^*\phi
  + \Psi^\+\sigma_+\Psi\phi + \Psi^\+\sigma_-\Psi\phi^*,
\end{equation}
where $\Psi=(\psi_\uparrow,\psi^\+_\downarrow)^T$ is a two-component
Nambu--Gor'kov field, and $\sigma_{1,2,3}$ and 
$\sigma_\pm=\frac12(\sigma_1\pm i\sigma_2)$ are the Pauli
matrices~\cite{Nambu,Gorkov}.  We define the average chemical potential
as $\mu=(\mu_\uparrow+\mu_\downarrow)/2$ and the chemical potential
difference as $H=(\mu_\uparrow-\mu_\downarrow)/2$. 

The ground state at finite density system (at least when $H=0$) 
is a superfluid state where
$\phi$ condenses: $\<\phi\>=\phi_0$ with $\phi_0$ being chosen to be 
real.  With that in mind we expand $\phi$ around its vacuum expectation
value $\phi_0$ as 
\begin{equation}\label{eq:coupling}
 \phi=\phi_0+ g\varphi, \qquad\quad 
  g = \frac{(8\pi^2\epsilon)^{1/2}}m
  \left(\frac{m\phi_0}{2\pi}\right)^{\epsilon/4}.
\end{equation}
Here we introduced the effective coupling $g\sim\epsilon^{1/2}$ in
Eq.~(\ref{eq:g}). The extra factor
$\left(m\phi_0/2\pi\right)^{\epsilon/4}$ was chosen so that $\varphi$
has the same dimension as a non-relativistic field~\footnote{The choice
of the extra factor is arbitrary, if it has the correct dimension, and does
not affect the final results because the difference can be absorbed into
the redefinition of the fluctuation field $\varphi$.  The particular
choice of $g$ in Eq.~(\ref{eq:coupling}) will simplify the form of loop
integrals in the intermediate steps.}.  Since the Lagrangian density in
Eq.~(\ref{eq:L}) does not have the kinetic term for the boson field
$\varphi$, we add and subtract its kinetic part by hand.  In other
words, we rewrite the Lagrangian density as a sum of three parts,
$\mathcal{L}=\mathcal{L}_0+\mathcal{L}_1+\mathcal{L}_2$, where
\begin{align}\label{eq:L_012}
  \mathcal{L}_0 & = \Psi^\+\left(i\d_t + H + \frac{\sigma_3\grad^2}{2m}
  + \sigma_+\phi_0 + \sigma_-\phi_0\right)\Psi
  + \varphi^*\left(i\d_t+\frac{\grad^2}{4m}\right)\varphi
  - \frac{\phi_0^{\,2}}{c_0}\,, \\
  \mathcal{L}_1 & = g\Psi^\+\sigma_+\Psi\varphi 
  + g\Psi^\+\sigma_-\Psi\varphi^* + \mu\Psi^\+\sigma_3\Psi
  + \left(2\mu-\frac{g^2}{c_0}\right)\varphi^*\varphi 
  - \frac{g\phi_0}{c_0}\varphi - \frac{g\phi_0}{c_0}\varphi^*\,, 
 \phantom{\frac{\frac\int\int}{\frac\int\int}}\hspace{-4.5mm} \\
 \mathcal{L}_2 & = -\varphi^*\left(i\d_t
 +\frac{\grad^2}{4m}\right)\varphi - 2\mu\varphi^*\varphi\,.
\end{align}
As we shall soon see, $\phi_0$ coincides, to leading order in $\epsilon$,
to the energy gap in the fermion spectrum.  In the unitarity limit, we have
$g^2/c_0=0$.  When $c_0$ is finite and negative, $-g^2/c_0\simeq\eb$
gives the binding energy of boson to the leading order in $\eps$ from
Eq.~(\ref{eq:eb-4d}).  Throughout this thesis, we consider the vicinity
of the unitarity point where $g^2/c_0$ can be treated as a small
quantity~\footnote{The precise meaning of the ``small quantity'' here is
that $g^2/c_0$ compared to $\phi_0$ is of the order $\eps$,
$g^2/c_0\sim\eps\phi_0$.}. 

\begin{figure}[tp]
 \begin{center}
  \includegraphics[width=0.95\textwidth,clip]{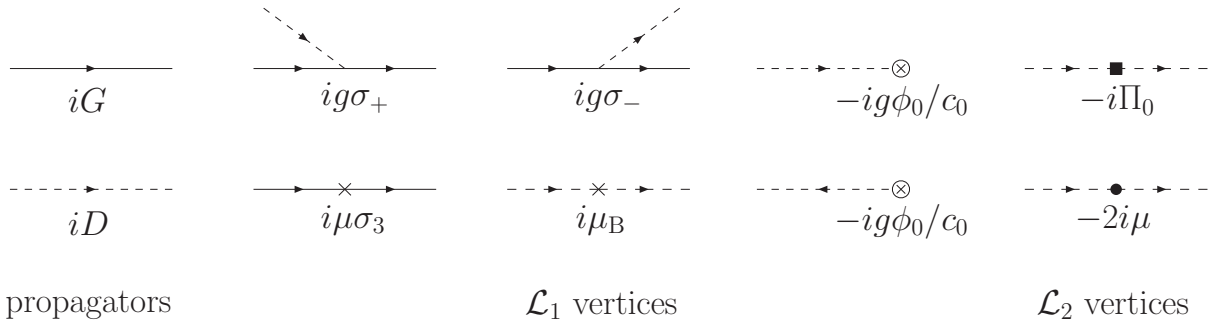}
  \caption{Feynman rules from the Lagrangian density in
  Eq.~(\ref{eq:L_012}).  The two vertices on the last column come 
  from $\mathcal{L}_2$, while the rest from $\mathcal{L}_1$. Solid
  (dotted) lines represent the fermion (boson) propagator $iG$ $(iD)$. 
  \label{fig:feynman_rules}} 
 \end{center}
\end{figure}

The part $\mathcal{L}_0$ represents the Lagrangian density of
non-interacting fermion quasiparticles and bosons with the mass $2m$, 
whose kinetic terms are introduced by hand in $\mathcal{L}_0$ and taken
out in $\mathcal{L}_2$. 
The propagators of fermion and boson are generated by
$\mathcal{L}_0$.  The fermion propagator $G$ is a $2\times2$ matrix,
\begin{equation}\label{eq:G}
  G(p_0,\p) = \frac1{(p_0+H)^2-E_\p^{\,2}+i\delta}
  \begin{pmatrix}
   p_0 + H + \ep & -\phi_0 \\
   -\phi_0 & p_0 + H -\ep
  \end{pmatrix},
\end{equation}
where $E_\p=\sqrt{\ep^{\,2}+\phi_0^{\,2}}$ is the excitation energy of
the fermion quasiparticle.  We use the $i\delta$ prescription where
$\delta=0^+$ for positive $p_0$ while $\delta=0^-$ for negative $p_0$.
The boson propagator $D$ is given by the same form as in
Eq.~(\ref{eq:D_vacuum}), 
\begin{equation}\label{eq:D}
  D(p_0,\p) = \left(p_0 
  - \frac{\ep}2 + i\delta\right)^{-1}.
\end{equation}

The parts $\mathcal{L}_1$ and $\mathcal{L}_2$ generate the vertices
which are depicted in Fig.~\ref{fig:feynman_rules}.  The first two terms
in $\mathcal{L}_1$ represent the fermion-boson interactions, whose
coupling is proportional to $g$ and small in the limit $\epsilon\to0$.
This coupling originates from the two-body scattering in vacuum in the
unitarity limit which was studied in the previous
Chapter (see Fig.~\ref{fig:scattering}).  The third and fourth terms are 
insertions of chemical potentials, $\mu$ and the boson chemical
potential 
\begin{equation}
 \mu_\mathrm{B}\equiv2\mu-\frac{g^2}{c_0},
\end{equation}
to the fermion and boson propagators, respectively.  We treat these two
terms as perturbations because $\mu$ will turn out to be small,
$\mu/\phi_0\sim\eps$, and we limit ourselves to the vicinity of the
unitarity point where we can regard $g^2/c_0$ to be so small
that $\mu_\mathrm{B}/\phi_0\sim\eps$.  The last two terms in
$\mathcal{L}_1$ give tadpoles to the $\varphi$ and $\varphi^*$ fields, 
which are proportional to $-ig\phi_0/c_0$.  The condition of
cancellation of tadpole diagrams will determine the value of the
condensate $\phi_0$. 

Finally, we treat the two terms in $\mathcal{L}_2$ as additional
vertices for the boson propagator as depicted in the last column of
Fig.~\ref{fig:feynman_rules}, which are proportional to $-i\Pi_0$ and
$-2i\mu$ where  
\begin{equation}\label{eq:Pi_0}
  \Pi_0(p_0,\p) = p_0 -\frac{\ep}2.
\end{equation}
These two vertices play roles of counter terms so as to avoid double
counting of certain types of diagrams which are already taken into
$\mathcal{L}_0$ and $\mathcal{L}_1$. 
In the latter part of this Chapter, we will consider the unpolarized
case with $H=0$ at zero temperature.

\begin{table}[tp]
 \begin{center}
  \begin{tabular}{l|l}
   $O(\eps^{-1})$ & $N$ (fermion density) \\\hline
   $O(\eps^{-1/2})$ & $\eF$ (Fermi energy) \\\hline
   $O(1)$ & $\phi_0$ (condensate),\quad $\Delta$ (energy gap)\\
   & $P$ (pressure),\quad $E$ (energy density) \\\hline
   $O(\eps)$ & $\mu,\,\muB$ (chemical potentials)\\
  \end{tabular}
  \caption{Summary of powers of $\eps$ in various quantities at zero
  temperature when the condensate $\phi_0$ is regarded as $O(1)$.
  \label{table:power}} 
 \end{center}
\end{table}

\section{Power counting rule of $\epsilon$ \label{sec:counting}}
Near unitarity, 
we can use the same power counting rule of $\epsilon$ developed in the
unitarity limit~\cite{Nishida-Son1}.  Let us first consider Feynman
diagrams constructed only from $\mathcal{L}_0$ and $\mathcal{L}_1$,
without the vertices from  $\mathcal{L}_2$.  We make a prior assumption
$\mu/\phi_0\sim\epsilon$, which will be checked later, and consider
$\phi_0$ to be $O(1)$.  Each pair of boson-fermion vertices is
proportional to $g^2\sim\eps$ and hence brings a factor of $\epsilon$. 
Also, each insertion of $\mu\sim\eps$ or $\muB=2\mu-g^2/c_0\sim\eps$
brings another factor of $\epsilon$.  Therefore the naive power of
$\epsilon$ for a given diagram is $N_g/2+N_\mu$, where $N_g$ is the
number of couplings $g$ from $\mathcal{L}_1$. $N_\mu=N_{\muF}+N_{\muB}$
is the sum of the number of $\mu$ insertions to the fermion line and
$\muB$ insertions to the boson line. 

However, this naive counting does not take into account the fact that
there might be inverse powers of $\epsilon$ coming from integrals 
which have ultraviolet divergences at $d=4$.  Each loop integral in
the ultraviolet region behaves as 
\begin{equation}
 \int dp_0d\p\sim \int d\p\,\ep\sim p^{6},
\end{equation}
while each fermion or boson propagator behaves, at worst, as 
$G(p)\sim p^{-2}$ or $D(p)\sim p^{-2}$. Therefore, a given diagram may 
diverges as $\sim p^\mathcal{D}$ with $\mathcal{D}$ being the
superficial degree of divergence given by  
\begin{equation}
 \mathcal{D}=6L-2P_\mathrm{F}-2P_\mathrm{B}.
\end{equation}
Here $L$ is the number of loop integrals and $P_\mathrm{F(B)}$ is the
number of fermion (boson) propagators.  From the similar argument to the 
relativistic field theories~\cite{Peskin:1995ev}, one can derive the
following relations including the number of external fermion (boson)
lines $E_\mathrm{F(B)}$: 
\begin{equation}\label{eq:loop}
 \begin{split}
  L&=(P_\mathrm{F}-N_\muF)+(P_\mathrm{B}-N_\muB)-N_g+1,
  \phantom{\frac{}{\int}}\\
  N_g&=(P_\mathrm{F}-N_\muF)+\frac{E_\mathrm{F}}2
  =2(P_\mathrm{B}-N_\muB)+E_\mathrm{B}.
 \end{split}
\end{equation}
With the use of these relations, the superficial degree of divergence
$\mathcal{D}$ is written as 
\begin{equation}\label{eq:estimate}
 \mathcal{D}=6-2(E_\mathrm{F}+E_\mathrm{B}+N_\mu), 
\end{equation}
which shows that the inverse powers of $\epsilon$ appear only in
diagrams with no more than three external lines and chemical potential
insertions.  This is similar to the situation in quantum electrodynamics 
where infinities occur only in electron and photon self-energies and the 
electron-photon triple vertex.

However, this estimation of $\mathcal{D}$ is actually an over-estimate:
for many diagrams the real degree of divergence is smaller than given in 
Eq.~(\ref{eq:estimate}).  To see that we split $G(p)$ into the retarded and
advanced parts, $G(p)=G^\mathrm{R}(p)+G^\mathrm{A}(p)$, where 
$G^\mathrm{R}$ ($G^\mathrm{A}$) has poles only in the lower (upper) half
of the complex $p_0$ plane:
\begin{equation}
 \begin{split}
  G_{11}(p) &=\frac{p_0+\ep}{p_0^{\,2}-\Ep^{\,2}+i\delta}
  =\frac{\Ep+\ep}{2\Ep\left(p_0-\Ep+i\delta\right)}
  +\frac{\Ep-\ep}{2\Ep\left(p_0+\Ep-i\delta\right)}, \\
  G_{22}(p) &=\frac{p_0-\ep}{p_0^{\,2}-\Ep^{\,2}+i\delta}
  =\frac{\Ep-\ep}{2\Ep\left(p_0-\Ep+i\delta\right)}
  +\frac{\Ep+\ep}{2\Ep\left(p_0+\Ep-i\delta\right)}, \\
  G_{12}(p)=G_{21}(p) &=\frac{-\phi_0}{p_0^{\,2}-\Ep^{\,2}+i\delta}
  =\frac{-\phi_0}{2\Ep\left(p_0-\Ep+i\delta\right)}
  +\frac{-\phi_0}{2\Ep\left(p_0+\Ep-i\delta\right)}.
 \end{split}
\end{equation}
From these expressions, it is easy to see that the ultraviolet behaviors
of different components of the propagators are different:
\begin{align}\label{eq:analytic}
  &G_{11}^\mathrm{R}(p)\sim G_{22}^\mathrm{A}(p)\sim 
  D^\mathrm{R}(p)\sim p^{-2}, \phantom{\frac{}{\int}}\\
  &G_{12}^\mathrm{R,A}(p)\sim G_{21}^\mathrm{R,A}(p)\sim p^{-4},\qquad 
 \notag  G_{11}^\mathrm{A}(p)\sim G_{22}^\mathrm{R}(p)\sim p^{-6}.
\end{align}
Note that the boson propagator $D(p)$ has a pole on the lower half plane of
$p_0$, we have only the retarded Green's function for boson,
$D^\mathrm{A}(p)=0$. 

\begin{figure}[tp]
 \begin{center}
  \includegraphics[width=0.98\textwidth,clip]{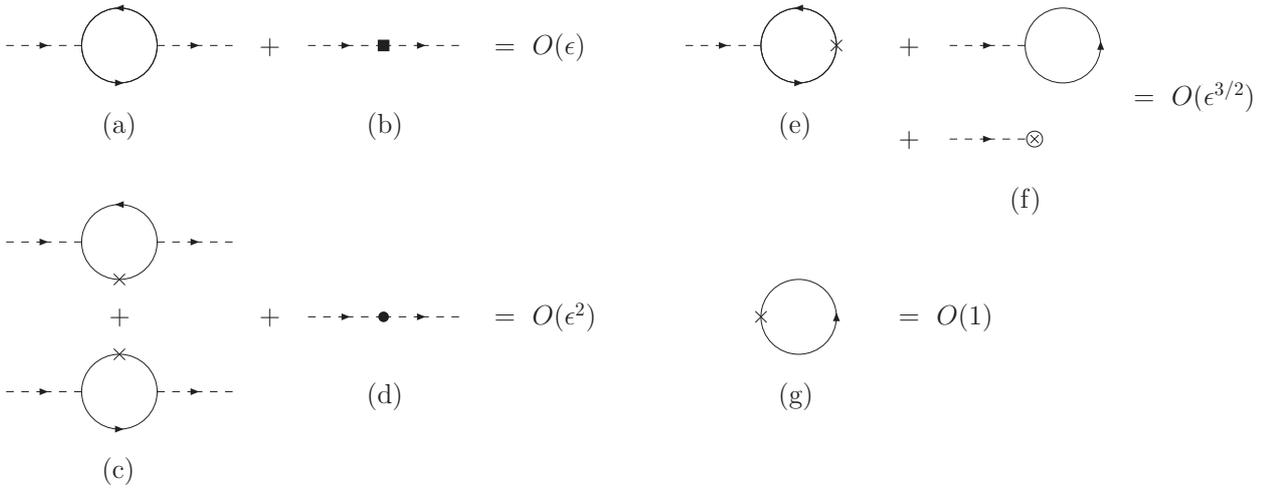}
  \caption{Four apparent exceptions of naive power counting rule of
  $\epsilon$, (a, c, e, g). The boson self-energy diagram, 
  (a) or (c), is combined with the vertex from $\mathcal{L}_2$, (b) or
  (d), to restore the naive $\epsilon$ counting. 
  The tadpole diagram in (e) cancels with other tadpole diagrams at the
  minimum of the effective potential. The vacuum diagram (g) is the only
  exception of the naive power counting rule of $\epsilon$, which is
  $O(1)$ instead of $O(\eps)$. \label{fig:cancel}} 
 \end{center}
\end{figure}

From these analytic properties of the propagators in the ultraviolet
region and the vertex structures in $\mathcal{L}_1$ as well as the
relation of Eq.~(\ref{eq:estimate}), one can show that there are only
four skeleton diagrams which have the $1/\epsilon$ singularity near four 
dimensions.  They are one-loop diagrams of the boson self-energy 
[Figs.~\ref{fig:cancel}(a) and \ref{fig:cancel}(c)], the $\varphi$ tadpole 
[Fig.~\ref{fig:cancel}(e)], and the vacuum diagram
[Fig.~\ref{fig:cancel}(g)].  We shall examine these apparent four
exceptions of naive power counting rule of $\eps$ one by one. 

The first diagram, Fig.~\ref{fig:cancel}(a), is the one-loop diagram
of the boson self-energy.  The frequency integral can be done
explicitly, yielding
\begin{align}\label{eq:Pi_a}
 -i\Pi_\mathrm{a}(p) &= -g^2\int\!\frac{dk}{(2\pi)^{d+1}}\, 
 G_{11}\!\left(k+\frac p2\right)G_{22}\!\left(k-\frac p2\right) \notag\\ 
 &= ig^2 \int\!\frac{d\k}{(2\pi)^d}\,
 \frac1{4E_{\k-\frac\p2}E_{\k+\frac\p2}} \\
 &\qquad \times\left[\frac{(E_{\k-\frac\p2}+\varepsilon_{\k-\frac\p2})
 (E_{\k+\frac\p2}+\varepsilon_{\k+\frac\p2})}
 {E_{\k-\frac\p2}+E_{\k+\frac\p2}-p_0-i\delta}
 +\frac{(E_{\k-\frac\p2}-\varepsilon_{\k-\frac\p2})
 (E_{\k+\frac\p2}-\varepsilon_{\k+\frac\p2})}
 {E_{\k-\frac\p2}+E_{\k+\frac\p2}+p_0-i\delta}\right]. \notag
\end{align}
The integral over $\k$ is ultraviolet divergent at $d=4$ and has a pole
at $\epsilon=0$.  Thus it is $O(1)$ by itself instead of 
$O(\epsilon)$ according
to the naive counting.  The residue at the pole is
\begin{equation}
 \begin{split}
  \Pi_\mathrm{a}(p) & = -g^2\int\!\frac{d\k}{(2\pi)^d}\,
  \left( 2\ek - p_0 + \frac{\ep}2 \right)^{-1} + \cdots\\
  &= -\left(p_0-\frac{\ep}2\right) + O(\epsilon),
 \end{split} 
\end{equation}
which is cancelled out exactly by adding the vertex $\Pi_0$ in
$\mathcal{L}_2$.  Therefore the diagram of the type in
Figs.~\ref{fig:cancel}(a), when combined with the vertex from
$\mathcal{L}_2$ in Fig.~\ref{fig:cancel}(b), conforms to the naive
$\epsilon$ power counting, i.e., is $O(\epsilon)$. 

Similarly, the diagram in Fig.~\ref{fig:cancel}(c) representing the boson
self-energy with one $\mu$ insertion is given by
\begin{equation}
  -i\Pi_\mathrm{c}(p)
  = 2\mu\,g^2\int\!\frac{dk}{(2\pi)^{d+1}}
  \left[G_{11}\!\left(k+\frac p2\right)^2
  -G_{12}\!\left(k+\frac p2\right)^2\right]
  G_{22}\!\left(k-\frac p2\right),
\end{equation}
which also contains a $1/\epsilon$ singularity, and is $O(\epsilon)$ instead
of the naive $O(\epsilon^2)$.  The leading part of this diagram is 
\begin{equation}
 \begin{split}
  \Pi_\mathrm{c}(p) &= -2\mu\,g^2\int\!\frac{d\k}{(2\pi)^d}\,
  \left( 2\ek \right)^{-2} + \cdots \\
  &= -2\mu + O(\epsilon^2),
 \end{split}
\end{equation}
and is cancelled out by the second vertex from $\mathcal{L}_2$. 
Then the sum of Figs.~\ref{fig:cancel}(c) and \ref{fig:cancel}(d) is again
$O(\epsilon^2)$, consistent with the naive power counting. 

The $\varphi$ tadpole diagram with one $\mu$ insertion
in Fig.~\ref{fig:cancel}(e) is
\begin{equation}
 \begin{split}
  \Xi_\mathrm{e} &= -\mu g\int\!\frac{dk}{(2\pi)^{d+1}}
  \left[G_{11}(k)-G_{22}(k)\right]G_{21}(k)\\
  &= ig\mu\phi_0\int\!\frac{d\k}{(2\pi)^d}\frac{\ek}{2E_k^{\,3}}
  =2i\frac{\mu\phi_0}g+O(\epsilon^{3/2}), 
 \end{split}
\end{equation}
which is $O(\epsilon^{1/2})$ instead of the naive $O(\epsilon^{3/2})$.  
This tadpole diagram should be cancelled by other tadpole diagrams of
order $O(\epsilon^{1/2})$ in Fig.~\ref{fig:cancel}(f), 
\begin{equation}
 \begin{split}
  \Xi_\mathrm{f} 
  &= g\int\!\frac{dk}{(2\pi)^{d+1}}G_{21}(k)-i\frac{g\phi_0}{c_0}\\
  &= -\frac{ig}2\left(\frac{m\phi_0}{2\pi}\right)^2-i\frac{g\phi_0}{c_0} 
  +O(\epsilon^{3/2}).
 \end{split}
\end{equation}
The condition of cancellation, $\Xi_\mathrm{e}+\Xi_\mathrm{f}=0$, gives
the gap equation to determine $\phi_0(\mu)$ to the leading order in
$\epsilon$.  The solution to the gap equation is
\begin{equation}
 \phi_0= \frac{2\mu}\eps
  -\frac2{c_0}\left(\frac{2\pi}m\right)^2 + O(\eps).
\end{equation}
When $c_0<0$, this solution can be written in terms of the binding
energy $\eb$ as $\phi_0=(2\mu+\eb)/\eps$.  Now the previously made
assumption $\mu/\phi_0=O(\epsilon)$ is checked.  The condition of
cancellation of tadpole diagrams is automatically satisfied by the
minimization of the effective potential as we will see later. 

Finally, the one-loop vacuum diagram with one $\mu$ insertion in
Fig.~\ref{fig:cancel}(g) also contains the $1/\epsilon$ singularity as 
\begin{equation}
 \begin{split}
  i\mu\int\!\frac{dk}{(2\pi)^{d+1}}\left[G_{11}(k)-G_{22}(k)\right]
  &=\mu\int\!\frac{d\k}{(2\pi)^d}\frac{\ek}{E_\k} \\
  &=-\frac\mu\epsilon\left(\frac{m\phi_0}{2\pi}\right)^2 +O(\epsilon). 
 \end{split}
\end{equation}
The leading part of this diagram is $O(1)$ instead of naive
$O(\epsilon)$ and can not be cancelled by any other diagrams.
Therefore, Fig.~\ref{fig:cancel}(g) $\sim O(1)$ is the only exception of
our naive power counting rule of $\epsilon$. 

Thus, we can now develop a diagrammatic technique for the Fermi gas near 
the unitarity limit where $g^2/|c_0|\sim\mu$ in terms of the systematic 
expansion of $\epsilon=4-d$.  The power counting rule of $\epsilon$ is
summarized as follow: 
\begin{enumerate}
 \item We consider $\mu/\phi_0\sim\eps$ and regard $\phi_0$ as $O(1)$.
 \item For any Green's function, we write down all Feynman diagrams
       according to the Feynman rules in Fig.~\ref{fig:feynman_rules}
       using the propagators from $\mathcal{L}_0$ and the vertices from 
       $\mathcal{L}_1$. 
 \item If there is any subdiagram of the type in Fig.~\ref{fig:cancel}(a) 
       or Fig.~\ref{fig:cancel}(c), we add a diagram where the subdiagram
       is replaced by a vertex from $\mathcal{L}_2$,
       Fig.~\ref{fig:cancel}(b) or Fig.~\ref{fig:cancel}(d).      
 \item The power of $\eps$ for the given Feynman diagram will be
       $O(\epsilon^{N_g/2+N_\mu})$, where $N_g$ is the number of
       couplings $g$ and $N_\mu$ is the number of chemical potential
       insertions.
 \item The only exception is the one-loop vacuum diagram with one $\mu$
       insertion in Fig.~\ref{fig:cancel}(g), which is $O(1)$ instead of
       the naive $O(\eps)$. 
\end{enumerate}
We note that the sum of all tadpole diagrams in Figs.~\ref{fig:cancel}(e)
and \ref{fig:cancel}(f) vanishes with the solution of the gap equation 
$\phi_0$.

\section{Effective potential to leading and next-to-leading orders}

\begin{figure}[tp]
 \begin{center}
  \includegraphics[width=0.6\textwidth,clip]{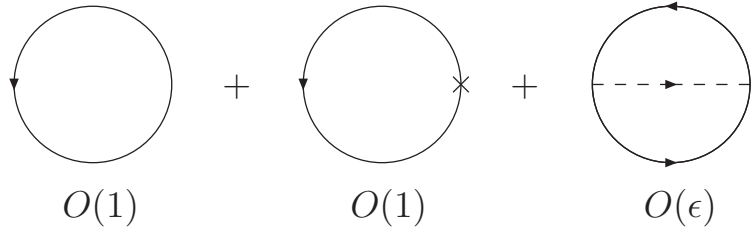}
  \caption{Vacuum diagrams contributing to the effective potential up to 
  the next-to-leading order in $\epsilon$. The second diagram is $O(1)$
  instead of naive $O(\epsilon)$ because of the $1/\epsilon$
  singularity.  \label{fig:potential}}
 \end{center}
\end{figure}

We now perform calculations to leading and next-to-leading orders in
$\eps$, employing the Feynman rules and the $\epsilon$ power counting
that have just been developed.  To find the dependence of $\phi_0$ on
$\mu$, we use the effective potential method~\cite{Peskin:1995ev} in
which this dependence follows from the the minimization of the effective 
potential $V_\mathrm{eff}(\phi_0)$.  
The effective potential at the tree level is given by
\begin{equation}\label{eq:V_0}
 V_0(\phi_0) = \frac{\phi_0^{\,2}}{c_0}. 
\end{equation}
Up to the next-to-leading order, the effective potential receives 
contributions from three vacuum diagrams drawn in
Fig.~\ref{fig:potential}: fermion loops with and without a $\mu$
insertion and a fermion loop with one boson exchange.  

The contributions from two one-loop diagrams are $O(1)$
and given by 
\begin{equation}
 \begin{split}
  V_1(\phi_0) &= i\int\!\frac{dp}{(2\pi)^{d+1}}
  \Tr\left[\ln G^{-1}(p) + \mu\sigma_3G(p)\right]\\
  &= -\int\!\frac{d\p}{(2\pi)^d}\left(E_\p-\mu\frac\ep{E_\p}\right).
 \end{split}
\end{equation}
By changing the integration variable to $z=(\ep/\phi_0)^2$ and with the
use of the formula
\begin{equation}
 \int_0^\infty\!dz\,\frac{z^{\alpha-1}}{(z+1)^\beta}
  =\frac{\Gamma(\alpha)\Gamma(\beta-\alpha)}{\Gamma(\beta)},
\end{equation}
we can perform the integration over $\p$ that results in
\begin{equation}
  V_1(\phi_0)= -\frac{\left(\frac{m\phi_0}{2\pi}\right)^{\frac d2}}
  {2\Gamma(\frac d2)}\left[\frac{\Gamma(\frac d4)
  \Gamma(-\frac12-\frac d4)}{\Gamma(-\frac12)}\phi_0
  -\frac{\Gamma(\frac d4+\frac12)\Gamma(-\frac d4)}{\Gamma(\frac12)}
  \mu\right].
\end{equation}
Substituting $d=4-\eps$ and expanding in terms of $\eps$ up to
$O(\eps)$, we obtain the effective potential from the one-loop diagrams
as 
\begin{equation}\label{eq:V_1}
  V_1(\phi_0) 
   = \frac{\phi_0}3\left[1+\frac{7-3(\gamma+\ln2)}6\epsilon\right] 
  \left(\frac{m\phi_0}{2\pi}\right)^{d/2}
  -\frac\mu\epsilon\left[1+\frac{1-2(\gamma-\ln2)}4\epsilon\right] 
  \left(\frac{m\phi_0}{2\pi}\right)^{d/2}\!+O(\eps^2),
\end{equation}
where $\gamma\approx0.57722$ is the Euler-Mascheroni constant.

The contribution of the two-loop diagram is $O(\eps)$ and given by
\begin{align}
  V_2(\phi_0) 
  & = g^2\int\!\frac{dp\,dq}{(2\pi)^{2d+2}}
  \Tr\left[G(p)\sigma_+G(q)\sigma_-\right]D(p-q) \notag\\
  & = g^2\int\!\frac{dp\,dq}{(2\pi)^{2d+2}}\,
  G_{11}(p)G_{22}(q)D(p-q).
\end{align}
Performing the integrations over $p_0$ and $q_0$, we obtain
\begin{align}
 V_2(\phi_0) 
 = -\frac{g^2}4\! \int\!\frac{d\p\,d\q}{(2\pi)^{2d}}\,
 \frac{(E_\p-\ep)(E_\q-\eq)}
 {E_\p E_\q (E_\p+E_\q+\varepsilon_{\p-\q}/2)}.
\end{align}
Since this is a next-to-leading order diagram and the integral converges
at $d=4$, we can evaluate it at $d=4$. 
Changing the integration variables to $x=\ep/\phi_0$,
$y=\eq/\phi_0$, and $\cos\theta=\hat\p\cdot\hat\q$, the integral can be
expressed by 
\begin{equation}\label{eq:V_2}
 V_2(\phi_0) = -\eps\left(\frac{m\phi_0}{2\pi}\right)^{d/2}
  \frac{\phi_0}\pi\int_0^\infty\!dx\int_0^\infty\!dy
  \int_0^\pi\!d\theta\,xy\sin^2\theta\,\frac{[f(x)-x][f(y)-y]}
  {f(x)f(y)\left[g(x,y)-\sqrt{xy}\cos\theta\right]}
\end{equation}
with $f(x)=\sqrt{x^2+1}$ and $g(x,y)=f(x)+f(y)+\frac12(x+y)$. The
integration over $\theta$ can be performed analytically to lead to
\begin{equation}
  V_2(\phi_0)=-C\epsilon\left(\frac{m\phi_0}{2\pi}\right)^{d/2}\phi_0,
\end{equation}
where the constant $C$ is given by a two-dimensional integral
\begin{equation}
  C = \int_0^\infty\!dx\int_0^\infty\!dy\, 
 \frac{[f(x)-x][f(y)-y]}{f(x)f(y)}
 \left[g(x,y) - \sqrt{g^2(x,y)-xy}\right].
\end{equation}
The numerical integrations over $x$ and $y$ result in
\begin{equation}\label{eq:C}
  C\approx 0.14424.  
\end{equation}

Now, gathering up Eqs.~(\ref{eq:V_0}), (\ref{eq:V_1}), and
(\ref{eq:V_2}), we obtain the effective potential up to the
next-to-leading order in $\eps$,
\begin{align}\label{eq:Veff}
 V_\mathrm{eff}(\phi_0)&=V_0(\phi_0)+V_1(\phi_0)+V_2(\phi_0)\\
 &= \frac{\phi_0^{\,2}}{c_0} + \left[\frac{\phi_0}3
 \left\{1+\frac{7-3(\gamma+\ln2)}6\epsilon-3C\epsilon\right\}
 -\frac\mu\epsilon\left\{1+\frac{1-2(\gamma-\ln2)}4\epsilon\right\}
 \right]\left(\frac{m\phi_0}{2\pi}\right)^{d/2} \notag\\
 &\qquad +O(\eps^2). \notag
\end{align}
The condition of the minimization of the effective potential in terms of
$\phi_0$ gives the gap equation; $\d V_\mathrm{eff}/\d\phi_0=0$.
The solution of the gap equation satisfies 
\begin{equation}
 \phi_0 = \frac{2\mu}\eps\left[1+\left(3C-1+\ln2\right)\eps\right]
  -\frac{2\phi_0^{\,\eps/2}}{c_0}\left(\frac{2\pi}{m}\right)^{d/2}
  \left[1+\left(3C-1+\frac{\gamma+\ln2}2\right)\eps\right]. 
\end{equation}
Using the relation of $c_0$ with the binding energy of boson $\eb$ in 
Eq.~(\ref{eq:eb}), we can rewrite the solution of the gap equation in
terms of $\eb$ up to the next-to-leading order in $\eps$ as
\begin{equation}\label{eq:phi_0}
 \phi_0 = \frac{2\mu}\eps\left[1+\left(3C-1+\ln2\right)\eps\right]
  +\frac\eb\eps \left[1+\left(3C-\frac12+\ln2
  -\frac12\ln\frac\eb{\phi_0}\right)\eps\right]. 
\end{equation}
We note that the leading term in Eq.~(\ref{eq:phi_0}) could be
reproduced using the mean field approximation, but the $O(\epsilon)$
corrections are not. 
The $O(\epsilon)$ correction proportional to $C$ is the result of the 
summation of fluctuations around the classical solution and is beyond 
the mean field approximation.

\section{Thermodynamic quantities near unitarity}
The value of the effective potential $V_\mathrm{eff}$ at its minimum
determines the pressure $P=-V_\mathrm{eff}(\phi_0)$ at a given chemical
potential $\mu$ and a given binding energy of boson $\eb$.  Substituting
the dependence of $\phi_0$ on $\mu$ and $\eb$ in Eq.~(\ref{eq:phi_0}),
we obtain the pressure as 
\begin{equation}\label{eq:pressure}
 P = \frac{\phi_0}6
  \left[ 1+ \left(\frac{17}{12}-3C-\frac{\gamma+\ln 2}2\right)\epsilon
   -\frac{3\eb}{4\phi_0}\right]\left(\frac{m\phi_0}{2\pi}\right)^{d/2}.
\end{equation}
The fermion number density $N$ is determined by differentiating the
pressure in terms of $\mu$ as
\begin{equation}\label{eq:N}
 N = \frac{\d P}{\d\mu} 
  = \frac1\epsilon \left[1+\frac{1-2\gamma+2\ln2}4\epsilon\right]
  \left(\frac{m\phi_0}{2\pi}\right)^{d/2}.
\end{equation}
Then the Fermi energy from the thermodynamic of free gas in $d$
spatial dimensions is given by
\begin{equation}\label{eq:eF}
  \eF = \frac{2\pi}m 
   \left[ \frac12\Gamma\left(\frac d2+1\right) N \right]^{2/d}
   = \frac{\phi_0}{\epsilon^{2/d}}\left(1-\frac{1-\ln2}4\epsilon\right).
\end{equation}
Note that non-trivial dependences on $\epsilon$ like $\sqrt\eps$ and
$\ln\eps$ appear by taking $N\sim\epsilon^{-1}$ to the power of $2/d$. 

From Eqs.~(\ref{eq:phi_0}) and (\ref{eq:eF}), we can determine the ratio
of the chemical potential and the Fermi energy $\mu/\eF$ near the
unitary limit as 
\begin{equation}\label{eq:mu}
  \frac\mu\eF =\frac{\eps^{3/2}}2 
  \exp\!\left(\frac{\epsilon\ln\epsilon}{8-2\epsilon}\right) 
  \left[1 - \left(3C -\frac54 (1-\ln2)\right)\eps\right]
  -\frac{\eb}{2\eF}
  \left[1+\frac\eps2+\frac{\eps\ln\eps}4-\frac\eps2\ln\frac\eb\eF\right].
\end{equation}
The logarithmic terms in the second line originates by introducing
$\phi_0=\eps^{1/2}\eF$ to the $\ln\eb/\phi_0$ term in
Eq.~(\ref{eq:phi_0}).  The first term in Eq.~(\ref{eq:mu}) gives the
universal parameter of the unitary Fermi gas
$\xi\equiv\left.\mu/\eF\right|_{\eb=0}$ as  
\begin{equation}\label{eq:xi}
 \begin{split}
  \xi &= \frac{\epsilon^{3/2}}2 \left[ 1 + \frac{\eps\ln\eps}8 
  - \left(3C -\frac54 (1-\ln2)\right)\epsilon\right]\\
  &= \frac12\epsilon^{3/2} + \frac1{16}\epsilon^{5/2}\ln\epsilon
  - 0.0246\,\epsilon^{5/2} + \cdots. 
 \end{split}
\end{equation}
Here we have substituted the numerical value for $C\approx0.14424$ in 
Eq.~(\ref{eq:C}). The smallness of the coefficient in front of
$\epsilon^{5/2}$ is a result of the cancellation between the two-loop 
correction and the subleading terms from the expansion of the one-loop
diagrams around $d=4$.  The $O(\eps^{7/2})$ correction to the
universal parameter $\xi$ was recently computed to find the large
correction $0.480\,\eps^{7/2}$~\cite{Arnold}. 
This may be related with the asymptotic nature of the $\eps$ expansion
and some sort of resummation will be necessary to go beyond the
next-to-leading order level.

Using Eqs.~(\ref{eq:pressure}), (\ref{eq:N}), and (\ref{eq:eF}), the
pressure near the unitarity limit normalized by $\eF N$ is given by
\begin{equation}\label{eq:P}
 \frac{P}{\eF N} = \frac2{d+2}\xi-\frac{\eb}{8\eF}\eps.
\end{equation}
Then the energy density $E=\mu N-P$ can be calculated from
Eqs.~(\ref{eq:mu}) and (\ref{eq:P}) as
\begin{equation}\label{eq:E}
 \frac{E}{\eF N} = \frac d{d+2}\xi
  -\frac{\eb}{2\eF}
  \left[1+\frac\eps4+\frac{\eps\ln\eps}4-\frac\eps2\ln\frac\eb\eF\right].
\end{equation}
The pressure and energy density in the unitarity limit are obtained 
from $\xi$ via the universal relations depending only on the
dimensionality of space.  Partial resummation of logarithmic terms of
the binding energy $\ln(\eb/\eF)$ change the exponent of $\eb/\eF$ to
$(\eb/\eF)^{1-\eps/2}$~\cite{Chen}.

\section{Quasiparticle spectrum \label{sec:spectrum}}
\begin{figure}[tp]
 \begin{center}
  \includegraphics[width=0.7\textwidth,clip]{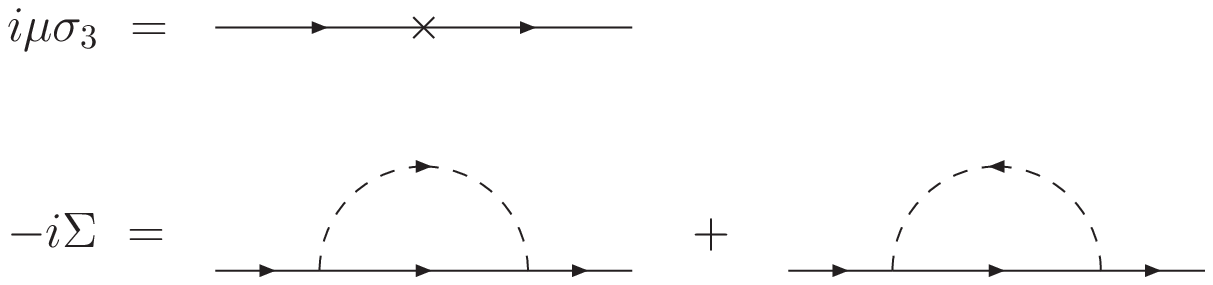}
  \caption{Corrections to the fermion self-energy of order $O(\epsilon)$;
  a $\mu$ insertion and one-loop diagrams. \label{fig:self_energy}}
 \end{center}
\end{figure}

The $\eps$ expansion we have developed is also useful for the 
calculations of physical observables other than the thermodynamic
quantities.  Here we shall look at the dispersion relation of fermion
quasiparticles.  To the leading order in $\eps$, the dispersion relation
is given by $\omega_\mathrm{F}(\p)=E_\p=\sqrt{\ep^{\,2}+\phi_0^{\,2}}$,
which has a minimum at zero momentum $\p=\bm0$ with the energy gap equal
to $\Delta=\phi_0=(2\mu+\eb)/\epsilon$.  The next-to-leading order
corrections to the dispersion relation come from three sources: from the 
correction of $\phi_0$ in Eq.~(\ref{eq:phi_0}), from the $\mu$ insertion 
to the fermion propagator, and from the one-loop self-energy diagrams,
$-i\Sigma(p)$, depicted in Fig.~\ref{fig:self_energy}.  

Using the Feynman rules, the one-loop diagram of the fermion self-energy
in Fig.~\ref{fig:self_energy} is given by 
\begin{equation}\label{eq:self-energy}
 -i\Sigma(p) = g^2\int\!\frac{dk}{(2\pi)^{d+1}}\,
  \left[\sigma_+G(k)\sigma_-D(p-k)+\sigma_-G(k)\sigma_+D(k-p)\right].
\end{equation}
There are corrections only to the diagonal elements of the self-energy
and each element is evaluated as
\begin{equation}\label{eq:Sigma11}
\begin{split}
  \Sigma_{11}(p) &= ig^2\int\!\frac{dk}{(2\pi)^{d+1}}\,
  G_{22}(k)D(p-k)\\
  &=-\frac{g^2}2\!\int\!\frac{d\k}{(2\pi)^d}\, \frac{E_\k-\ek}
  {E_\k (E_\k + \varepsilon_{\k-\p}/2-p_0 -i\delta)},
\end{split}
\end{equation}
and
\begin{equation}
\begin{split}
  \Sigma_{22}(p) &= ig^2\int\!\frac{dk}{(2\pi)^{d+1}}\,
  G_{11}(k)D(k-p)\\
  &=\frac{g^2}2\!\int\!\frac{d\k}{(2\pi)^d}\, \frac{E_\k-\ek}
  {E_\k (E_\k + \varepsilon_{\k-\p}/2+p_0 -i\delta)}.
\end{split}
\end{equation}

The dispersion relation of the fermion quasiparticle
$\omega_\mathrm{F}(\p)$ is obtained as a pole of the fermion propagator 
$\det[G^{-1}(\omega,\p)+\mu\sigma_3-\Sigma(\omega,\p)]=0$, which reduces
to the following equation: 
\begin{equation}\label{eq:dispersion}
 \begin{vmatrix}
   \omega-\ep+\mu-\Sigma_{11}(\omega,\p)&\phi_0\\
   \phi_0&\omega+\ep-\mu-\Sigma_{22}(\omega,\p)
 \end{vmatrix}=0.
\end{equation}
To find the $O(\eps)$ correction to the dispersion relation, we only
have to evaluate the self-energy $\Sigma(\omega,\p)$ with $\omega$
given by the leading order solution $\omega=\Ep$.  Denoting
$\Sigma_{11}(\Ep,\p)$ and $\Sigma_{22}(\Ep,\p)$ simply by $\Sigma_{11}$
and $\Sigma_{22}$ and solving Eq.~(\ref{eq:dispersion}) in terms of
$\omega$, we obtain the dispersion relation of the fermion quasiparticle
as
\begin{equation}\label{eq:omega_F}
 \omega_\mathrm{F}(\p)=\Ep+\frac{\Sigma_{11}+\Sigma_{22}}2
  +\frac{\Sigma_{11}-\Sigma_{22}-2\mu}{2\Ep}\ep +O(\eps^2). 
\end{equation}
Since the minimum of the dispersion relation will appear at small
momentum $\ep\sim\mu$, we can expand $\Sigma(\Ep,\p)$ around zero
momentum $\p=0$ as
$\Sigma(\Ep,\p)=\Sigma^0(\phi_0,\bm0)+\Sigma'(\phi_0,\bm0)\,\ep/\phi_0$
to find the energy gap of the fermion quasiparticles.  Performing the
integration over $\k$ in $\Sigma$ analytically, we have
\begin{align}
 \Sigma_{11}(\Ep,\p)&=\epsilon\left(2-8\ln3+8\ln2\right)\phi_0
  +\epsilon\left(-\frac83+8\ln3-8\ln2\right)\ep,
\intertext{and}
 \Sigma_{22}(\Ep,\p)&=\epsilon\left(-2-8\ln3+16\ln2\right)\phi_0
  +\epsilon\left(-\frac73-8\ln3+16\ln2\right)\ep.
\end{align}
Introducing these expressions into Eq.~(\ref{eq:omega_F}), we find 
the fermion dispersion relation around its minimum has the following
form
\begin{equation}\label{eq:omega}
 \omega_\mathrm{F}(\p)
  \simeq\Delta+\frac{(\ep-\varepsilon_0)^2}{2\phi_0} 
  \simeq\sqrt{(\ep-\varepsilon_0)^2+\Delta^2}. 
\end{equation}

\begin{figure}[tp]
 \begin{center}
  \includegraphics[scale=1.2,clip]{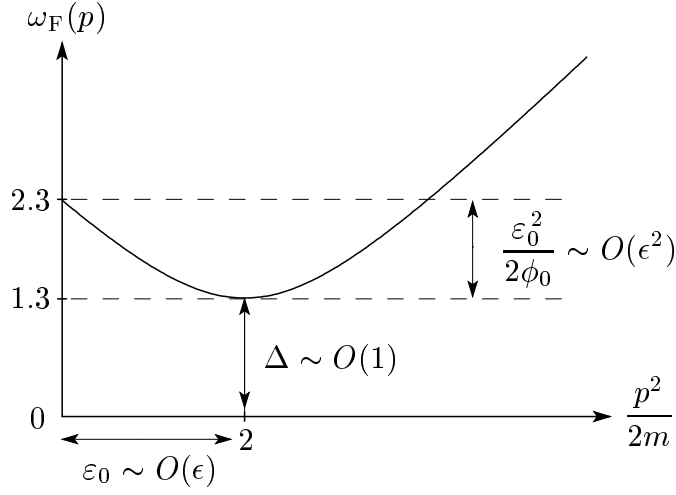}
  \caption{Illustration of the dispersion relation of the fermion
  quasiparticle in the unitarity limit from the $\eps$ expansion. 
  $\omega_\mathrm{F}(\p)=\sqrt{(\ep-\varepsilon_0)^2+\Delta^2}$ in
  Eq.~(\ref{eq:omega}) is plotted as a function of
  $\ep=p^2/2m$.  Values on both axes are extrapolated results to $\eps=1$
  in units of the chemical potential $\mu$. \label{fig:dispersion}}  
 \end{center}
\end{figure}

Here $\Delta$ is the energy gap of the fermion quasiparticle, which is
given by 
\begin{equation}\label{eq:gap}
 \begin{split}
  \Delta &= \phi_0 + \frac{\Sigma_{11}^0+\Sigma_{22}^0}2 \\
  &= \left[1-\left(8\ln3-12\ln2\right)\eps+O(\eps^2)\right]\phi_0.
 \end{split}
\end{equation}
The minimum of the dispersion curve is located at a nonzero value of
momentum, $|\p|=(2m\varepsilon_0)^{1/2}$, where
\begin{equation}\label{eq:loc}
 \begin{split}
  \varepsilon_0 &= \mu + \frac{\Sigma_{22}^0-\Sigma_{11}^0}2 
  - \frac{\Sigma_{11}'+\Sigma_{22}'}2 \\
  &= \mu + \frac{\eps\phi_0}2 +O(\eps^2).
 \end{split}
\end{equation}
Then, introducing the solution of the gap equation (\ref{eq:phi_0}), 
the energy gap $\Delta$ as a function of the chemical potential and the
binding energy is given by
\begin{align}\label{eq:gap_mu}
 \Delta &= \frac{2\mu}\eps\left[1+(3C-1-8\ln3+13\ln2)\eps\right]
 +\frac\eb\eps\left[1+\left(3C-\frac12-8\ln3+13\ln2
 -\frac12\ln\frac\eb{\Delta}\right)\eps\right] \notag\\
 &= \frac{2\mu}\eps\left(1-0.345\,\eps\right)  
 + \frac\eb\eps\left(1+0.155\,\eps-\frac{\eps\ln\eps}2\right) 
 +\frac\eb2\ln\!\left(1+\frac{2\mu}\eb\right), 
\end{align}
while $\varepsilon_0$ is given by
\begin{equation}\label{eq:e_0}
 \varepsilon_0=2\mu+\frac\eb2.
\end{equation}
Note the difference with the mean field approximation, in which
$\varepsilon_0=\mu$.  When $\varepsilon_0$ is positive, the fermion
dispersion curve has its minimum at nonzero value of momentum as in the
BCS limit, while the minimum is located at zero momentum when
$\varepsilon_0$ is negative as in the BEC limit.  We find the former
$(\varepsilon_0=2\mu>0)$ is the case in the unitarity limit.  The
dispersion curve of the fermion quasiparticle $\omega_\mathrm{F}(\p)$ in
the unitarity limit $\eb=0$ is illustrated Fig~\ref{fig:dispersion}.  
In particular, the difference between the fermion quasiparticle energy
at zero momentum and at its minimum is given by
$\varepsilon_0^{\,2}/2\phi_0=\eps\mu\sim O(\eps^2)$.

\section{Location of the splitting point}
Let us consider the situation where the binding energy $\eb$ is
increasing from zero while the number density $N$ is kept fixed.  Then 
$\Delta>0$ is held fixed but $\mu$ is changing.  Since to the leading 
order in $\eps$, the chemical potential as a function of the energy gap 
is given by $2\mu=\eps\Delta-\eb$, the location of the minimum of the 
dispersion curve can be written as 
\begin{equation}\label{eq:e0}
 \varepsilon_0=\eps\Delta-\frac\eb2.
\end{equation}
Therefore, $\varepsilon_0$ decreases as $\eb$ increases. 
When the binding energy reaches $\eb=2\eps\Delta$, the minimum of the
dispersion curve sits exactly at zero momentum $\p=\bm0$.  This point is
referred to as a \textit{splitting point}~\cite{son05}.
We find the splitting point is located at the BEC side of the unitarity
limit where the binding energy is positive $\eb=2\eps\Delta>0$ and the 
chemical potential is negative $2\mu=-\eps\Delta$.
This splitting point will play an important role to determine the phase 
structure of the polarized Fermi gas in the unitary regime as we will
study in Chap.~\ref{sec:polarization}.

\section{Momentum distribution function}
Other interesting observables which can be measured in experiments are a
momentum distribution function of fermion
quasiparticles~\cite{Virerit,Tan,Regal-momentum,Astrakharchik} and a 
condensate fraction in the fermion
density~\cite{Regal04,Zwierlein04,Astrakharchik,Ortiz,Salasnich}.  The
momentum distribution functions, $n_{\p\uparrow}$ and
$n_{\p\downarrow}$, can be computed from the fermion propagator
$\mathcal G(p)$ through
\begin{equation}
 n_{\p\uparrow}=\int\!\frac{dp_0}{2\pi i}\,
  e^{ip_00^+}\mathcal G_{11}(p_0,\p)
  \quad\qquad\text{and}\qquad\quad
  n_{\p\downarrow}=-\int\!\frac{dp_0}{2\pi i}\,
  e^{-ip_00^+}\mathcal G_{22}(p_0,\p),
\end{equation}
where the fermion propagator up to the order $\eps$ is given by
\begin{equation}\label{eq:full_G}
 \mathcal G(p)=G(p)
  +G(p)\left[\Sigma(p)-\mu\sigma_3\right]G(p)+O(\eps^2). 
\end{equation}
With the use of the definition of the bare fermion propagator $G(p)$ in
Eq.~(\ref{eq:G}) and the one-loop fermion self-energy $\Sigma(p)$ in
Eq.~(\ref{eq:self-energy}), the diagonal elements of the fermion
propagator becomes
\begin{equation}\label{eq:full_G11}
 \begin{split}
  \mathcal G_{11}(p)
  &=G_{11}(p)+G_{11}(p)\left[\Sigma_{11}(p)-\mu\right]G_{11}(p)
  +G_{12}(p)\left[\Sigma_{22}(p)+\mu\right]G_{21}(p) \\
  &=\frac{p_0+\ep}{p_0^{\,2}-E_\p^{\,2}+i\delta}
  +\left(\frac{p_0+\ep}{p_0^{\,2}-E_\p^{\,2}+i\delta}\right)^2
  \left[\Sigma_{11}(p)-\mu\right]
  +\left(\frac{\phi_0}{p_0^{\,2}-E_\p^{\,2}+i\delta}\right)^2
  \left[\Sigma_{22}(p)+\mu\right]
 \end{split}
\end{equation}
and $\mathcal G_{22}(p_0,\p)=-\mathcal G_{11}(-p_0,\p)$.  Since we are
considering a symmetric case with $\mu_\uparrow=\mu_\downarrow$ and
$m_\uparrow=m_\downarrow$, we can confirm the relationship
$n_\p=n_{\p\uparrow}=n_{\p\downarrow}$. 

From the first term in Eq.~(\ref{eq:full_G11}), the $O(1)$ part of the
momentum distribution function $n_\p^{(0)}$ is easily obtained as
\begin{equation}
 n_\p^{(0)}=\int\!\frac{dp_0}{2\pi i}\,e^{ip_00^+}
  \frac{p_0+\ep}{p_0^{\,2}-E_\p^{\,2}+i\delta}=\frac{E_\p-\ep}{2E_\p}.
\end{equation}
The last two terms in Eq.~(\ref{eq:full_G11}) contribute to the momentum
distribution function to the order of $\eps$.  Since
$\Sigma_{11}(p_0,\p)$ in Eq.~(\ref{eq:Sigma11}) is analytic at
$\mathrm{Im}[p_0]>0$, the $p_0$ integral of the second term can be
evaluated as 
\begin{equation}
 \begin{split}
  \int\!\frac{dp_0}{2\pi i}\,
  \left(\frac{p_0+\ep}{p_0^{\,2}-E_\p^{\,2}+i\delta}\right)^2
  \left[\Sigma_{11}(p)-\mu\right] &=\left.\frac{\d}{\d p_0}
  \left[\left(\frac{p_0+\ep}{p_0-E_\p}\right)^2
  \left[\Sigma_{11}(p)-\mu\right]\right]\,\right|_{p_0=-\Ep} \\
  &=-\frac{\phi_0^{\,2}}{4E_\p^{\,3}}\left[\Sigma_{11}(-\Ep,\p)-\mu\right]
  +\frac{(E_p-\ep)^2}{4E_\p^{\,2}}\Sigma'_{11}(-\Ep,\p). 
 \end{split}
\end{equation}
The prime in $\Sigma'_{11}(p_0,\p)$ represents the derivative with
respect to $p_0$.  Similarly, using the relationship
$\Sigma_{22}(p_0,\p)=-\Sigma_{11}(-p_0,\p)$, the $p_0$ integral of the
third term in Eq.~(\ref{eq:full_G11}) can be evaluated as
\begin{equation}
 \begin{split}
  \int\!\frac{dp_0}{2\pi i}\,
  \left(\frac{\phi_0}{p_0^{\,2}-E_\p^{\,2}+i\delta}\right)^2
  \left[\Sigma_{22}(p)+\mu\right] &=-\int\!\frac{dp_0}{2\pi i}\,
  \left(\frac{\phi_0}{p_0^{\,2}-E_\p^{\,2}+i\delta}\right)^2
  \left[\Sigma_{11}(p)-\mu\right] \\ &=\left.-\frac{\d}{\d p_0}
  \left[\left(\frac{\phi_0}{p_0-E_\p}\right)^2
  \left[\Sigma_{11}(p)-\mu\right]\right]\,\right|_{p_0=-\Ep} \\
  &=-\frac{\phi_0^{\,2}}{4E_\p^{\,3}}\left[\Sigma_{11}(-\Ep,\p)-\mu\right]
  -\frac{\phi_0^{\,2}}{4E_\p^{\,2}}\Sigma'_{11}(-\Ep,\p).
 \end{split}
\end{equation}
From these two contributions, we obtain the next-to-leading order
correction to the momentum distribution function as
\begin{equation}
 n_\p^{(1)}=\frac{\phi_0^{\,2}}{2\Ep^{\,3}}\mu
  -\frac{\phi_0^{\,2}}{2\Ep^{\,3}}\Sigma_{11}(-\Ep,\p)
  -\frac{\ep(\Ep-\ep)}{2\Ep^{\,2}}\Sigma'_{11}(-\Ep,\p).
\end{equation}
Substituting the solution of the gap equation in the unitarity limit 
$\mu=\eps\phi_0/2+O(\eps^2)$ and the expression for $\Sigma_{11}(p)$ in
Eq.~(\ref{eq:Sigma11}), the $O(\eps)$ part of the momentum distribution
function $n_\p^{(1)}$ is given by
\begin{equation}
 n_\p^{(1)} = \eps\frac{\phi_0^{\,3}}{4\Ep^{\,3}}
  +\frac{g^2}4\!\int\!\frac{d\k}{(2\pi)^4}\,
  \frac{(\Ek-\ek)}{\Ep\Ek(\Ep+\Ek+\varepsilon_{\k-\p}/2)}
  \left[\frac{\phi_0^{\,2}}{\Ep^{\,2}}
  +\frac{\ep(\Ep-\ep)}{\Ep(\Ek+\varepsilon_{\k-\p}/2+\Ep)}\right].
\end{equation}

\begin{figure}[tp]
 \begin{center}
  \includegraphics[width=0.7\textwidth,clip]{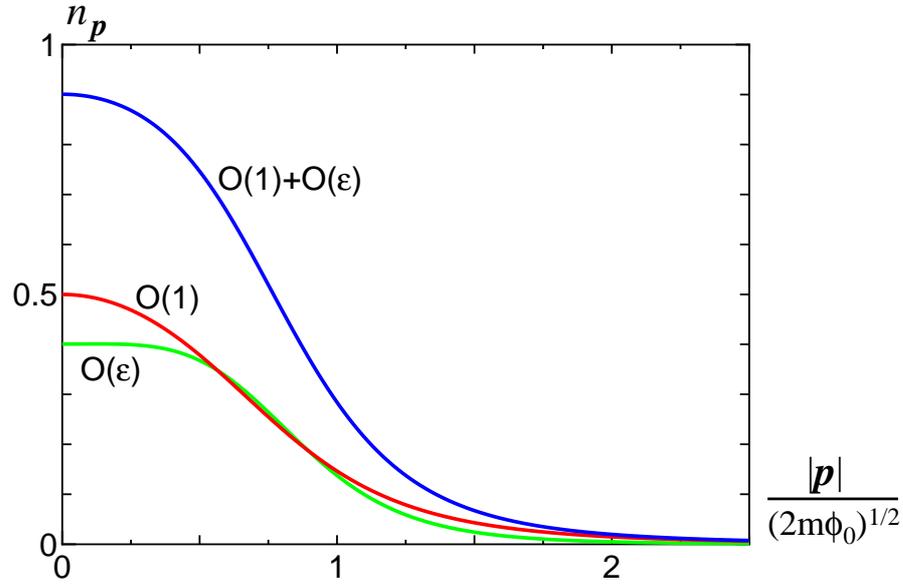}
  \caption{Momentum distribution function $n_\p$ of the fermion
  quasiparticles in the unitarity limit as a function of the momentum
  $|\p|/\sqrt{2m\phi_0}$ to the leading and next-to-leading orders in
  $\eps$.  The $O(1)$ part $n_\p^{(0)}$ (red curve), the $O(\eps)$ part
  $n_\p^{(1)}$ (green curve), and their sum $n_\p=n_\p^{(0)}+n_\p^{(1)}$
  (blue curve) at $\eps\to1$ are respectively shown.
  \label{fig:distribution}}
 \end{center}
\end{figure}

The momentum distribution function of the fermion quasiparticles
$n_\p=n_\p^{(0)}+n_\p^{(1)}$ to the leading and next-to-leading orders
in $\eps$ is plotted as a function of the momentum
$|\p|/\sqrt{2m\phi_0}$ in Fig.~\ref{fig:distribution}.  Since the
condensate $\phi_0$ is related to the Fermi energy $\eF=\kF^{\,2}/2m$ by
Eq.~(\ref{eq:eF}), 
\begin{equation}
  \eF = \frac{\phi_0}{\eps^{2/d}}\left(1-\frac{1-\ln2}4\eps\right), 
\end{equation}
the Fermi momentum in the horizontal axis of Fig.~\ref{fig:distribution}
corresponds to
\begin{equation}
 \frac{\kF}{\sqrt{2m\phi_0}}=\eps^{-1/d}\left(1-\frac{1-\ln2}8\eps\right)
  \approx0.962\qquad(\eps\to1).  
\end{equation}
We can find overall agreement of the momentum distribution function
computed by the $\eps$ expansion up to the next-to-leading order in
$\eps$ with that from the Monte Carlo simulation~\cite{Astrakharchik}.

\section{Condensate fraction}
The fermion density in the condensation $N_0$ is related with the
off-diagonal element of the fermion propagator $\mathcal G_{12}(p)$
through~\cite{Astrakharchik,Ortiz,Salasnich,Yang}
\begin{equation}\label{eq:N0}
 N_0=2\int\!\frac{d\p}{(2\pi)^{d}}
  \left[\int\!\frac{dp_0}{2\pi i}\,\mathcal G_{12}(p_0,\p)\right]^2.
\end{equation}
Since $\mathcal G_{12}(p)$ up to the order $\eps$ is given from
Eq.~(\ref{eq:full_G}) as 
\begin{equation}\label{eq:full_G12}
 \begin{split}
  \mathcal G_{12}(p)
  &=G_{12}(p)+G_{11}(p)\left[\Sigma_{11}(p)-\mu\right]G_{12}(p)
  +G_{12}(p)\left[\Sigma_{22}(p)+\mu\right]G_{22}(p) \\
  &=-\frac{\phi_0}{p_0^{\,2}-E_\p^{\,2}+i\delta}
  -\frac{\phi_0\left(p_0+\ep\right)}
  {\left(p_0^{\,2}-E_\p^{\,2}+i\delta\right)^2}
  \left[\Sigma_{11}(p)-\mu\right]
  -\frac{\phi_0\left(p_0-\ep\right)}
  {\left(p_0^{\,2}-E_\p^{\,2}+i\delta\right)^2}
  \left[\Sigma_{22}(p)+\mu\right],
 \end{split}
\end{equation}
the $p_0$ integration of each terms results in
\begin{equation}
 -\int\!\frac{dp_0}{2\pi i}\,\frac{\phi_0}{p_0^{\,2}-E_\p^{\,2}+i\delta}
  =\frac{\phi_0}{2\Ep}
\end{equation}
and
\begin{align}
  -\int\!\frac{dp_0}{2\pi i}\,\frac{\phi_0\left(p_0+\ep\right)}
  {\left(p_0^{\,2}-E_\p^{\,2}+i\delta\right)^2}
  \left[\Sigma_{11}(p)-\mu\right]
  &=-\int\!\frac{dp_0}{2\pi i}\,\frac{\phi_0\left(p_0-\ep\right)}
  {\left(p_0^{\,2}-E_\p^{\,2}+i\delta\right)^2}
  \left[\Sigma_{22}(p)+\mu\right] \notag\\
  &=\left.-\phi_0\frac{\d}{\d p_0}
  \left[\frac{p_0+\ep}{\left(p_0-E_\p\right)^2}
  \left[\Sigma_{11}(p)-\mu\right]\right]\,\right|_{p_0=-\Ep} \\
  &=-\phi_0\frac{\ep}{4E_\p^{\,3}}\left[\Sigma_{11}(-\Ep,\p)-\mu\right]
  +\phi_0\frac{E_p-\ep}{4E_\p^{\,2}}\Sigma'_{11}(-\Ep,\p). \notag
\end{align}
Here we used the relationship
$\Sigma_{22}(p_0,\p)=-\Sigma_{11}(-p_0,\p)$ and the fact that
$\Sigma_{11}(p_0,\p)$ in Eq.~(\ref{eq:Sigma11}) is analytic at
$\mathrm{Im}[p_0]>0$. 

Therefore, the fermion density in the condensation $N_0$ up to the order
$\eps$ is given by
\begin{equation}
 N_0=\phi_0^{\,2}\int\!\frac{d\p}{(2\pi)^{d}}
  \left[\frac{1}{2\Ep^{\,2}}+\frac{\ep}{E_\p^{\,4}}\mu
   -\frac{\ep}{E_\p^{\,4}}\Sigma_{11}(-\Ep,\p)
   +\frac{E_p-\ep}{E_\p^{\,3}}\Sigma'_{11}(-\Ep,\p)\right].
\end{equation}
The first term provides the leading contribution to $N_0$, while the
other three terms are $O(\eps)$ corrections.  
Substituting the solution of the gap equation in the unitarity limit 
$\mu=\eps\phi_0/2+O(\eps^2)$ and performing the integration over $\p$,
we obtain $N_0$ as 
\begin{equation}
 N_0
  =\left[\frac{\Gamma\left(1-\frac\eps4\right)\Gamma\left(\frac\eps4\right)}
    {4\Gamma\left(2-\frac\eps2\right)}+\frac\pi8\eps+0.127741\,\eps
    +O(\eps^2)\right]\left(\frac{m\phi_0}{2\pi}\right)^{\frac d2}.
\end{equation}
Note that because the $\p$ integration of the leading term is
logarithmically divergent at $d=4$, the leading contribution to $N_0$
becomes $O(1/\eps)$ at $d=4-\eps$. 
The fermion density in the condensation should be compared to the total
fermion density $N$ to the same order in $\eps$~\cite{Arnold}:
\begin{equation}
 N=\left[\frac{\Gamma\left(\frac32-\frac\eps4\right)
    \Gamma\left(1+\frac\eps4\right)}
    {2\sqrt\pi\,\Gamma\left(2-\frac\eps2\right)}+\frac12\eps+0.258352\,\eps
    +O(\eps^2)\right]\left(\frac{m\phi_0}{2\pi}\right)^{\frac d2}.
\end{equation}
Taking the ratio of $N_0$ to $N$, we find the condensate fraction
$N_0/N$ to be
\begin{equation}\label{eq:fraction}
 \frac{N_0}N=1-0.0966\,\eps-0.2423\,\eps^2+O(\eps^3).
\end{equation}
The condensate fraction can take a value from zero (BCS limit) to one
(BEC limit).  The lowest term in the $\eps$ expansion is one as in the
BEC limit and thus all fermion are in the condensation at four spatial
dimensions.  Below $d=4$, the condensate fraction decreases and the
naive extrapolation to $\eps=1$ gives $N_0/N\approx0.661$.

\section{Extrapolation to $\epsilon=1$}
Finally, we discuss the extrapolation of the series expansion to the
physical case at three spatial dimensions.  
Although the formalism is based on the smallness of $\epsilon$, we see
that the corrections are reasonably small even at $\epsilon=1$.  If we
naively use only the leading and next-to-leading order results for $\xi$ 
in Eq.(\ref{eq:xi}), $\Delta$ in Eq.(\ref{eq:gap_mu}), 
$\varepsilon_0$ in Eq.(\ref{eq:e_0}), and $N_0/N$ in
Eq.~(\ref{eq:fraction}) in the unitarity limit, their extrapolations to
$\epsilon=1$ give for three spatial dimensions
\begin{equation}\label{eq:extrapolation}
 \xi \approx 0.475, \qquad \frac{\Delta}{\mu}\approx 1.31, 
  \qquad \frac{\varepsilon_0}{\mu}\approx 2, 
  \qquad \frac{N_0}N\approx0.661.  
\end{equation}
They are reasonably close to the results of recent Monte Carlo 
simulations, which yield $\xi\approx0.42$, $\Delta/\mu\approx1.2$, 
$\varepsilon_0/\mu\approx1.9$~\cite{Carlson:2005kg}, and
$N_0/N\approx0.58$~\cite{Astrakharchik}.  They are also consistent with
recent experimental measurements of $\xi$, where
$\xi=0.51\pm0.04$~\cite{Kinast05} and
$\xi=0.46\pm0.05$~\cite{Partridge06}.  These agreements can be taken as
a strong indication that the $\epsilon$ expansion is useful even at
$\epsilon=1$.  In Table~\ref{tab:comparison}, we show results on $\xi$,
$\Delta/\eF$, and $N_0/N$ from other analytic and numerical
calculations.  The self-consistent approach~\cite{Zwerger} quoted in
Table~\ref{tab:comparison} and Table~\ref{tab:comparison_Tc} is based on
the Luttinger--Ward and DeDominicis--Martin
formalism~\cite{Luttinger-Ward,DeDominicis-Martin} where the potential
functional is self-consistently approximated by ladder
diagrams~\cite{Baym-Kadanoff,Baym}.  We can see some improvement of our
results compared to naive mean-field approximations and other analytic
approaches.

\begin{table}[tp]
 \begin{center}
  \begin{tabular}{l|l|l|l}
   & $\xi=\mu/\eF$ & $\Delta/\mu$ & $N_0/N$ \\\hline
   mean-field approximation & $0.5906$ & $1.1622$ &
   $0.6994$~\cite{Salasnich}\\ 
   $\eps$ expansion (NLO) & $0.475$ & $1.31$ & $0.661$\\
   $1/N$ expansion (NLO)~\cite{Veillette} & $0.279$ & $1.50$ & --- \\
   self-consistent approach~\cite{Zwerger} & $0.360$ & --- & --- \\
   Monte Carlo simulation & $0.42$~\cite{Carlson:2005kg} &
   $1.2$~\cite{Carlson:2005kg} & $0.58$~\cite{Astrakharchik}
  \end{tabular}
  \caption{Comparison of the results by the $\eps$ expansion with other
  analytic and numerical calculations in the unitarity limit.
  \label{tab:comparison}}
 \end{center}
\end{table}

Our $\eps$ expansion predicts the behavior of the thermodynamic
quantities near the unitarity limit.  From Eqs.~(\ref{eq:P}),
(\ref{eq:E}), and (\ref{eq:mu}), the extrapolations to the three spatial 
dimensions $\epsilon=1$ lead to
\begin{align}
 \frac{P}{\eF N} &\approx \frac2{d+2}\,\xi -\frac18\frac\eb\eF, \\
 \frac{E}{\eF N} &\approx \frac{d}{d+2}\,\xi 
 - \frac14\frac\eb\eF\left(\frac52-\ln\frac\eb\eF\right), 
 \phantom{\frac{\frac\int\int}{\frac\int\int}}\hspace{-4.5mm} \\ 
 \frac{\mu}{\eF} &\approx \xi 
 - \frac14\frac\eb\eF\left(3-\ln\frac\eb\eF\right).
\end{align}
The pressure, energy density, and chemical potential at fixed density
are decreasing function of the binding energy $\eb$ with the slope shown
above.  
In Chap.~\ref{sec:matching}, we will make a discussion to improve the
extrapolation of the series expansion by imposing the exact result at
$d=2$ as a boundary condition. 

We can also try to determine the location of the splitting point.  At
this point,
\begin{equation}
  \frac\eb\eF = \frac2{\eps^{2/d-1}}[1+O(\eps)] \to 2 \qquad
  \text{at}\ \  d=3
\end{equation}
and comparing with Eq.~(\ref{eq:ebeF}), one finds that at the splitting
point $a\kF\approx 1$.  Since this result is known only to leading order 
in $\eps$, one must be cautious with the numerical value.  However,
certain qualitative features are probably correct: the splitting point
is located on the BEC side of the unitarity limit ($a>0$) and that the
chemical potential $\mu/\eF\approx-0.5$ is negative at this point.


\chapter{Phase structure of polarized Fermi gas \label{sec:polarization}}

\section{Background and proposed phase diagram}
Here we apply the $\eps$ expansion developed in the previous Chapter to
the unitary Fermi gas with unequal densities of two components
(polarization).  We put special emphasis on investigating the phase
structure of the polarized Fermi gas in the unitary regime, which has a
direct relation with the recent measurements in atomic
traps~\cite{Zwierlein05,Partridge06,Zwierlein06,Shin06,Partridge06-2}.
Furthermore, it may be possible in future to realize the Feshbach
resonances between two different species of fermionic atoms to study a
Fermi gas with finite mass difference between different fermion species.
Such asymmetric systems of fermions with density and mass imbalances
will be also interesting as a prototype of high density quark matter in
the core of neutron stars, where the density and mass asymmetries exist
among different quark
flavors~\cite{Alford99,Shovkovy,Huang,Gubankova03,Alford03}.  Extension
of the $\eps$ expansion to the polarized Fermi gas with unequal masses
will be studied in the next Chapter and we concentrate on the equal mass
case here.

\begin{figure}[tp]
 \begin{center}
  \includegraphics[width=0.65\textwidth,clip]{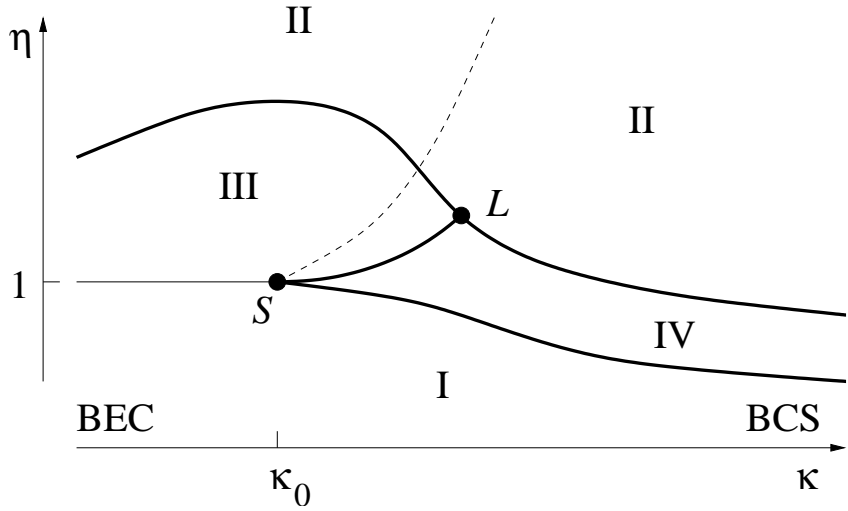}
  \caption{Proposed phase diagram in the plane of the diluteness
  parameter $\kappa=-1/a\kF$ and the polarization chemical potential
  $\eta=H/\Delta_{H=0}$ by Son and Stephanov~\cite{son05}.  I is the
  unpolarized BEC/BCS phase, II is the normal phase, III is the gapless
  superfluid phase, and IV is a region of
  Fulde-Ferrell-Larkin-Ovchinikov phases~\cite{FF,LO}.  The dashed line
  divides phases II and III into regions with one (on the left) and two
  (on the right) Fermi surfaces.  Region IV must be divided into phases
  with different patterns of breaking of the rotational symmetry (not
  shown).  \label{fig:son-stephanov}}
 \end{center}
\end{figure}

Reliable facts on the phase structure of the polarized Fermi gas
is known only in the weak-coupling BCS and BEC limits.  If the
polarization chemical potential $H=(\mu_\uparrow-\mu_\downarrow)/2$ is
sufficiently small, the ground state is the unpolarized superfluid state
both in the BCS and BEC limits and these two limits are considered to be
smoothly connected as a function of $1/a\kF$ (phase I in
Fig.~\ref{fig:son-stephanov}).  When $H$ is increased, we will have
different situations in the two limits.  In the BCS limit, the BCS
superfluid state becomes unstable toward to the
Fulde-Ferrell-Larkin-Ovchinikov (FFLO) states where Cooper pairs form
with a nonzero momentum and the superfluid order parameter varies in
space~\cite{FF,LO} (phase IV).  For the sufficiently large $H$, the
Cooper paring is unfavorable and the system goes to a polarized normal
Fermi gas (phase II).  On the other hand, when $H$ is increased in the
BEC limit, unbound fermions are created on the top of the molecular BEC
ground state.  The system becomes a homogeneous mixture of the polarized
fermions with the condensed molecules (phase III).  For the sufficiently
large $H$, all bound molecules disappear and the ground state is a fully
polarized normal Fermi gas (phase II).  Since the FFLO phase where the
rotational symmetry is spontaneously broken does not appear in the BEC
limit, such a phase should terminates somewhere between the BCS and BEC
limits. 

From these knowledge in the two limits and assuming there is a phase
transition line between the phases III and IV instead of I and II, Son
and Stephanov proposed a global phase diagram of the polarized Fermi gas
in the BCS-BEC crossover as shown in
Fig.~\ref{fig:son-stephanov}~\cite{son05}.  The point $S$ where the FFLO 
phase terminates is called the \textit{splitting point} and the phase
structure around it is studied based on the effective field
theory~\cite{son05}.  The purpose of this Chapter is to study the phase
structure of the polarized Fermi gas in the unitary regime including the
splitting point from the microscopic point of view using the $\eps$
expansion. 

\section{Effective potential at finite superfluid velocity}
In order to take into account the possibility of the FFLO state in the
phase diagram, we generalize our formalism to allow a spatially varying
condensate where $\<\phi(x)\>=e^{2im\vs\cdot\bm x}\phi_0$ with $\vs$
being the superfluid velocity.  The factor $e^{2im\vs\cdot\bm x}$ in
front of $\phi_0$ in the Lagrangian density can be absorbed by making
the Galilean transformation on the fermion field as
$\psi_\sigma(x)\to e^{im\vs\cdot\bm x}\psi_\sigma(x)$ and the boson
field as $\varphi(x)\to e^{2im\vs\cdot\bm x}\varphi(x)$. Accordingly,
the fermion propagator in Eq.~(\ref{eq:G}) is modified as
\begin{equation}\label{eq:G_vs}
 \begin{split}
  G(p_0,\p) &= \frac1{(p_0+H-\p\cdot\vs)^2
  -(\ep+\varepsilon_{m\vs})^2-\phi_0^{\,2}+i\delta} \\ 
  &\qquad\qquad\qquad\qquad\times
  \begin{pmatrix}
   p_0 + H + \varepsilon_{\p-m\vs} & -\phi_0 \\
   -\phi_0 & p_0+ H - \varepsilon_{\p+m\vs}
  \end{pmatrix},
 \end{split}
\end{equation}
while the boson propagator in Eq.~(\ref{eq:D}) as
\begin{equation}
 D(p_0,\p) = \left(p_0 - \frac{\varepsilon_{\p+2m\vs}}2 
	      + i\delta\right)^{-1}.
\end{equation}

The value of the superfluid velocity $\vs$ is determined from
minimizing the $\vs$ dependent part of the effective potential $V_H(\vs)$.  
As far as the polarization $H$ is sufficiently small compared to the
energy gap $\Delta$ of the fermion quasiparticle, $V_H(\vs)$ to the
leading order in $\eps$ is given by the fermion one-loop diagram with
the propagator in Eq.(\ref{eq:G_vs}): 
\begin{equation}
 V_{H<\Delta}(\vs)=-\int\!\frac{d\p}{(2\pi)^d}
  \sqrt{\left(\ep+\ev\right)^2+\phi_0^{\,2}}.
\end{equation}
Since it will turn out that $\ev$ is $O(\eps^{11})$, we can expand the 
integrand in terms of $\ev/\phi_0$ to lead to 
\begin{equation}
 V_{H<\Delta}(\vs)=-\int\!\frac{d\p}{(2\pi)^d}\frac\eb{E_\p}\ev
  \simeq\frac{\ev}\eps\left(\frac{m\phi_0}{2\pi}\right)^2.
\end{equation}
Using the definition of the fermion number density $N$ in
Eq.~(\ref{eq:N}), this part can be rewritten as
$V_{H<\Delta}(\vs)=N\ev$, which represents the energy cost due to the
presence of the superfluid flow. 

If $H-\p\cdot\vs$ reaches the bottom of the fermion quasiparticle
spectrum $\Delta+(\ep+\ev-\varepsilon_0)^2/2\phi_0$, $V_H(\vs)$ receives 
an additional contribution from the filled fermion quasiparticles. 
Since $\ev\sim\eps^{11}$ is small, we can neglect it in the
quasiparticle energy.  Then the $\vs$ dependent part of the effective  
potential is given by
\begin{equation}
 V_H(\vs)=N\ev-\int\!\frac{d\p}{(2\pi)^d}
  \left(H-\p\cdot\vs-\omega_\mathrm{F}(\p)\right)_>,
\end{equation}
where we have introduced a notation $(x)^y_>=x^y\,\theta(x)$ and
$\omega_\mathrm{F}(\p)$ is the fermion quasiparticle spectrum
$\omega_\mathrm{F}(\p)=\Delta+(\ep-\varepsilon_0)^2/2\phi_0$ derived in 
Eq.~(\ref{eq:omega}). The effective potential $V_H(\vs)$ to the leading
order in $\eps$ has the same form as that studied based on the effective
field theory in~\cite{son05}.

\section{Critical polarizations}
Now we evaluate the effective potential $V_H(\vs)$ at $d=4$ as a
function of $\vs$ and $H$.  Changing the integration variables to
$z=\ep/\phi_0$ and $w=\hat\p\cdot\hat{\bm{v}}_\mathrm{s}$, we have 
\begin{align}
 V_H(\vs) &=\left(\frac{m\phi_0}{2\pi}\right)^2 \\
 &\quad\times\left[\frac\ev{\eps}
 -\frac2\pi\int_0^\infty\!dz\!\int_{-1}^1\!dw\,z\sqrt{1-w^2}
 \left(H-\Delta-\frac{(z-z_0)^2}{2}\phi_0
 -2\sqrt{z\phi_0\ev}w\right)_>\right], \notag
\end{align}
where $z_0=\varepsilon_0/\phi_0\sim\eps$.
We can approximate $\sqrt{z\phi_0\ev}$ by $\sqrt{z_0\phi_0\ev}$ because
the difference will be $O(\eps^8)$ and negligible compared to itself 
$\sqrt{z_0\phi_0\ev}\sim\eps^6$.  Then the integration over $z$ leads to 
\begin{equation}
  V_H(\vs)=\left(\frac{m\phi_0}{2\pi}\right)^2\phi_0
  \left[\frac\ev{\eps\phi_0}-\frac{32}{3\pi}z_0
  \left(\frac\ev{\phi_0}z_0\right)^{3/4} 
  \int_{-1}^1dw\,\sqrt{1-w^2}\,
  \biggl(\frac{H-\Delta}{2\sqrt{\ev{\phi_0}z_0}}-w\biggr)^{3/2}_>\right].
\end{equation}
By introducing the dimensionless variables $x$ and $h$ as
\begin{equation}
 \frac\ev{\phi_0}=x^2\left(\frac{32}{3\pi}\eps\right)^4z_0^{\,7}
  \qquad\quad\text{and}\qquad\quad
  \frac{H-\Delta}{\phi_0}=2h\left(\frac{32}{3\pi}\eps\right)^2z_0^{\,4},
\end{equation}
we can rewrite the effective potential $V_H(\vs)=V_h(x)$ in the simple 
form as  
\begin{equation}
  V_h(x)=\phi_0\left(\frac{m\phi_0}{2\pi}\right)^2
  \left(\frac{32}{3\pi}\right)^4\eps^3z_0^{\,7}
  \left[x^2-x^{3/2}\int_{-1}^1dw\,\sqrt{1-w^2}
  \left(\frac hx-w\right)^{3/2}_>\right]. 
\end{equation}
Now we see that if the superfluid velocity exists, $\ev\sim\eps^{11}$
and the $\vs$ dependent part of the effective potential is
$O(\eps^{10})$. 

Numerical studies on the effective potential $V_h(x)$ as a
function of $x$ show that there exists a region of $h$, $h_1<h<h_2$,
where $V_h(x)$ has its minimum at finite $x$.  These two critical values
are numerically given by 
\begin{equation}
 h_1=-0.00366\qquad\quad\text{and}\qquad\quad h_2=0.0275.
\end{equation}
Correspondingly, we obtain the critical polarizations normalized by the
energy gap as 
\begin{align}
 \frac{H_1}\Delta &= 1-0.0843\,\eps^2
 \left(\frac{\varepsilon_0}\Delta\right)^{4},
\intertext{and}
 \frac{H_2}\Delta &= 1+0.634\,\eps^2
 \left(\frac{\varepsilon_0}\Delta\right)^{4}.
\end{align}
Here we have replaced $\phi_0$ by $\Delta$ because they only differ by
$O(\eps)$ [Eq.~(\ref{eq:gap})]. 
The region for the phase with spatially varying condensate
is $H_2-H_1\sim\eps^6$, where the superfluid velocity $\vs$ is finite at
the ground state. 

As the polarization increases further $H>H_2$, the superfluid velocity
disappears.  If $H<\omega_\mathrm{F}(\bm0)$, fermion quasiparticles 
which have momentum $\omega_\mathrm{F}(\p)<H$ are filled and hence there
exist two Fermi surfaces, while there is only one Fermi surface for
$H>\omega_\mathrm{F}(\bm0)$ [see Fig.~\ref{fig:dispersion}]. 
Therefore, from the quasiparticle spectrum derived in
Eq.~(\ref{eq:omega}), the polarization for the disappearance of the inner
Fermi surface $H_3$ is given by
\begin{equation}
 \frac{H_3}\Delta=\frac{\omega_\mathrm{F}(\bm0)}\Delta
  =1+\frac12\left(\frac{\varepsilon_0}\Delta\right)^2.
\end{equation}
Here the location of minimum in the fermion quasiparticle spectrum
$\varepsilon_0$ is related with the binding energy $\eb$ near the
unitarity limit via $\varepsilon_0=\eps\Delta-\eb/2$
[Eq.~(\ref{eq:e0})].  The critical polarizations $H_1/\Delta$ and
$H_2/\Delta$ as functions of $\eb/\eps\Delta$ are illustrated in
Fig.~{\ref{fig:polarization}}.

\section{Phase transition to normal Fermi gas}
Next we turn to the phase transition to the normal Fermi
gas which occurs at $H-\Delta\sim\eps$.  Since the region for the phase 
with spatially varying condensate is $H_{1,2}-\Delta\sim\eps^6$, we can
neglect the superfluid velocity $\vs$ here.  We can also neglect the
possibility to have two Fermi surfaces where $H_3-\Delta\sim\eps^2$.
Then the contribution of the polarized quasiparticles to the effective
potential is evaluated as
\begin{equation}
 \begin{split}
  V_H(\bm0) &= -\int\!\frac{d\p}{(2\pi)^d}
  \left(H-\omega_\mathrm{F}(\p)\right)_> \\
  &\simeq -\frac{(H-\Delta)_>^2}{2\phi_0}
  \left(\frac{m\phi_0}{2\pi}\right)^{d/2},
 \end{split}
\end{equation}
where we have neglected the higher order corrections due to the shift of
the location of minimum in the dispersion relation
$\varepsilon_0\sim\eps$.  According to the modification of the effective
potential in Eq.~(\ref{eq:Veff}) to $\Veff(\phi_0)+V_H$, the solution of
the gap equation in Eq.~(\ref{eq:phi_0}) becomes
$\phi_0\to\phi_0+\phi_H$, where 
\begin{equation}
\phi_H = -(H-\Delta)_> + \frac{(H-\Delta)^2_>}{2\phi_0}.
\end{equation}
Then the pressure of polarized Fermi gas in the superfluid state is
given by $P=-\Veff(\phi_0+\phi_H)-V_H$, which results in 
\begin{equation}\label{eq:P_H}
 P = \frac{\phi_0}6 \left[ 1 + \left(\frac{17}{12}-3C
 -\frac{\gamma+\ln 2}2\right)\epsilon -\frac{3\eb}{4\phi_0} \right]
 \left(\frac{m\phi_0}{2\pi}\right)^{d/2} 
 + O\left(\,\eps(H-\Delta)^2_>,\,(H-\Delta)^3_>\,\right).
\end{equation}
The polarization dependent part of the pressure $P_H\sim\eps^3$ is small
and negligible in the region considered here $H-\Delta\sim\eps$. 

The pressure of the superfluid state should be compared to that of the
normal state with the same chemical potentials.  Since the phase
transition to the normal state happens at
$H\sim\phi_0\gg\mu\sim\eps\phi_0$, only one-component of fermions exists
in the normal state.  Therefore, the interaction is completely
suppressed in the normal Fermi gas and its pressure $P_\mathrm{n}$ is
simply given by that of a free Fermi gas with a single component:
\begin{equation}\label{eq:Pn}
  P_\mathrm{n} = \int\!\frac{d\bm p}{(2\pi)^d}(\mu_\uparrow-\ep)_>
  = \frac{(H+\mu)^{\frac d2+1}}{\Gamma\!\left(\frac d2+2\right)}
  \left(\frac{m}{2\pi}\right)^{\frac d2}.
\end{equation}

The phase transition of the superfluid state to the normal state occurs
at $H=\Hc$ where the two pressures coincide $P=P_\mathrm{n}$.  From 
Eqs.~(\ref{eq:P_H}) and (\ref{eq:Pn}), the critical polarization $\Hc$
satisfies the following equation, 
\begin{equation}\label{eq:Hc}
 \begin{split}
  \Hc &=\left[1+\frac\eb{4\phi_0}-C\eps-\frac{2+\ln2}6\eps\right]\phi_0\\
  &=\left[1+\frac\eb{4\Delta}-C\eps-\frac{2+\ln2}6\eps
  +(8\ln3-12\ln2)\eps\right]\Delta,
 \end{split}
\end{equation}
where we have substituted the relation between the condensate $\phi_0$ 
and the energy gap $\Delta$ at zero polarization in Eq.~(\ref{eq:gap}). 
Defining a number $\sigma\sim O(1)$ by
\begin{equation}\label{eq:sigma}
  \sigma=C+\frac{2+\ln2}6-(8\ln3-12\ln2) \approx 0.12197. 
\end{equation}
the critical polarization normalized by the energy gap at zero
polarization is written as 
\begin{equation}
 \frac\Hc\Delta = 1-\eps\sigma+\frac{\eb}{4\Delta} + O(\eps^2). 
\end{equation}
If the binding energy is large enough $\eb/\eps\Delta>4\sigma=0.488$,
the superfluid state remains above $H=\Delta$. 
The superfluid state at $\Delta<H<\Hc$ is referred to as a
\textit{gapless} superfluid state.  The fermion number densities of two
difference components are asymmetric there and the fermionic excitation
does not have the energy gap. 

In particular, at the unitarity limit
where $\eb/\Delta=0$, the critical polarization is given by 
\begin{equation}
  \left.\frac{H_\mathrm{c}}\Delta\right|_\mathrm{UL}
  =1-\eps\sigma =1-0.122\,\eps, 
\end{equation}
At the splitting point where $\eb/\Delta=2\eps$, it is 
\begin{equation}
  \left.\frac{H_\mathrm{c}}\Delta\right|_\mathrm{SP}
  =1-\eps\sigma+\frac\eps2 =1+0.378\,\eps. 
\end{equation}
The extrapolations to three spatial dimensions $\eps=1$ give the
critical polarizations at the two typical points as
$\Hc/\Delta\,|_\mathrm{UL}=0.878$ and
$\Hc/\Delta\,|_\mathrm{SP}=1.378$.  At unitarity, hence, there is no
gapless superfluid phase.  On the other hand, near the splitting point
the normal phase is not competitive compared to the gapless phases.
The phase boundary between the
superfluid and normal phases $\Hc/\Delta$ as a function of
$\eb/\eps\Delta$ is illustrated in Fig.~{\ref{fig:polarization}}.

\section{Phase structure near the unitarity limit}

\begin{figure}[tp]
 \begin{center}
  \includegraphics[scale=1.2,clip]{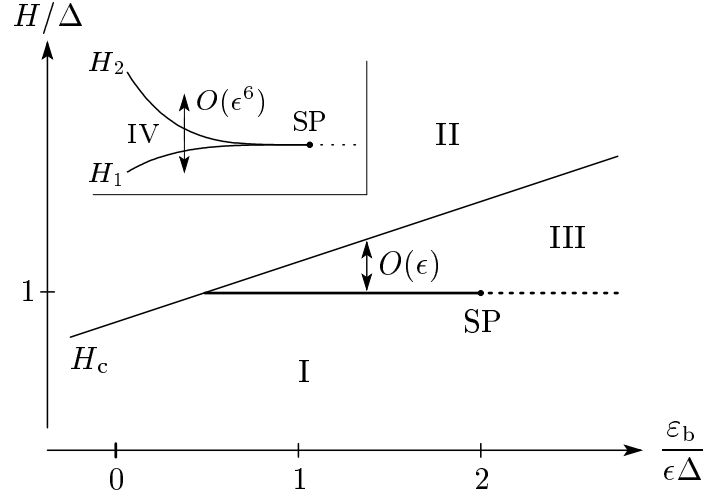}
  \caption{Schematic phase diagram of the polarized Fermi gas near the 
  unitarity limit from the $\eps$ expansion. The phase diagram can be
  divided into four phases, I: the gapped superfluid phase, II: the
  polarized normal phase, III: the gapless superfluid phase, and IV: the
  phase with spatially varying condensate.  The inset is the
  magnification of the region around the splitting point (SP).  The
  phase IV appears in the narrow region represented by the thick line
  between the phases I and III.  \label{fig:polarization}}
 \end{center}
\end{figure}

The schematic phase diagram of the polarized Fermi gas in the unitary
regime is shown in Fig.~\ref{fig:polarization} in the plane of $H$ and
$\eb/\eps$ for the fixed energy gap $\Delta$ at zero polarization.  The
critical polarization $\Hc/\Delta$ divides the phase diagram into two
regions; the superfluid phase at $H<\Hc$ (I and III) and the normal
phase at $H>\Hc$ (II).  The Fermi gas in the normal phase is fully
polarized near the unitarity limit because $H\gg\mu$.  The phase
transition of the superfluid state to the normal state is of the first
order, because of the discontinuity in the fermion number density which
is $O(\eps^{-1})$ in the superfluid phase while $O(1)$ in the normal
phase. 

The superfluid phase can be divided further into two regions; the gapped 
superfluid phase at $H<\Delta$ (I) and the gapless superfluid phase at
$\Delta<H<\Hc$ (III).  At the BEC side of the splitting point (SP) where 
$\eb/\eps\Delta>2$, the phase transition from the gapped phase to the 
gapless phase is of the second order because a discontinuity appears in
the second derivative of the pressure 
$\d^2 P/\d H^2\sim\eps\,\theta(H-\Delta)$ [see Eq.~(\ref{eq:P_H})].  
On the other hand, at the BCS side of the splitting point where
\begin{equation}\label{eq:range-eb}
 4\sigma<\frac\eb{\eps\Delta}<2, 
\end{equation}
there exists the superfluid phase with spatially varying condensate (IV)
at $H_1<H<H_2$ between the gapped and gapless phases.  This phase
appears only in the narrow region where $H_2-H_1\sim\eps^6$.  The phase
transitions at $H=H_1$, $H_2$ are of the first order. 

In actual experiments using fermionic atoms, unitary Fermi gases are
trapped in the optical potential
$V(r)$~\cite{Zwierlein05,Partridge06,Zwierlein06,Shin06,Partridge06-2}.
In such cases, the polarization $H$ and the scattering length or $\eb$
are constant, while the effective chemical potential $\mu-V(r)$ decrease
from the center to the peripheral of the trapping potential.  For the
purpose of comparison of our results with the experiments on the
polarized Fermi gases, it is convenient to visualize the phase diagram
in the plane of $H$ and $\mu$ for the fixed $\eb$. 

\begin{figure}[tp]
 \begin{center}
  \includegraphics[scale=1.2,clip]{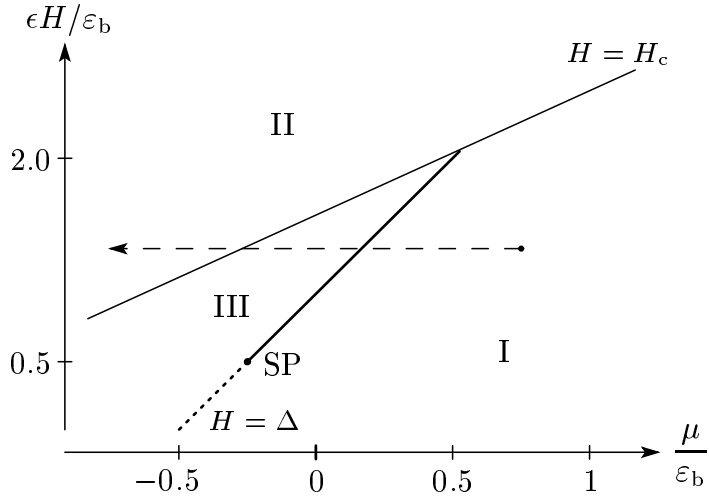}
  \caption{Schematic phase diagram of the polarized Fermi gas near the 
  unitarity limit in the $H$-$\mu$ plane for the fixed binding
  energy $\eb>0$.  The phases I, II, and III are same as in
  Fig.~\ref{fig:polarization}.  The horizontal dashed line from right to
  left tracks the chemical potentials $(\mu,H)$ in trapped Fermi gases 
  from the center to the peripheral.  \label{fig:polarization2}}
 \end{center}
\end{figure}

In Fig.~\ref{fig:polarization2}, $\eps\Hc/\eb$ and $\eps\Delta/\eb$ are
plotted as functions of $\mu/\eb$ for the fixed binding energy $\eb>0$.
The superfluid phase and the normal phase are separated by the line of
the critical polarization $H=\Hc$, while the gapped and gapless
superfluid phases are separated by the line of the energy gap at zero
polarization $H=\Delta$.  The intersection of two lines, $\Hc=\Delta$,
is located at 
$(\mu/\eb,\eps H/\eb)=((1-4\sigma)/8\sigma,1/4\sigma)=(0.524,2.05)$,
while the splitting point in the phase diagram is at
$(\mu/\eb,\eps H/\eb)=(-1/4,1/2)$. 
The horizontal dashed line from right to left tracks the chemical
potential $\mu$ in the trapped Fermi gas from the center to the
peripheral.  If the polarization is small enough compared to the binding
energy $\eps H/\eb<2.05$, there exists the gapless superfluid phase
(III) between the gapped superfluid phase (I) and the normal phase
(II). When the polarization is above the splitting point 
$0.5<\eps H/\eb<2.05$, the phase with spatially varying condensate
(thick line between I and III) will appear between the gapped and
gapless superfluid phases. 

Assuming that the picture remains qualitatively valid in three
dimensions, we thus found that the physics around the splitting point is
the same as argued in Ref.~\cite{son05}.  However, due to the
competition with the normal phase, the gapless phases disappear at some
point, probably before the unitarity is reached.  The point where this
happens can be estimated from our calculations to be
\begin{equation}
  (a\kF)^{-1} = \sqrt{2\sigma} = 0.494. 
\end{equation}
The gapless phase with spatially varying condensate exists in a finite range 
$0.494<(a\kF)^{-1}<1$.  Thus the $\eps$ expansion does not support the
hypothesis of Ref.~\cite{son05} that the region of the gapless state
with spatially varying condensate is connected with the FFLO region on
the BCS side (see Fig.~\ref{fig:son-stephanov}).

\chapter{Fermions with unequal masses \label{sec:unequal}}

\section{Two-body scattering in vacuum}
Now we generalize our discussion on the unitary Fermi gas in the $\eps$
expansion to the fermions with unequal masses $m_\uparrow\neq
m_\downarrow$, especially to study the phase structure of unitary Fermi
gas with unequal densities and masses.  As we mentioned in the previous
Chapter, such asymmetric systems of fermions with density and mass
imbalances will be interesting as a prototype of high density quark
matter in the core of neutron stars, where the density and mass
asymmetries exist among different quark
flavors~\cite{Alford99,Shovkovy,Huang,Gubankova03,Alford03}.  So far the
phase structure of the two-component Fermi gas with density and mass
imbalances in the BCS-BEC crossover has been studied within the
mean-field approximation~\cite{iskin06,yip06,lin06,parish06}, which is
not a controllable approximation in the strong-coupling unitary regime.

First we reconsider the two-fermion scattering in vacuum studied in
Chap.~\ref{sec:vacuum}.  As a result of the summation of the geometric
series of bubble diagrams, the inverse $T$-matrix of the two-body
scattering is given by
\begin{equation}\label{eq:T^-1_unequal}
 T(p_0,\p)^{-1}=\frac1{c_0}+\int\!\frac{d\k}{(2\pi)^d}\,
  \frac1{p_0-\frac{\left(\frac\p2+\k\right)^2}{2m_\uparrow}
  -\frac{\left(\frac\p2-\k\right)^2}{2m_\downarrow}+i\delta}.
\end{equation}
The limit of infinite scattering length corresponds to $1/c_0\to0$ as
before.  Introducing the total mass $M=m_\uparrow+m_\downarrow$ and the
reduced mass $\mr=m_\uparrow m_\downarrow/(m_\uparrow+m_\downarrow)$,
the inverse $T$-matrix at $d=4-\eps$ spatial dimensions to the
leading order in $\eps$ can be written as 
\begin{equation}
 T(p_0,\p)^{-1}=\frac1{c_0}-\frac2\eps\left(\frac{\mr}{2\pi}\right)^2
  \left(p_0-\frac{\p^2}{2M}+i\delta\right)+\cdots.
\end{equation}
This result suggests that the boson propagator is given by
\begin{equation}\label{eq:D_unequal}
  D(p_0,\p) = \left(p_0 - \frac{\p^2}{2M} + i\delta\right)^{-1}
\end{equation}
with the fermion-boson effective coupling
\begin{equation}
 g^2 \simeq \frac{2\pi^2\eps}{\mr^{\,2}}.
\end{equation}

The binding energy $\eb>0$ is obtained as the location of the pole of
the $T$-matrix at zero external momentum: $T(-\eb,\bm0)^{-1}=0$.  From
Eq.~(\ref{eq:T^-1_unequal}), we see $c_0$ is related with $\eb$ via the
following equation 
\begin{equation}\label{eq:eb_unequal}
 \frac1{c_0}=\Gamma\!\left(1-\frac d2\right)
  \left(\frac{\mr}{2\pi}\right)^{\frac d2}\eb^{\,\frac d2-1}.
\end{equation}
Near the four spatial dimensions, the relationships between $\eb$ and
$c_0$ to the next-to-leading order in $\eps=4-d$ is given by
\begin{equation}
 \frac1{c_0}\simeq-\frac{\eb}{2\eps}\left(\frac{\mr}{\pi}\right)^2
  \left[1+\frac\eps2\left(1-\gamma\right)\right]
  \left(\frac{\mr\eb}{2\pi}\right)^{-\frac\eps2}.
\end{equation}

\section{$\eps$ expansion with unequal masses}
Now we consider the finite density system.  The formalism of the $\eps$
expansion developed in the equal mass case in Chap.~\ref{sec:4d} holds
just by replacing the boson propagator $D(p)$ by Eq.~(\ref{eq:D_unequal})
and the fermion-boson effective coupling $g$ by
\begin{equation}
 g = \frac{(2\pi^2\eps)^{1/2}}\mr
  \left(\frac{\mr\phi_0}{\pi}\right)^{\eps/4}.
\end{equation}
Then the Lagrangian density under consideration is given by 
\begin{align}
 \mathcal{L}_0 & = \Psi^\+\left(i\d_t + H 
 + \frac{\sigma_3+\kappa}{4\mr}\,\grad^2 
 + \sigma_+\phi_0 + \sigma_-\phi_0\right)\Psi
 + \varphi^*\left(i\d_t+\frac{\grad^2}{2M}\right)\varphi
 - \frac{\phi_0^{\,2}}{c_0}\,, \\
 \mathcal{L}_1 & = g\Psi^\+\sigma_+\Psi\varphi 
 + g\Psi^\+\sigma_-\Psi\varphi^* + \mu\Psi^\+\sigma_3\Psi 
 + \left(2\mu-\frac{g^2}{c_0}\right)\varphi^*\varphi
 - \frac{g\phi_0}{c_0}\varphi - \frac{g\phi_0}{c_0}\varphi^*\,, 
  \phantom{\frac{\frac\int\int}{\frac\int\int}}\hspace{-4.5mm} \\
 \mathcal{L}_2 & = -\varphi^*\left(i\d_t
 +\frac{\grad^2}{2M}\right)\varphi - 2\mu\varphi^*\varphi\,.
\end{align}
where $\mu_\sigma=\mu\pm H$ and
$1/2m_\sigma=(1\pm\kappa)/4\mr$.  We introduced the dimensionless
parameter $-1<\kappa<1$ defined by
\begin{equation}
 \kappa\equiv-\frac{m_\uparrow-m_\downarrow}{m_\uparrow+m_\downarrow},
\end{equation}
which measures the mass difference between two fermions. 

The Lagrangian density $\mathcal{L}_0$ generates the fermion propagator
given by 
\begin{equation}
 G(p_0,\p) = \frac1{(p_0+H-\kappa\ep)^2-\Ep^{\,2}+i\delta}
  \begin{pmatrix}
   p_0+H -\kappa\ep + \ep & -\phi_0 \\
   -\phi_0 & p_0+H -\kappa\ep - \ep
  \end{pmatrix},
\end{equation}
where $\ep=\p^2/4\mr$ and $\Ep=\sqrt{\ep^{\,2}+\phi_0^{\,2}}$, and the
boson propagator given by 
\begin{equation}
  D(p_0,\p) = \left(p_0 - \frac{\p^2}{2M} + i\delta\right)^{-1}
    = \left(p_0 - \frac{1-\kappa^2}2\ep + i\delta\right)^{-1}.
\end{equation}
We can use the same Feynman rule and the same power counting rule of
$\eps$ as the equal mass case developed in Sec.~\ref{sec:counting}, as
far as the mass difference is not so large.  When the mass difference is
as large as $|\kappa|\simeq1-O(1/\eps)$, the power counting rule of
$\eps$ breaks down because of the existence of another large parameter
$m_\mathrm{heavy}/m_\mathrm{light}\sim1/\eps$.  Accordingly, the
fermion-boson scattering leads to the three-body bound states (Efimov
states) in vacuum when the mass ratio is given by 
$m_\mathrm{heavy}/m_\mathrm{light}=4/\eps+O(1)$~\cite{Efimov,Braaten,Nielsen}.
In this Chapter, we will only consider the case where the mass
difference is at most $\kappa\sim O(1)$ and free from the Efimov states.
First we will determine the pressure and the fermion quasiparticle
spectrum to the leading and next-to-leading orders in $\eps$ at the
unpolarized ground state where $H=0$.  Then we investigate the phase
structure of polarized Fermi gas with the small mass difference
$\kappa\sim\eps$.

\section{Effective potential and pressure}
The computation of the effective potential with finite $\kappa$ is 
straightforward according to the systematic expansion over $\eps$.  
Since $\kappa\ep<\Ep$ for any momentum $\p$, $\kappa$ does not appear at
the one-loop level.  Therefore, the effective potential from the tree
and one-loop diagrams depends only on the reduced mass $\mr$ as 
\begin{align}
 &V_0(\phi_0)+V_1(\phi_0) \notag\\ &\qquad = \frac{\phi_0^{\,2}}{c_0}
 -\int\!\frac{d\p}{(2\pi)^d}\left(E_\p+\mu\frac\ep{E_\p}\right) \\
 &\qquad =\frac{\phi_0^{\,2}}{c_0}
 +\left[\frac{\phi_0}3\left\{1+\frac{7-3(\gamma+\ln2)}6\eps\right\}
 -\frac\mu\eps\left\{1+\frac{1-2(\gamma-\ln2)}4\eps\right\}\right]
 \left(\frac{\mr\phi_0}{\pi}\right)^{d/2}. \notag
\end{align}
The effective potential at two-loop level is given by the same diagram 
in Fig.~\ref{fig:potential} as
\begin{equation}
 \begin{split}
  V_2(\phi_0) &= g^2\int\!\frac{dp\,dq}{(2\pi)^{2d+2}}\,
  G_{11}(p)G_{22}(q)D(p-q) \\
  &= -\frac{g^2}4\! \int\!\frac{d\p\,d\q}{(2\pi)^{2d}}\,
  \frac{(E_\p-\ep)(E_\q-\eq)}{E_\p E_\q 
  \left[E_\p+E_\q+\frac{1-\kappa^2}2\varepsilon_{\p-\q}
  -\kappa\ep+\kappa\eq\right]}.
 \end{split}
\end{equation}
Changing the integration variables to $x=\ep/\phi_0$,
$y=\eq/\phi_0$, and $\cos\theta=\hat\p\cdot\hat\q$, the integral can be
expressed by 
\begin{equation}
 V_2(\phi_0) = -\eps\left(\frac{\mr\phi_0}{\pi}\right)^2
  \frac{\phi_0}\pi\int_0^\infty\!dx\int_0^\infty\!dy
  \int_0^\pi\!d\theta\,xy\sin^2\theta\,\frac{[f(x)-x][f(y)-y]}
  {f(x)f(y)\left[g_\kappa(x,y)-(1-\kappa^2)\sqrt{xy}\cos\theta\right]}
\end{equation}
with $f(x)=\sqrt{x^2+1}$ and
$g_\kappa(x,y)=f(x)+f(y)+\frac{1-\kappa^2}2(x+y)-\kappa(x-y)$. 
The integration over $\theta$ can be performed analytically to lead to 
\begin{equation}
  V_2(\phi_0) = -C_\kappa\eps\left(\frac{\mr\phi_0}{\pi}\right)^2\phi_0
\end{equation}
where $C_\kappa$ is a function of $\kappa$ given by a two-dimensional
integral 
\begin{equation}\label{eq:C_k}
 C_\kappa = \int_0^\infty\!dx\int_0^\infty\!dy\,
  \frac{[f(x)-x][f(y)-y]}{f(x)f(y)}\,\frac{g_\kappa(x,y)
  -\sqrt{g_\kappa(x,y)^2-(1-\kappa^2)^2xy}}{(1-\kappa^2)^2}.
\end{equation}
$C_\kappa$ is symmetric under $\kappa\to-\kappa$ and numerically we find
$C_\kappa$ is an increasing function of $\kappa$ at $0\leq\kappa<1$.  In
the equal mass limit $\kappa=0$, we reproduce $C_0\approx0.14424$ in
Eq.~(\ref{eq:C}). 

Consequently, we obtain the effective potential $\Veff=V_0+V_1+V_2$ up
to the next-to-leading order in $\eps$, 
\begin{equation}
 \Veff(\phi_0) = \frac{\phi_0^{\,2}}{c_0} + \left[\frac{\phi_0}3
  \left\{1+\frac{7-3(\gamma+\ln2)}6\eps-3C_\kappa\eps\right\}
  -\frac\mu\eps\left\{1+\frac{1-2(\gamma-\ln2)}4\eps\right\}
  \right]\left(\frac{\mr\phi_0}{\pi}\right)^{d/2}.
\end{equation}
Comparing $\Veff(\phi_0)$ with the effective potential in the equal mass
case in Eq.~(\ref{eq:Veff}), we see the solution of the gap equation 
$\phi_0$ is simply given by replacing the fermion mass $m$ by $2\mr$ and
the constant $C$ by $C_\kappa$.  Using the relation of $c_0$ with the
binding energy of boson $\eb$ in Eq.~(\ref{eq:eb_unequal}), we obtain
the condensate as 
\begin{equation}
 \phi_0 =
  \frac{2\mu}\eps\left[1+\left(3C_\kappa-1+\ln2\right)\eps\right] 
  +\frac\eb\eps \left[1+\left(3C_\kappa-\frac12+\ln2
  -\frac12\ln\frac\eb{\phi_0}\right)\eps\right].
\end{equation}
Then the pressure up to the next-to-leading order in $\eps$ is given by 
\begin{equation}\label{eq:P_unequal}
 P=-\Veff(\phi_0) =\frac{\phi_0}6
  \left[1+\left(\frac{17}{12}-3C_\kappa-\frac{\gamma+\ln2}2\right)\eps 
   -\frac{3\eb}{4\phi_0}\right]
  \left(\frac{\mr\phi_0}{\pi}\right)^{d/2}.
\end{equation}
When the mass difference $\kappa$ is as small as $\eps$ which is the
case we will concentrate on later, we can neglect the $\kappa$
dependence in $C_\kappa$ to give $C\approx 0.14424$.  In this case, the
pressure of the superfluid state up to the order $O(\eps)$ depends only
on $\mr$ and not on $\kappa$ at all.

\section{Self-energy and dispersion relation}
Let us turn to the computation of the fermion quasiparticle spectrum
with finite $\kappa$.  To the leading order in $\eps$, the dispersion
relation of fermion quasiparticles are given by
\begin{equation}\label{eq:kappa}
 \omega_\mathrm{F}(\p)=\Ep\pm\kappa\ep. 
\end{equation}
The lighter fermion has the energy gap $\Delta=\phi_0$ at $\ep=0$, while
the heavier fermion has the energy gap
$\Delta=\sqrt{1-\kappa^2}\,\phi_0$ at
$\ep=|\kappa|\,\phi_0/\sqrt{1-\kappa^2}$. 
The next-to-leading order corrections to the dispersion relation come
from the $\mu$ insertion to the fermion propagator and the one-loop
self-energy diagrams, $-i\Sigma(p)$, depicted in
Fig.~\ref{fig:self_energy}.  The one-loop diagrams in
Fig.~\ref{fig:self_energy} give corrections to the diagonal elements of
the fermion self-energy, which is evaluated as 
\begin{equation}
 \begin{split}
  \Sigma_{11}(p) 
  &= ig^2\int\!\frac{dk}{(2\pi)^{d+1}}\,G_{22}(k)D(p-k) \\
  &=-\frac{g^2}2\!\int\!\frac{d\k}{(2\pi)^d}\, \frac{E_\k-\ek}
  {E_\k \left[E_\k + \kappa\ek + \frac{1-\kappa^2}2
  \varepsilon_{\k-\p}-p_0\right]},
 \end{split}
\end{equation}
and
\begin{equation}
 \begin{split}
  \Sigma_{22}(p) 
  &= ig^2\int\!\frac{dk}{(2\pi)^{d+1}}\,G_{11}(k)D(k-p) \\
  &=\frac{g^2}2\!\int\!\frac{d\k}{(2\pi)^d}\, \frac{E_\k-\ek}
  {E_\k \left[E_\k - \kappa\ek + \frac{1-\kappa^2}2
  \varepsilon_{\k-\p}+p_0\right]}.
 \end{split}
\end{equation}

From now on, we concentrate on the case where the mass difference is as
small as $\kappa\sim\eps$, where we will see interesting physics in the
phase diagram.  In this case, we can neglect the $\kappa$ dependence
in $\Sigma_{11}$ and $\Sigma_{22}$ to the order $O(\eps)$ because
$\Sigma$ is already small by the factor $g^2\sim\eps$.  Therefore, the
modification of the fermion quasiparticle spectrum due to the finite
$\kappa$ comes from Eq.~(\ref{eq:kappa}).  Following the calculations in
Sec.~\ref{sec:spectrum}, we find that the dispersion relation of the
fermion quasiparticle around its minimum has the following form 
\begin{equation}
 \omega_\mathrm{F}(\p)
  \simeq\Delta+\frac{(\ep\pm\kappa\phi_0-\varepsilon_0)^2}{2\phi_0}
  \simeq\sqrt{(\ep\pm\kappa\phi_0-\varepsilon_0)^2+\Delta^2}.
\end{equation}
Here $\Delta$ is the energy gap of the fermion quasiparticle defined in
Eq.~(\ref{eq:gap}) and $\varepsilon_0$ is given by Eq.~(\ref{eq:loc}).
Note that the energy gap is not affected by $\kappa$ up to the order
$O(\eps)$.  The minimum of the dispersion curve is located at a nonzero
value of momentum $|\p|$ satisfying $\ep=\varepsilon_0-|\kappa|\phi_0$
for the lighter fermion and $\ep=\varepsilon_0+|\kappa|\phi_0$ for the
heavier fermion when those are positive.

\section{Critical polarizations and phase diagram}
\begin{figure}[tp]
 \begin{center}
  \includegraphics[scale=1.2,clip]{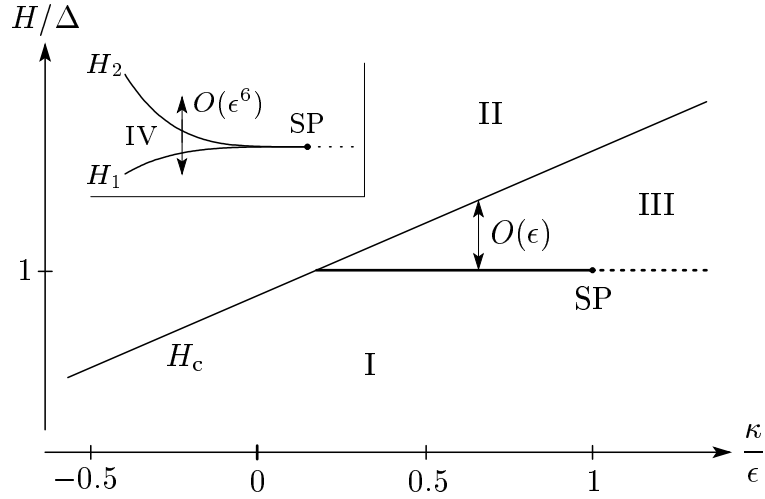}
  \caption{Schematic phase diagram of the polarized Fermi gas with
  unequal masses in the unitarity limit from the $\eps$ expansion.
  $\kappa>0\ (<0)$ corresponds to the system where the majority is the
  lighter (heavier) fermions and an increase of $\kappa$ means a
  decrease of major fermion's mass.  The phase diagram can be divided
  into four phases, I: the gapped superfluid phase, II: the polarized
  normal phase, III: the gapless superfluid phase, and IV: the phase
  with spatially varying condensate.  The inset is the magnification of
  the region around the splitting point (SP).  The phase IV appears in
  the narrow region represented by the thick line between the phases I
  and III.  \label{fig:unequal-mass}}
 \end{center}
\end{figure}

Since the result should be symmetric under $H\to-H$ and
$\kappa\to-\kappa$, we can choose $H>0$ without losing generality. 
Accordingly, $\kappa>0\ (<0)$ corresponds to the system where the
majority is the lighter (heavier) fermions and an increase of $\kappa$
means a decrease of major fermion's mass.  When $H$ is increased, the
phase transition to the normal Fermi gas occurs at $H-\Delta\sim\eps$.
The critical polarization $\Hc$ is given where the pressure of the
superfluid state coincides with that of the normal state with the same
chemical potentials.  The pressure of the superfluid state is given by
Eq.~(\ref{eq:P_unequal}) because the contribution of the polarized
quasiparticles to the pressure is $O(\eps^3)$ and negligible [see
Eq.~(\ref{eq:P_H})].  On the other hand, since
$H\sim\phi_0\gg\mu\sim\eps\phi_0$, the pressure of the normal state is
given by that of the fully polarized free Fermi gas
\begin{equation}
 P_\mathrm{n} 
  = \int\!\frac{d\bm p}{(2\pi)^d}
  \left(\mu_\uparrow-\frac{\p^2}{2m_\uparrow}\right)_>
  = \frac{(H+\mu)^{\frac d2+1}}{\Gamma\!\left(\frac d2+2\right)}
  \left(\frac{\mr/\pi}{1+\kappa}\right)^{\frac d2}.
\end{equation}
The condition $P=P_\mathrm{n}$ for the phase transition to the normal
state gives the critical polarization $H=\Hc$ normalized by the energy
gap at zero polarization $\Delta$ in Eq.~(\ref{eq:gap}) as
\begin{equation}\label{eq:Hc_unequal}
 \frac{H_\mathrm{c}}\Delta 
  = 1-\eps\sigma+\frac{\eb}{4\Delta} + \frac23\kappa + O(\eps^2).
\end{equation}
The number $\sigma\approx0.12197$ is defined in Eq.~(\ref{eq:sigma}). 
When $\Hc>\Delta$, the gapless superfluid state appears at
$\Delta<H<\Hc$.  In particular, in the unitarity limit $\eb=0$, the
gapless superfluid state is possible when the majority is the lighter
fermions and the mass difference is as large as
$\kappa/\eps>3\sigma/2=0.183$.  The phase boundary between the
superfluid and normal phases $\Hc/\Delta$ in the unitarity limit as a
function of $\kappa/\eps$ is illustrated in
Fig.~{\ref{fig:unequal-mass}}. 

The minimum of the dispersion curve for the major fermions is located at
a nonzero value of momentum $|\p|$ satisfying 
\begin{equation}
 \varepsilon_0-\kappa\phi_0
  =\left(\eps-\kappa\right)\Delta-\frac{\eb}2+O(\eps^2).
\end{equation}
Therefore, we find the splitting point where the minimum of the
dispersion curve sits exactly at zero momentum exists at 
\begin{equation}\label{eq:SP}
 \frac\eb{2\Delta}+\kappa=\eps. 
\end{equation}
The phase structure around the splitting point is determined by the same
discussion given in Chap.~\ref{sec:polarization} just by replacing 
$\varepsilon_0$ by $\varepsilon_0-\kappa\phi_0$.  There exists the phase
with spatially varying condensate (or finite superfluid velocity) at
$H_1<H<H_2$, where
\begin{align}
 \frac{H_1}\Delta &= 1-0.0843\,\eps^2
 \left(\frac{\varepsilon_0}\Delta-\kappa\right)^{4},
\intertext{and}
 \frac{H_2}\Delta &= 1+0.634\,\eps^2
 \left(\frac{\varepsilon_0}\Delta-\kappa\right)^{4}.
\end{align}
Furthermore, the polarization for the disappearance of the inner
Fermi surface $H_3$ is given by
\begin{equation}
 \frac{H_3}\Delta=\frac{\omega_\mathrm{F}(\bm0)}\Delta
  =1+\frac12\left(\frac{\varepsilon_0}\Delta-\kappa\right)^2.
\end{equation}
If $\Delta<H<H_3$, fermion quasiparticles which have momentum
$\omega_\mathrm{F}(\p)<H$ are filled and hence there exist two Fermi
surfaces, while there is only one Fermi surface for $H_3<H$.  Since
$\varepsilon_0-\kappa\phi_0=\left(\eps-\kappa\right)\Delta$ in the
unitarity limit $\eb=0$, the splitting point is located at
$\kappa=\eps$.  The critical polarization for the phase transition to
the normal state at this point is $\Hc/\Delta=1+0.545\eps>0$.
Therefore, the phase with spatially varying condensate and the phase
with two Fermi surfaces are stably exist in the unitarity limit with
finite mass difference given by
\begin{equation}\label{eq:range-kappa}
 \frac32\sigma<\frac\kappa\eps<1. 
\end{equation}
The critical polarizations $H_1/\Delta$ and
$H_2/\Delta$ in the unitarity limit as functions of $\kappa/\eps$ are
illustrated in Fig.~{\ref{fig:unequal-mass}}.

\begin{figure}[tp]
 \begin{center}
  \includegraphics[width=0.65\textwidth,clip]{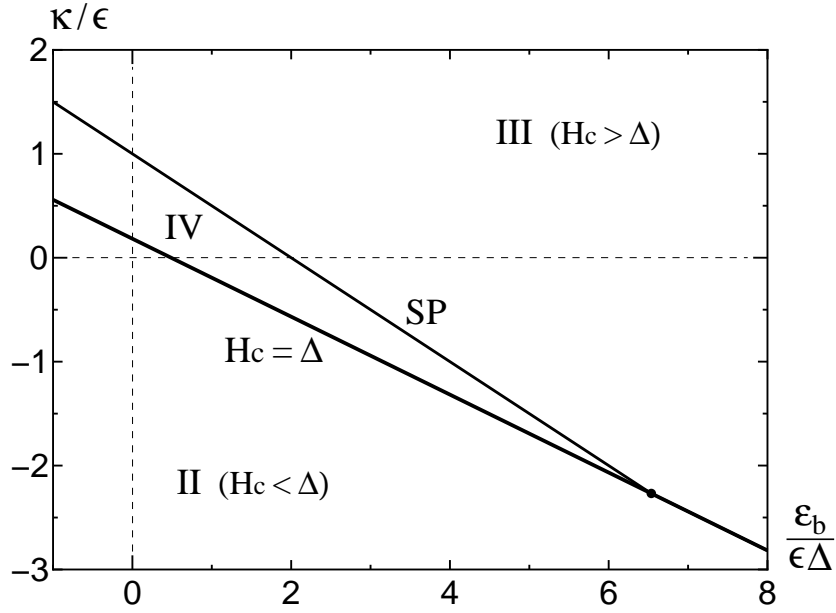}
  \caption{Phase diagram of the polarized Fermi gas in the plane of
  binding energy $\eb$ and mass difference $\kappa$ from the $\eps$
  expansion.  Two solid lines show $\Hc=\Delta$ in
  Eq.~(\ref{eq:Hc=Delta}) and the location of the splitting point (SP)
  in Eq.~(\ref{eq:SP}).  The intersection of these two lines
  is given by Eq.~(\ref{eq:intersect}).  The phases II, III, and IV
  represents the ground state slightly above $H=\Delta$.  The phase
  diagrams in the equal mass limit $\kappa=0$ and in the unitarity limit
  $\eb=0$ with $H$-axis are already illustrated in
  Fig.~\ref{fig:polarization} and Fig.~\ref{fig:unequal-mass},
  respectively.  \label{fig:eb-kappa}} 
 \end{center}
\end{figure}

Finally, we discuss the whole phase structure of the polarized Fermi gas
with unequal masses near the unitarity limit.  $\Hc(\eb,\kappa)$ in
Eq.~(\ref{eq:Hc_unequal}) is a function of the binding energy $\eb$ and
the mass difference $\kappa$.  Hence $H=\Hc$ forms a single plane in the
three-dimensional phase diagram in the space of $\eb/\Delta$, $\kappa$,
and $H/\Delta$, which separates the superfluid phase at $H<\Hc$ from the 
polarized normal phase at $H>\Hc$.  The splitting point given by
Eq.~(\ref{eq:SP}) with $H=\Delta$ forms a ``line'' in such a
three-dimensional phase diagram.  The superfluid phase with spatially
varying condensate emerges from the SP line, which occupies a finite
region in the phase diagram between the gapped superfluid phase at
$H<H_1$ and the gapless superfluid phase at $H_2<H<\Hc$. 

Fig.~\ref{fig:eb-kappa} shows a part of such a three-dimensional phase
diagram in the space of $\eb/\Delta$, $\kappa$, and $H/\Delta$, sliced
slightly above $H=\Delta$.  When $\Hc<\Delta$ for a given set of $\eb$
and $\kappa$ (lower half of the $\Hc=\Delta$ line in
Fig.~\ref{fig:eb-kappa}), the superfluid state is unstable at $H=\Delta$
and there exists the phase transition from the gapped superfluid state
to the polarized normal state (II).  On the other hand, when
$\Hc>\Delta$ (upper half of the $\Hc=\Delta$ line in
Fig.~\ref{fig:eb-kappa}), the superfluid is stable even at $H>\Delta$ 
and the gapless superfluid state (III) appears there.  The boundary 
$\Hc=\Delta$ dividing the phase diagram in the plane of $\eb$ and
$\kappa$ into these two regions is given from Eq.~(\ref{eq:Hc_unequal})
by 
\begin{equation}\label{eq:Hc=Delta}
 \frac\eb{4\Delta}+\frac23\kappa=\eps\sigma
\end{equation}
As the mass of major fermions is increased (or $\kappa$ is decreased),
the boundary shifts to the larger value of $\eb$ because the pressure of
the normal state increases.  Consequently, the polarized normal state
occupies the wider domain near the unitarity limit. 

\begin{figure}[tp]
 \begin{center}
  \includegraphics[width=\textwidth,clip]{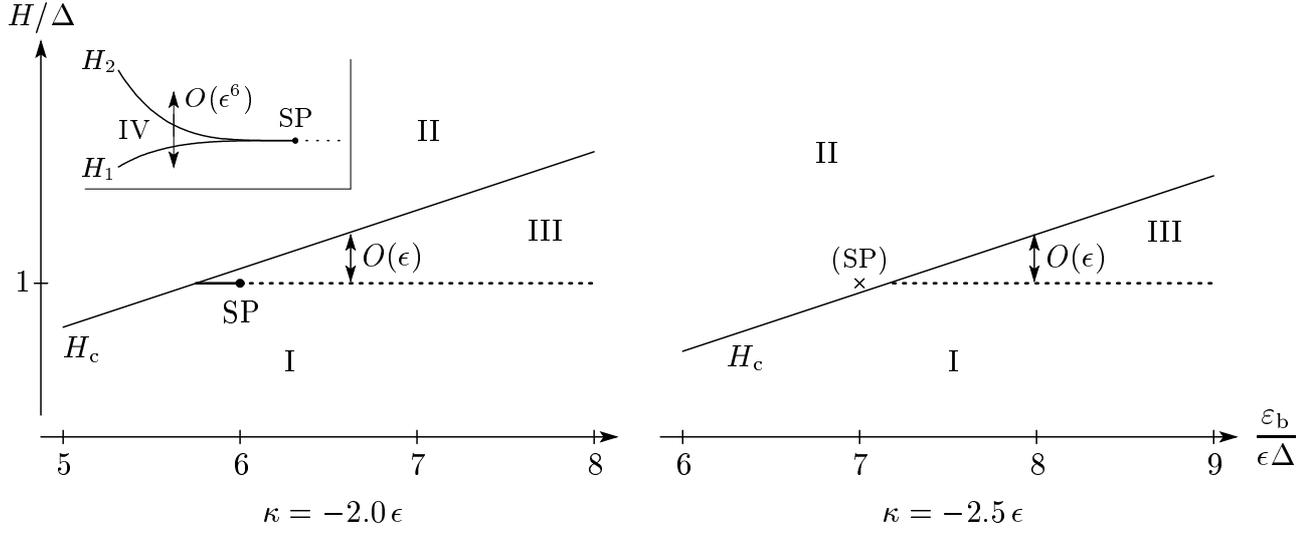}
  \caption{Schematic phase diagram of the polarized Fermi gas in the
  $H$-$\eb$ plane near the disappearance of the splitting point at
  $\kappa=-2.0\,\eps$ and $-2.5\,\eps$.  The phases I, II, III, and VI
  are same as in Fig.~\ref{fig:unequal-mass}.  If $\kappa>-2.27\,\eps$
  (left panel), the splitting point and the phase with spatially varying
  condensate stably exist, while those become unstable when
  $\kappa<-2.27\,\eps$ (right panel).  \label{fig:disappear}}
 \end{center}
\end{figure}

The region where $\Hc>\Delta$ can be further divided into two regions
according to whether the minimum of the dispersion curve is located at
a nonzero value of momentum $\varepsilon_0-\kappa\phi_0>0$ or not.  If
$\varepsilon_0-\kappa\phi_0>0$ (below the SP line in
Fig.~\ref{fig:eb-kappa}), there exist the superfluid state with
spatially varying condensate (IV) between the normal state and the
gapless superfluid state.  The boundary $\varepsilon_0=\kappa\phi_0$
represents the line of the splitting point given by Eq.~(\ref{eq:SP}).
As the mass of major fermions is increased, the region for the
superfluid phase with spatially varying condensate shrinks.  Eventually,
such a phase disappears at the following point where the two lines
$\Hc=\Delta$ and $\varepsilon_0=\kappa\phi_0$ merges
\begin{equation}\label{eq:intersect}
 \begin{split}
  \frac\eb\Delta &= 4\left(2-3\sigma\right)\eps =6.54\,\eps, 
  \phantom{\frac{}\int}\\
  \kappa &= -3\left(1-2\sigma\right)\eps =-2.27\,\eps. 
 \end{split}
\end{equation}
Therefore, if the mass difference is as large as $\kappa<-2.27\,\eps$
with major heavier fermions, the splitting point becomes unstable
towards the polarized normal state and the superfluid phase with
spatially varying condensate never appears for any value of the binding
energy $\eb$.  The schematic phase diagram in the plane of $\eb$ and $H$
near the disappearance of the splitting point is illustrated in
Fig.~\ref{fig:disappear} by taking the typical values at
$\kappa=-2.0\,\eps$ and $-2.5\,\eps$.

\section{Summary on the phase structure of polarized Fermi gas}
In Chapters \ref{sec:polarization} and \ref{sec:unequal}, the phase
structure of the polarized Fermi gas with equal and unequal fermion
masses has been studied in the unitary regime based on the $\eps$
expansion around four spatial dimensions.  Although our results are
valid only near four spatial dimensions where $\eps\ll1$, we can draw
the following conclusions assuming that the picture remains
qualitatively valid in three dimensions.  At unitarity in the equal mass
limit, there is a first-order phase transition from the unpolarized
superfluid state to the fully polarized normal state. On the BEC side of
the unitarity point, the gapless superfluid phase and the superfluid
phase with spatially varying condensate stably exist between the gapped
superfluid phase and the polarized normal phase in a certain range of
the binding energy and the mass difference.  In the equal mass limit
(Fig.~\ref{fig:polarization}), our study gives a microscopic foundation
to the phase structure around the splitting point, which has been
proposed on the BEC side of the unitarity point using the effective
field theory~\cite{son05}.  Moreover, we found the gapless phases
terminate at some point before the unitarity is reached due to the
competition with the polarized normal state.  The naive extrapolation of
our result (\ref{eq:range-eb}) with the use of Eq.~(\ref{eq:ebeF})
estimates the range of the superfluid phase with spatially varying
condensate to be $0.494\lesssim(a\kF)^{-1}\lesssim1$ at three dimensions
$\eps\to1$. 

We also found the splitting point and the same phase structure
around it in the unitarity limit but with finite mass difference between
two fermion species when the majority is the lighter fermions
($\kappa>0$ in Fig.~\ref{fig:unequal-mass}).  The range of the
superfluid phase with spatially varying condensate can be estimated from
Eq.~(\ref{eq:range-kappa}) to be  
$1.37\lesssim m_\mathrm{minority}/m_\mathrm{majority}\lesssim3$ at
$\eps\to1$.  Such splitting points form a smooth line on the $H=\Delta$
plane of the three-dimensional phase diagram in the space of
$\eb/\Delta$, $\kappa$, and $H/\Delta$.  It gets away from the unitarity
with increasing the mass of major fermions (or decreasing $\kappa$ in
Fig.~\ref{fig:eb-kappa}).  Eventually the splitting point becomes
unstable due to the competition with the polarized normal state at the
point in Eq.~(\ref{eq:intersect}), which we estimate to be
$(a\kF)^{-1}\approx1.81$ and  
$m_\mathrm{majority}/m_\mathrm{minority}\approx5.54$.  
Accordingly, the region for the superfluid phase with spatially varying
condensate shrinks and eventually disappears.  The superfluid state with
finite superfluid velocity is never realized for any value of the
binding energy $\eb$ for a sufficiently large mass difference where the
majority is the heavier fermions
$m_\mathrm{majority}/m_\mathrm{minority}\gtrsim5.54$. 
While our quantitative results are not reliable at three dimensions, it
is interesting to point out that our estimates on the mass ratios are
well below the critical value at $d=3$, 
$m_\mathrm{heavy}/m_\mathrm{light}=13.6$, where the instability occurs
due to the Efimov effect~\cite{Efimov,Braaten,Nielsen}.  Further study
will be worthwhile to confirm these possibilities. 


\chapter{Expansion around two spatial dimensions \label{sec:2d}}

\section{Lagrangian and Feynman rules}
In this Chapter, we formulate the systematic expansion for the unitary
Fermi gas around two spatial dimensions in a similar way as we have done
for the $\eps=4-d$ expansion.  Here we start with the Lagrangian density
given in Eq.~(\ref{eq:L}) limited to the unpolarized Fermi gas in the
unitarity limit where $H=0$ and $1/c_0=0$:
\begin{equation}
  \mathcal{L} = \Psi^\+\left(i\d_t + \frac{\sigma_3\grad^2}{2m}
  + \mu\sigma_3\right)\Psi
  + \Psi^\+\sigma_+\Psi\phi + \Psi^\+\sigma_-\Psi\phi^*.
\end{equation}
Then we expand the field $\phi$ around its vacuum expectation value
$\phi_0$ as 
\begin{equation}\label{eq:coupling_2d}
 \phi=\phi_0 + \bar g\varphi, \qquad\quad 
  \bar g = \left(\frac{2\pi\bar\eps}m\right)^{1/2}
  \left(\frac{m\mu}{2\pi}\right)^{-\bar\epsilon/4},
\end{equation}
where the effective coupling $\bar g\sim\bar\eps^{1/2}$ in
Eq.~(\ref{eq:g-bar}) was introduced.  The extra factor
$\left(m\mu/2\pi\right)^{-\bar\eps/4}$ was chosen so that the product of
fields $\varphi^*\varphi$ has the same dimension as the Lagrangian
density~\footnote{The choice of the extra factor is arbitrary, if it has
the correct dimension, and does not affect the final results because the
difference can be absorbed into the redefinition of the fluctuation
field $\varphi$.  The particular choice of $\bar g$ in
Eq.~(\ref{eq:coupling_2d}) will simplify the form of loop integrals in
the intermediate steps.}.

Then we rewrite the Lagrangian density as a sum of three parts, 
$\mathcal{L}=\mathcal{\bar L}_0+\mathcal{\bar L}_1+\mathcal{\bar L}_2$, 
where 
\begin{align}\label{eq:L-2d}
 \mathcal{\bar L}_0 & = \Psi^\+\left(i\d_t + \frac{\sigma_3\grad^2}{2m}
 + \mu\sigma_3 + \sigma_+\phi_0 + \sigma_-\phi_0\right)\Psi \,, \\ 
 \mathcal{\bar L}_1 & = - \varphi^*\varphi + \bar g
 \Psi^\+\sigma_+\Psi\varphi + \bar g\Psi^\+\sigma_-\Psi\varphi^*,
 \phantom{\frac\int\int}\hspace{-4.5mm} \\
 \mathcal{\bar L}_2 & = \varphi^*\varphi\,.
\end{align}
The part $\mathcal{\bar L}_0$ represents the gapped fermion
quasiparticle, whose propagator is given by 
\begin{equation}
  \bar G(p_0,\p) = \frac1{p_0^{\,2}-\bar E_\p^{\,2}+i\delta}
  \begin{pmatrix}
   p_0 + \ep -\mu & -\phi_0 \\
   -\phi_0 & p_0 -\ep + \mu
  \end{pmatrix},
\end{equation}
with $\bar E_\p=\sqrt{(\ep-\mu)^2+\phi_0^{\,2}}$ being the usual gapped
quasiparticle spectrum in the BCS theory.

\begin{figure}[tp]
 \begin{center}
  \includegraphics[width=0.8\textwidth,clip]{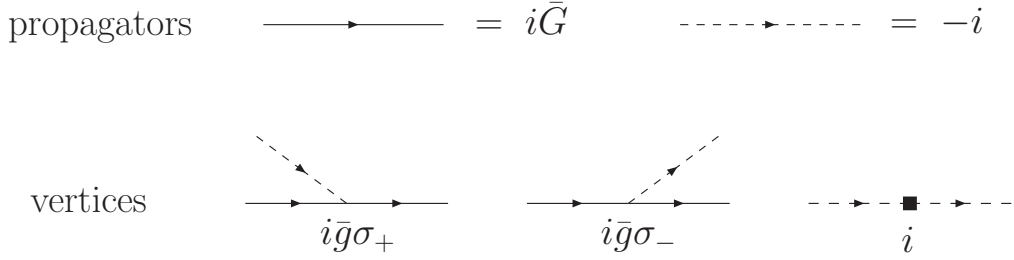}
 \caption{Feynman rules for the expansion around two spatial
 dimensions from the Lagrangian density in Eq.~(\ref{eq:L-2d}).  The
 first line gives propagators, while the second line gives vertices.
 \label{fig:feynman_rules-2d}} 
 \end{center}
\end{figure}

The second part $\mathcal{\bar L}_1$ represents the interaction between 
fermions induced by the auxiliary field $\varphi$.  The first term
in $\mathcal{\bar L}_1$ gives the propagator of the auxiliary field
$\varphi$ 
\begin{equation}
 \bar D(p_0,\p) = -1, 
\end{equation} 
and the last two terms give vertices coupling two fermions and
$\varphi$.  If we did not have the part $\mathcal{\bar L}_2$, we could 
integrate out the auxiliary fields $\varphi$ and $\varphi^*$ to lead to 
\begin{equation}
 \mathcal{\bar L}_1\to \bar g^2\Psi^\+\sigma_+\Psi\,\Psi^\+\sigma_-\Psi
  =\bar g^2\psi^\dagger_\uparrow\psi^\dagger_\downarrow
  \psi_\downarrow\psi_\uparrow,
\end{equation}
which gives the contact interaction of fermions with the small coupling 
$\bar g^2\sim\bar\eps$ as depicted in Fig.~\ref{fig:scattering}.  The
vertex in the third part $\mathcal{\bar L}_2$ plays a role of a counter
term so as to avoid double counting of a certain type of diagram which
is already taken into $\mathcal{\bar L}_1$ as we will see below. 
The Feynman rules corresponding to these Lagrangian densities are
summarized in Fig.~\ref{fig:feynman_rules-2d}. 

\section{Power counting rule of $\bar\eps$}
We can construct the similar power counting rule of $\bar\eps$ as in the
case of the expansion around four spatial dimensions.  First, we make a
prior assumption $\phi_0/\mu\sim e^{-1/\bar\eps}$, which will be checked 
later.  Since $e^{-1/\bar\eps}$ is exponentially small compared to any
powers of $\bar\eps$, we can neglect the contributions of $\phi_0$ when
we expand physical observables in powers of $\bar\eps$.  Then, since
each pair of fermion and $\varphi$ vertices brings a factor of
$\bar\eps$, the naive power of $\bar\eps$ for a given diagram is
$N_{\bar g}/2$, where $N_{\bar g}$ is the number of couplings 
$\bar g$ from $\mathcal{\bar L}_1$. 

\begin{figure}[tp]
 \begin{center}
  \includegraphics[width=0.65\textwidth,clip]{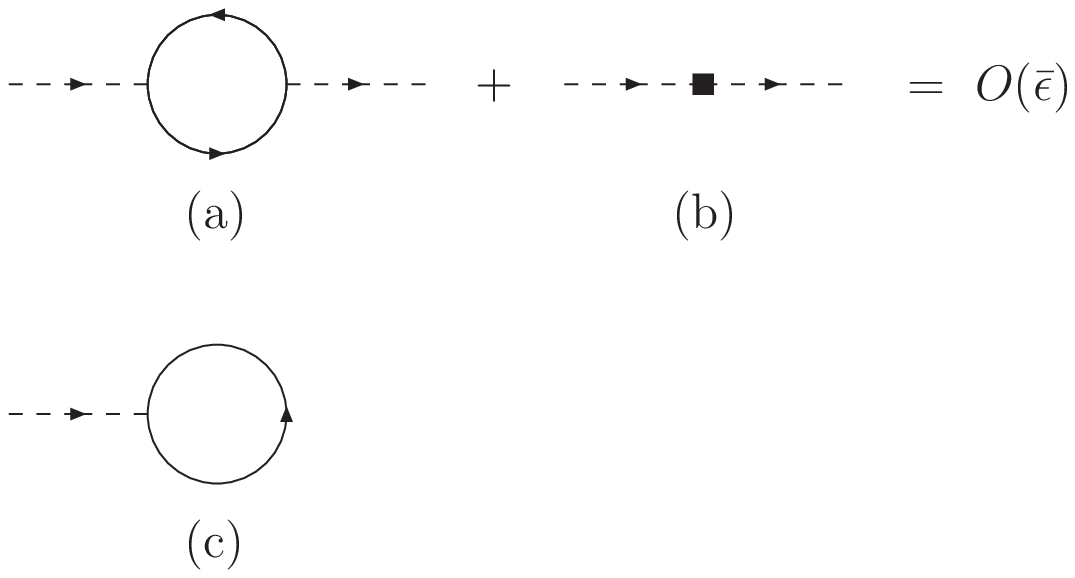}
 \caption{Two apparent exceptions of naive power counting rule of
 $\bar\eps$, (a, c).  The boson self-energy diagram (a) is combined
 with the vertex from $\mathcal{\bar L}_2$ (b) to restore the naive 
 $\bar\eps$ counting.  The condition of disappearance of the tadpole
 diagram (c) gives the gap equation to determine the value of condensate
 $\phi_0$.  \label{fig:cancel-2d}} 
 \end{center}
\end{figure}

However, this naive counting does not take into account the fact that
there might be inverse powers of $\bar\epsilon$ that come from integrals 
which have logarithmic divergences at $d=2$.  Each loop integral in
the ultraviolet region behaves as 
\begin{equation}
 \int dp_0d\p\sim \int d\p\,\ep\sim p^{4},
\end{equation}
while each fermion propagator behaves, at worst, as 
$\bar G(p)\sim p^{-2}$.  Therefore, a given diagram may 
diverges as $\sim p^\mathcal{D}$ with $\mathcal{D}$ being the
superficial degree of divergence given by  
\begin{equation}
 \mathcal{D}=4L-2P_\mathrm{F}.
\end{equation}
Here $L$ is the number of loop integrals and $P_\mathrm{F}$ is the
number of fermion propagators.  
Using the similar relations to Eq.~(\ref{eq:loop}), 
\begin{equation}
 \begin{split}
  L&=P_\mathrm{F}+P_\mathrm{B}-N_{\bar g}+1, \phantom{\frac{}{\int}}\\
  N_g&=P_\mathrm{F}+\frac{E_\mathrm{F}}2=2P_\mathrm{B}+E_\mathrm{B},
 \end{split}
\end{equation}
the superficial degree of divergence is written in
terms of the number of external fermion (auxiliary field) lines,
$E_\mathrm{F(B)}$, as 
\begin{equation}
 \mathcal{D}=4-E_\mathrm{F}-2E_\mathrm{B}.
\end{equation}
Therefore, the inverse power of $\bar\eps$ is possible only in diagrams
which satisfies $E_\mathrm{F}+2E_\mathrm{B}\leq4$. 
Moreover from the analytic properties of $G(p)$ in the
ultraviolet region discussed in Eq.~(\ref{eq:analytic}), one can show
that there are only two skeleton diagrams which have the $1/\bar\eps$ 
singularity near two dimensions.  They are one-loop diagrams of the
boson self-energy [Figs.~\ref{fig:cancel-2d}(a)], and the $\varphi$
tadpole diagram [Fig.~\ref{fig:cancel-2d}(c)]. 

The boson self-energy diagram in Fig.~\ref{fig:cancel-2d}(a) is
evaluated as 
\begin{equation}
 \begin{split}
  -i\bar\Pi_\mathrm{a}(p) 
  &= -\bar g^2\int\!\frac{dk}{(2\pi)^{d+1}}\, 
  G_{11}\!\left(k+\frac p2\right)G_{22}\!\left(k-\frac p2\right) \\ 
  &=i\bar g^2\int\!\frac{d\k}{(2\pi)^{d}}
  \frac{\theta(\varepsilon_{\k+\frac\p2}-\mu)-\theta
  (\mu-\varepsilon_{\k-\frac\p2})}{2\ek-(p_0-\frac12\ep+2\mu+i\delta)},
 \end{split}
\end{equation}
where we have neglected the contribution of $\phi_0$. 
The integral over $\k$ is logarithmic divergent at $d=2$ and has a pole
at $\bar\eps=0$.  Thus it is $O(1)$ instead of $O(\bar\eps)$ according
to the naive counting.  The residue at the pole can be computed as
\begin{align}
  \bar\Pi_\mathrm{a}(p) & = -\bar g^2\int\!\frac{d\k}{(2\pi)^d}
  \frac1{2\ek-(p_0-\frac12\ep+2\mu+i\delta)} + \cdots \notag\\
  &= 1 + O(\bar\eps),
\end{align}
which is cancelled out exactly by adding the vertex
$\bar\Pi_0=-1$ in $\mathcal{\bar L}_2$.  Therefore the diagram
of the type in Fig.~\ref{fig:cancel-2d}(a) should be combined with the 
vertex from $\mathcal{\bar L}_2$ in Fig.~\ref{fig:cancel-2d}(b) to restore 
the naive $\bar\eps$ power counting result, $O(\bar\eps)$. 

Similarly, the tadpole diagram in Fig.~\ref{fig:cancel-2d}(c) also
contains the $1/\bar\eps$ singularity.  
The requirement that this tadpole diagram should vanish by itself
gives the gap equation to the leading order to determine the condensate
$\phi_0$ as wee will see in the succeeding section.


\section{Gap equation}
The condensate $\phi_0$ as a function of the chemical potential $\mu$ is
determined by the gap equation which is obtained by the condition of the
disappearance of all tadpole diagrams.  The leading contribution to the
gap equation is the one-loop diagram drawn in Fig.~\ref{fig:cancel-2d}(c),
which is given by 
\begin{equation}
 \bar\Xi_1 = \bar g\int\!\frac{dk}{(2\pi)^{d+1}}\,G_{21}(k)
  = i\bar g\int\!\frac{d\k}{(2\pi)^d}\,\frac{\phi_0}{2\bar E_\p}.
\end{equation}
By changing the integration variable to $z=\ep/\mu$, we obtain 
\begin{equation}
 \begin{split}
  \bar\Xi_1 &= \frac{i\bar g\phi_0}{2\mu}
  \frac{\left(\frac{m\mu}{2\pi}\right)^{\frac d2}}
  {\Gamma\!\left(\frac d2\right)} 
  \int_0^\infty\!dz\,\frac{z^{\bar\eps/2}}{\sqrt{(z-1)^2+(\phi_0/\mu)^2}}.
 \end{split}
\end{equation}
The integration over $z$ in the dimensional regularization can be
performed to lead to 
\begin{equation}
 \bar\Xi_1 = \frac{i\bar g\phi_0}{\mu}
  \frac{\left(\frac{m\mu}{2\pi}\right)^{\frac d2}}
  {\Gamma\!\left(\frac d2\right)}\left[\ln\frac{2\mu}{\phi_0}
	    -\frac1{\bar\eps}+O(\bar\eps^2)\right].
\end{equation}
The first term $\ln2\mu/\phi_0$ originates from the singularity
around the Fermi surface as is well known as the Cooper instability, 
while the second term $1/\bar\eps$ is from the logarithmic singularity
of the $\k$ integration at $d=2$ in $\bar\Xi_1$.  Solving the gap
equation $\bar\Xi_1=0$,  we obtain the condensate as
$\phi_0=2\mu\,e^{-1/\bar\eps}$.  Note that this result is equivalent to
that obtained by the mean field BCS theory. 

\begin{figure}[tp]
 \begin{center}
  \includegraphics[width=0.4\textwidth,clip]{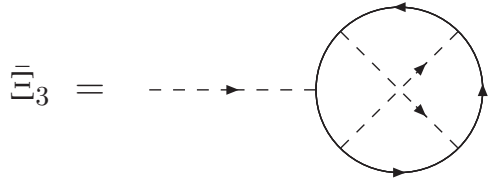}
 \caption{The tadpole diagram which gives the medium-effect
 correction to the gap equation.  \label{fig:tadpole-2d}} 
 \end{center}
\end{figure}

It is known that the pre--exponential factor in the mean field result
$\phi_0=2\mu\,e^{-1/\bar\eps}$ is modified due to the effects of
medium~\cite{gorkov,Heiselberg}.  In the language of the tadpole
diagrams, the corresponding modification to the gap equation comes from
the three-loop diagram $\bar\Xi_3$ depicted in
Fig.~\ref{fig:tadpole-2d}.  The diagram, which seems proportional to
$g^4\sim\bar\eps^2$, gives the $O(1)$ correction to the gap equation. 

Using the Feynman rules, the tadpole diagram in
Fig.~\ref{fig:tadpole-2d} is given by 
\begin{equation}
 \begin{split}
  \bar\Xi_3 &=-\bar g^5\int\!\frac{dk\,dp\,dq}{(2\pi)^{3d+3}}
  \Tr\left[\bar G\sigma_+\bar G\sigma_-
  \bar G\sigma_+\bar G\sigma_+\bar G\sigma_-\right] \\
  &=-\bar g^5\int\!\frac{dk\,dp\,dq}{(2\pi)^{3d+3}}\,
  \bar G_{11}(p)\,\bar G_{22}(p)\,
  \bar G_{11}(p-k)\,\bar G_{21}(q)\,\bar G_{22}(q+k).
 \end{split}
\end{equation}
Since the second term in the product,
\begin{equation}
 \bar G_{11}(p)\,\bar G_{22}(p)
  =\frac1{p_0^{\,2}-\bar E_\p^{\,2}}
  +\frac{\phi_0^{\,2}}{(p_0^{\,2}-\bar E_\p^{\,2})^2},
\end{equation}
gives only the $O(\bar\eps^2)$ correction to the gap equation, we can
neglect it for the current purpose.  Then we perform the integration
over $p_0$ and $q_0$ to result in
\begin{equation}
 \begin{split}
  \bar\Xi_3
  =-\bar g^5\int\!\frac{dk\,d\p\,d\q}{(2\pi)^{3d+1}}\,
  \frac{\phi_0}{4\bar E_\p \bar E_\q}
  &\left[\frac{\bar E_\p-k_0+\varepsilon_{\p-\k}-\mu}
  {(\bar E_\p-k_0)^2-\bar E_{\p-\k}^{\,2}}
  +\frac{\bar E_\p+\varepsilon_\p-\mu}
  {(k_0+\bar E_\p)^2-\bar E_{\p+\k}^{\,2}}\right]\\
  &\times\left[\frac{k_0+\bar E_\q-\varepsilon_{\q+\k}+\mu}
  {(k_0+\bar E_\q)^2-\bar E_{\q+\k}^{\,2}}
  +\frac{\bar E_\q-\varepsilon_\q+\mu}
  {(k_0-\bar E_\q)^2-\bar E_{\q-\k}^{\,2}}\right].
 \end{split}
\end{equation}
Because of the factor $1/\bar E_\p \bar E_\q$ in the integrand, the
integrations over $\p$ and $\q$ are dominated around the Fermi surface
where $\varepsilon_{\p (\q)}\sim\mu$ and hence 
$\bar E_{\p (\q)}\sim\phi_0$.  Keeping only the dominant part in the
integrand, we can write the 
integral as
\begin{equation}
  \bar\Xi_3
  \simeq\bar g^5\int\!\frac{dk\,d\p\,d\q}{(2\pi)^{3d+1}}
  \left.\frac{\phi_0}{4\bar E_\p \bar E_\q}
  \frac{1}{(k_0+\varepsilon_{\p-\k}-\mu)(k_0+\varepsilon_{\q+\k}-\mu)}
  \right|_{\varepsilon_{\p(\q)}=\mu}.
\end{equation}
Now the integration over $k_0$ can be performed easily to lead to
\begin{equation}
  \bar\Xi_3 = -i\frac{\bar g^5}2\phi_0
  \int\!\frac{d\p\,d\q}{(2\pi)^{2d}}\,\frac{1}{\bar E_\p \bar E_\q}
  \int\!\frac{d\k}{(2\pi)^{d}}\left.
  \frac{\theta(\varepsilon_{\q+\k}-\mu)\,\theta(\mu-\varepsilon_{\p-\k})}
  {\varepsilon_{\q+\k}-\varepsilon_{\p-\k}}\right|_{\varepsilon_{\p(\q)}=\mu}.
\end{equation}
If we evaluated the $\k$ integration at $d=3$, we would obtain the
static Lindhard function representing the medium-induced
interaction~\cite{Fetter-Walecka}.

Here we shall evaluate $\bar\Xi_3$ at $d=2$.  Changing the
integration variables to $\ep$, $\eq$, $\ek$,
$\cos\chi_p=\hat\k\cdot\hat\p$, $\cos\chi_q=\hat\k\cdot\hat\q$, and
performing the integrations over $\ep$ and $\eq$, we obtain
\begin{equation}
  \bar\Xi_3 \simeq -i\frac{\bar g^5}2\phi_0
  \left(\frac m{2\pi}\right)^3\left(2\ln\frac\mu{\phi_0}\right)^2
  \int\!\frac{d\ek\,d\chi_p\,d\chi_q}{\pi^2} \left.
  \frac{\theta(\varepsilon_{\q+\k}-\mu)\,\theta(\mu-\varepsilon_{\p-\k})}
  {\varepsilon_{\q+\k}-\varepsilon_{\p-\k}}\right|_{\varepsilon_{\p(\q)}=\mu}.
\end{equation}
The range of the integrations over $\chi_p$ and $\chi_q$ are from $0$ to
$\pi$.  Since $\phi_0/\mu\propto e^{-1/\bar\eps}$, we have 
$(2\ln\mu/\phi_0)^2\sim(2/\bar\eps)^2$ which cancels $\bar\eps^2$ coming
from the four vertex couplings $\bar g^4$. 
Finally the integrations can be performed as follows:
\begin{equation}
 \begin{split}
  &\int\!\frac{d\ek\,d\chi_p\,d\chi_q}{\pi^2}\,
  \frac{\theta(\varepsilon_{\q+\k}-\mu)\,\theta(\mu-\varepsilon_{\p-\k})}
  {\varepsilon_{\q+\k}-\varepsilon_{\p-\k}} \\
  &\qquad\qquad =\int_0^{4\mu\cos^2\!\chi_p}\!d\ek
  \int_0^{\pi/2}\!\frac{d\chi_p}{\pi}\int_0^{\pi/2}\!\frac{d\chi_q}{\pi}
  \frac1{2\sqrt{\mu\ek}\left(\cos\chi_p+\cos\chi_q\right)} \\
  &\qquad\qquad\qquad 
  +\int_{4\mu\cos^2\!\chi_q}^{4\mu\cos^2\!\chi_p}\!d\ek\int_0^{\pi/2}\!
  \frac{d\chi_p}{\pi}\int_{\pi/2}^{\pi-\chi_p}\!\frac{d\chi_q}{\pi}\,
  \frac1{2\sqrt{\mu\ek}\left(\cos\chi_p+\cos\chi_q\right)} \\
  &\qquad\qquad =\frac12\,,
 \end{split}
\end{equation}
which gives $\bar\Xi_3$ as
\begin{equation}
 \bar\Xi_3\simeq-i\bar g\phi_0\frac{m}{2\pi}.
\end{equation}

Consequently, the gap equation $\bar\Xi_1+\bar\Xi_3=0$, which receives the
$O(1)$ correction due to $\bar\Xi_3$, is modified as
\begin{equation}
 \ln\frac{2\mu}{\phi_0}-\frac1{\bar\eps}-1+O(\bar\eps)=0.
\end{equation}
The solution of the gap equation becomes
\begin{equation}
 \phi_0=2\mu\,\exp\!\left[-\frac1{\bar\eps}-1+O(\bar\eps)\right]
  =\frac{2\mu}{e}\left[1+O(\bar\eps)\right]e^{-1/\bar\eps},
\end{equation}
where the value of condensate is reduced by the factor 
$e\approx2.71828$. 
The reduction of the pre--exponential factor due to the medium effects
is known as the Gor'kov correction at $d=3$
theories~\cite{gorkov,Heiselberg}.

\section{Thermodynamic quantities}
\begin{figure}[tp]
 \begin{center}
  \includegraphics[width=0.4\textwidth,clip]{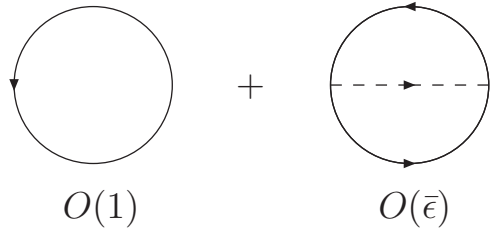}
 \caption{Vacuum diagrams contributing to the pressure up to 
 next-to-leading order in $\bar\eps$.  \label{fig:potential-2d}}  
 \end{center}
\end{figure}

The value of the effective potential $V_\mathrm{eff}$ at its minimum
determines the pressure $P=-\bar V_\mathrm{eff}(\phi_0)$ at a given
chemical potential $\mu$.  Since the energy gain due to the
superfluidity $\phi_0^{\,2}\sim e^{-2/\bar\eps}$ is exponentially small
compared to any power series of $\bar\eps$, we can simply neglect the
contributions of $\phi_0$ to the pressure.  To the next-to-leading
order, the effective potential receives contribution from two vacuum
diagrams drawn in Fig.~\ref{fig:potential-2d}: fermion loops without and
with an exchange of the auxiliary field.  The one-loop diagram at
$\phi_0=0$ is $O(1)$ and given by 
\begin{equation}\label{eq:P_free}
 \bar V_1(0) = -2\int\!\frac{d\p}{(2\pi)^d}(\mu-\ep)_> 
  =-\frac{2\mu\left(\frac{m\mu}{2\pi}\right)^{\frac d2}}
  {\Gamma\!\left(\frac d2+2\right)}
  \equiv -P_\mathrm{free},
\end{equation}
which represents the contributions of free fermions to the pressure. 
The two-loop diagram at $\phi_0=0$ is $O(\bar\eps)$, which represents
the density-density correlation as
\begin{equation}\label{eq:P_2loop}
 \begin{split}
  \bar V_2(0) &= \bar g^2\int\!\frac{dp\,dq}{(2\pi)^{2d+2}}\,
  \bar G_{11}(p)\bar G_{22}(q)\\  &= -\bar g^2
  \left[\int\!\frac{d\p}{(2\pi)^d}\theta(\mu-\ep)\right]^2
  =-\bar\eps\frac{\mu\left(\frac{m\mu}{2\pi}\right)^{\frac d2}}
  {\Gamma\!\left(\frac d2+1\right)^2}.
 \end{split}
\end{equation}
Thus we obtain the pressure up to the next-to-leading order in
$\bar\eps$ as
\begin{equation}
 P=\left(1+\bar\eps\right)P_\mathrm{free}.
\end{equation}

Accordingly, the fermion number density is given by
$N=\d P/\d\mu=\left(1+\bar\eps\right)N_\mathrm{free}$.  The Fermi
energy is obtained from the thermodynamics of free gas in $d$ spatial
dimensions as 
\begin{equation}\label{eq:eF-2d}
  \eF=\frac{2\pi}m
  \left[\frac12\,\Gamma\!\left(\frac d2+1\right)N\right]^{2/d}
  =\left(1+\bar\eps\right)\mu,
\end{equation}
which yields the universal parameter of the unitary Fermi gas from the
$\bar\eps$ expansion as
\begin{equation}\label{eq:xi-2d}
 \xi=\frac\mu\eF=1-\bar\eps+O(\bar\eps^2).
\end{equation}

\section{Quasiparticle spectrum}
\begin{figure}[tp]
 \begin{center}
  \includegraphics[width=0.75\textwidth,clip]{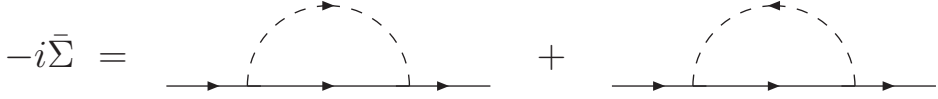}
 \caption{One-loop diagrams contributing to the fermion self-energy in 
 order $O(\bar\epsilon)$. \label{fig:self_energy-2d}}
 \end{center}
\end{figure}

To the leading order in $\bar\eps$, the dispersion relation of the
fermion quasiparticle is given by 
$\omega_\mathrm{F}(\p)=\bar E_\p=\sqrt{(\ep-\mu)^{2}+\phi_0^{\,2}}$,
which has the same form as that in the mean field BCS theory. 
There exist the next-to-leading order corrections to the fermion
quasiparticle spectrum from the one-loop self-energy diagrams, 
$-i\bar\Sigma(p)$, depicted in Fig.~\ref{fig:self_energy-2d}.
These corrections are only to the diagonal elements of the self-energy
and each element is evaluated as
\begin{equation}
 \begin{split}
  \bar\Sigma_{11}(p) 
  &= -i\bar g^2\int\!\frac{dk}{(2\pi)^{d+1}}\,\bar G_{22}(k)\\
  &=-\bar g^2\int\!\frac{d\k}{(2\pi)^{d}}\,\theta(\mu-\ek)=-\bar\eps\mu
 \end{split}
\end{equation}
and
\begin{equation}
 \begin{split}
  \bar\Sigma_{22}(p) 
  &= -i\bar g^2\int\!\frac{dk}{(2\pi)^{d+1}}\,\bar G_{11}(k)\\
  &=\bar g^2\int\!\frac{d\k}{(2\pi)^{d}}\,\theta(\mu-\ek)=\bar\eps\mu.
 \end{split}
\end{equation}
To this order, the self-energy is momentum independent, which
effectively shifts the chemical potential due to the interaction with
the other component of fermions. 

By solving the equation
$\det[\bar G^{-1}(\omega,\p)-\bar\Sigma]=0$ in terms of $\omega$, 
the dispersion relation of the fermion quasiparticle up to the
next-to-leading order is given by 
\begin{equation}
  \omega_\mathrm{F}(\p)=\sqrt{(\ep-\mu-\bar\eps\mu)^2+\phi_0^{\,2}}. 
\end{equation}
The minimum of the dispersion curve is located at a nonzero value of
momentum, $|\p|=(2m\varepsilon_0)^{1/2}$, where
\begin{equation}\label{eq:e_0-2d}
 \varepsilon_0=\left(1+\bar\eps\right)\mu.
\end{equation}
The location of the minimum coincides with the Fermi energy in
Eq.~(\ref{eq:eF-2d}), $\varepsilon_0=\eF$, in agreement with the
Luttinger theorem~\cite{luttinger}.  The energy gap $\Delta$ of the
fermion quasiparticle is given by the condensate, 
\begin{equation}\label{eq:gap-2d}
 \Delta=\phi_0 =\frac{2\mu}{e}\,e^{-1/\bar\eps}.
\end{equation}

\section{Extrapolation to $\bar\eps$=1}
Now we discuss the extrapolation of the expansion over $\bar\eps=d-2$ to 
the physical case at three spatial dimensions.  In contradiction to
the case of $\eps=4-d$ expansion, the coefficients of $O(\bar\eps)$
corrections are not small.  If we naively extrapolate the leading and
next-to-leading order results for $\xi$ in Eq.~(\ref{eq:xi-2d}),
$\Delta$ in Eq.~(\ref{eq:gap-2d}), and $\varepsilon_0$ in
Eq.~(\ref{eq:e_0-2d}) to $\bar\eps=1$, we would have
\begin{equation}
 \xi \approx 0, \qquad \frac{\Delta}{\mu}\approx 0.271,  
  \qquad \frac{\varepsilon_0}{\mu}\approx 2,
\end{equation}
which are not as good as the extrapolations in the expansions over
$\eps$ in Eq.~(\ref{eq:extrapolation}). 
Thus, instead of naively extrapolating the $\bar\eps$ expansions to
$d=3$, we use them as boundary conditions to improve the series
summation of the $\eps$ expansions in Chap.~\ref{sec:matching}.

\section{NNLO correction for $\xi$}

\begin{figure}[tp]
 \begin{center}
  \includegraphics[width=0.35\textwidth,clip]{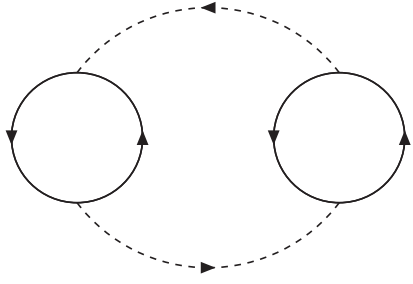}
 \caption{Vacuum diagram contributing to the pressure to the order
  $\bar\eps^2$.  The counter vertex $-i\bar\Pi_0=i$ for each bubble
  diagram is understood implicitly.  The other $O(\bar\eps^2)$ diagrams
  identically vanish.  \label{fig:nnlo-2d}}  
 \end{center}
\end{figure}

The order $\bar\eps^2$ contribution to the effective potential or the
pressure is solely given by the three-loop diagram depicted in
Fig.~\ref{fig:nnlo-2d}, which is written as
\begin{equation}
 \bar V_3(0) = \frac i2\int\!\frac{dk}{(2\pi)^{d+1}}
 \left[1+\bar g^2\!\int\!\frac{d\p}{(2\pi)^{d}}
 \frac{\theta(\varepsilon_{\p+\frac\k2}-\mu)-\theta
 (\mu-\varepsilon_{\p-\frac\k2})}{2\ep-(k_0-\frac12\ek+2\mu+i\delta)}
 \right]^2. 
\end{equation}
where $p_0$ integrations in each bubble diagram are already performed.
Note that $+1$ in the bracket comes from the counter vertex
$-\bar\Pi_0=1$.  We can see that the other $O(\bar\eps^2)$ diagrams
identically vanish.  The integration over $k_0$ in $\bar V_3$ leads to
\begin{equation}
 \bar V_3(0) = -\bar g^2\int\!\frac{d\k d\p}{(2\pi)^{2d}}\,
 \theta(\mu-\varepsilon_{\p+\frac\k2})\theta(\mu-\varepsilon_{\p-\frac\k2})
 \left[1+\bar g^2\!\int\!\frac{d\q}{(2\pi)^{d}}
 \frac{\theta(\varepsilon_{\q+\frac\k2}-\mu)\theta
 (\varepsilon_{\q-\frac\k2}-\mu)}{2\eq-2\ep}\right].
\end{equation}
Due to the $\theta$-functions, the range of integral over $\ek$ is
limited to $0\leq\ek\leq4\mu$ and the range of integral over $\ep$ is
limited to $0\leq\ep\leq\Lambda_p$, where
\begin{equation}
 \sqrt{\Lambda_p}=\frac{-|\cos\chi_p|\sqrt\ek+\sqrt{4\mu-\ek\sin^2\chi_p}}2
\end{equation}
with $\cos\chi_p=\hat\k\cdot\hat\p$.  Similarly, the range of integral
over $\eq$ is limited to $\Lambda_q\leq\eq$ where
\begin{equation}
  \sqrt{\Lambda_q}=\frac{|\cos\chi_q|\sqrt\ek+\sqrt{4\mu-\ek\sin^2\chi_q}}2
\end{equation}
with $\cos\chi_q=\hat\k\cdot\hat\q$.  Then the integration over $\eq$ is
performed to result in 
\begin{equation}
 \bar V_3(0)
  = \bar\eps^2\frac{m}{4\pi}\int_0^{4\mu}\!d\ek
  \int_0^\pi\frac{d\chi_p}{\pi}\int_0^\pi\frac{d\chi_q}{\pi}
  \int_0^{\Lambda_p}\!d\ep\left[\gamma+
  \ln\!\left(\frac{\Lambda_q-\ep}\mu\right)\right]
  +O(\bar\eps^3).
\end{equation}
Finally, introducing the dimensionless variable $z=\ek/\mu$ and
performing the integration over $\ep$, we obtain the following
expression for $V_3$, 
\begin{equation}
 \bar V_3(0) = \bar\eps^2\frac{m\mu^2}{4\pi}\left[\gamma+\int_0^{4}\!dz
 \int_0^{\pi}\frac{d\chi_p}{\pi}\int_0^{\pi}\frac{d\chi_q}{\pi}
 \left\{\tilde\Lambda_q\ln\tilde\Lambda_q-(\tilde\Lambda_q-\tilde\Lambda_p)
 \ln(\tilde\Lambda_q-\tilde\Lambda_p)-\tilde\Lambda_p\right\}\right],
\end{equation}
where $\tilde\Lambda_{p(q)}=\Lambda_{p(q)}/\mu$. 
The numerical integrations over $z$, $\chi_p$, and $\chi_q$ give
\begin{equation}
 \bar V_3(0)=\bar\eps^2\frac{m\mu^2}{2\pi}
  \left(\frac\gamma2+0.0568528\right).
\end{equation}

Therefore, combining this result with Eqs.~(\ref{eq:P_free}) and
(\ref{eq:P_2loop}), we obtain the pressure up to the
next-to-next-to-leading order (NNLO) in $\bar\eps$ as 
\begin{equation}
 P=P_\mathrm{free}
  \left[1+\bar\eps+\left(\frac\gamma2-\frac14\right)\bar\eps^2
   -\left(\frac\gamma2+0.0568528\right)\bar\eps^2\right]
  =P_\mathrm{free}
  \left[1+\bar\eps-0.3068528\,\bar\eps^2\right].
\end{equation}
Then $\xi$ up to the order $\bar\eps^2$ is found to be given by 
\begin{equation}
 \xi=\frac\mu\eF
  =\left[1+\bar\eps-0.3068528\,\bar\eps^2\right]^{-\frac2{2+\bar\eps}}
  =1-\bar\eps+1.80685\,\bar\eps^2+O(\bar\eps^3).
\end{equation}
Although the $O(\bar\eps^2)$ correction to the pressure is small,
the NNLO correction to $\xi$ turns out to be large because of the large
$O(\bar\eps)$ correction in the pressure.

\chapter{Matching of expansions around $d=4$ and $d=2$
\label{sec:matching}} 
As we have mentioned previously, we shall match two expansions around
$d=4$ and $d=2$, which are studied in Chapters~\ref{sec:4d} and
\ref{sec:2d} respectively, in order to extract results at $d=3$.  
We use the results around two spatial dimensions as boundary
conditions which should be satisfied by the series summations of the
expansions over $\eps=4-d$.  Because we do not yet have a precise
knowledge on the large order behavior of the expansion around four
spatial dimensions, we assume its Borel summability and employ Pad\'e
approximants. 

Let us demonstrate the matching of two expansions by taking $\xi$ as an
example.  In Ref.~\cite{nussinov04}, the linear interpolation between
exact values at $d=2$ $(\xi=1)$ and $d=4$ $(\xi=0)$ was discussed to
yield $\xi=0.5$ at $d=3$.  Now we have series expansions around these
two exact limits. 
The expansion of $\xi$ in terms of $\eps=4-d$ is obtained in
Eq.~(\ref{eq:xi}).  Assuming the Borel summability of the $\eps$
expansion, we write $\xi$ as a function of $\eps$ in the form of the
Borel transformation, 
\begin{equation}\label{eq:borel}
 \xi(\eps)=\frac{\eps^{3/2}}2
 \exp\!\left(\frac{\epsilon\ln\epsilon}{8-2\epsilon}\right)
 \int_0^\infty\!dt\, e^{-t}B_\xi(\eps t),
\end{equation}
where we factorized out the non-trivial dependence on $\epsilon$
explicitly.  $B_\xi(t)$ is the Borel transform of the power series in
$\xi(\eps)$, whose Taylor coefficients at origin is given from the
$\eps$ expansion of $\xi$ as 
\begin{equation}\label{eq:borel_sum}
 B_\xi(t)=1 - \left(3C -\frac54 (1-\ln2)\right) t +\cdots.
\end{equation}

In order to perform the integration over $t$ in Eq.~(\ref{eq:borel}),
the analytic continuation of the Borel transform $B_\xi(t)$ to the real 
positive axis of $t$ is necessary.  Here we employ the Pad\'e
approximant, where $B_\xi(t)$ is replaced by the following
rational functions 
\begin{equation}
 B_\xi(t) = \frac{1+p_1 t+\cdots+p_M t^M}{1+q_1 t+\cdots+q_N t^N}\,.
\end{equation}
From Eq.~(\ref{eq:borel_sum}), we require that the Pad\'e approximants
satisfy $p_1-q_1=-3C+\frac54 (1-\ln2)$.  Furthermore, we incorporate the
results around two spatial dimensions in Eq.~(\ref{eq:xi-2d}) by
imposing 
\begin{equation}
 \xi(2-\bar\eps)=1-\bar\eps+\cdots
\end{equation}
on the Pad\'e approximants as a boundary condition.
Since we have three known coefficients from the two expansions, the
Pad\'e approximants $[M/N]$ satisfying $M+N=3$ are possible. 
Since we could not find a solution satisfying the boundary condition 
$\xi(2-\bar\eps)=1-\bar\eps$ for $[M/N]=[2/1]$, we adopt other three 
Pad\'e approximants with $[M/N]=[3/0],\,[1/2],\,[0/3]$, whose
coefficients $p_m$ and $q_n$ are determined uniquely by the above
conditions. 

\begin{figure}[tp]
 \begin{center}
  \includegraphics[width=0.6\textwidth,clip]{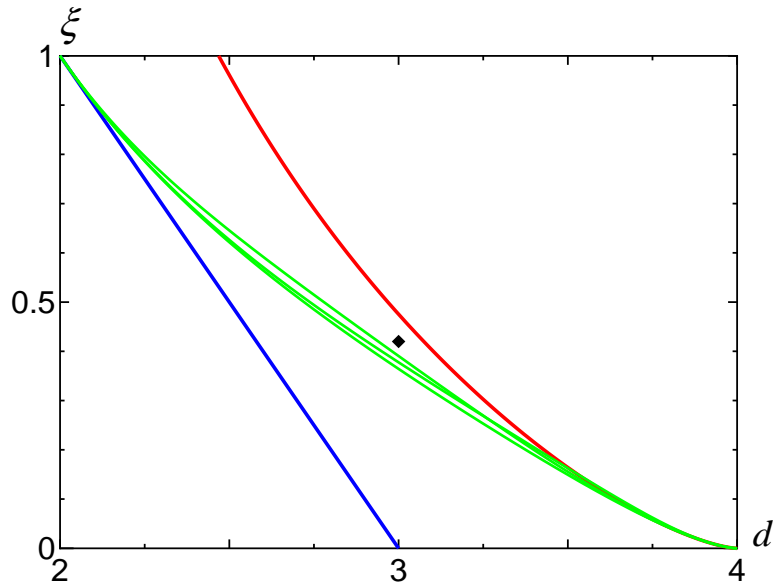}
 \caption{The universal parameter $\xi$ as a function of the spatial
 dimensions $d$.  The upper solid curve is from the expansion around
 $d=4$ in Eq.~(\ref{eq:xi}), while the lower solid line is from the
 expansion around $d=2$ in Eq.~(\ref{eq:xi-2d}).  The middle three
 curves show the different Borel--Pad\'e approximants connecting the
 two expansions.  The diamond indicates the result $\xi\approx0.42$ from
 the Monte Carlo simulation~\cite{Carlson:2005kg}. \label{fig:xi}}
 \end{center}
\end{figure}

Fig.~\ref{fig:xi} shows the universal parameter $\xi$ as a function of
the spatial dimensions $d$.  The middle three curves show $\xi$ in the
different Pad\'e approximants connecting the two expansions around $d=4$
and $d=2$.  These Borel--Pad\'e approximants give $\xi=0.391$,
$0.364$, and $0.378$ at $d=3$,  
which are small compared to the naive extrapolation of the
$\eps$ expansion to $d=3$ $(\xi\to0.475)$. 

The Pad\'e approximant above, however, almost certainly needs
serious modification, since it is known that in the expansion of
$\xi(\eps)$, there exist non-analytic terms at sufficiently higher 
orders~\cite{Arnold}, 
\begin{equation}
 \xi(\eps) = \frac{\eps^{3/2}}2 
  \exp\!\left[\frac{\eps\ln\eps}{8-2\eps}\right]
  \left(1 - 0.0492\eps + \# \eps^2 + \# \eps^3\ln\eps + \cdots\right).
\end{equation}
An understanding of the structure of high-order terms in the
perturbation theory around $d=4$ is currently lacking. 


\chapter{Thermodynamics below $\Tc$ \label{sec:below-Tc}}
Now we investigate the thermodynamics of the Fermi gas at finite
temperature near the unitarity limit.  At zero temperature, we found that
there exist two difference energy scales in the system; the scale of
condensate $\phi_0$ and that of chemical potential
$\mu\sim\eps\phi_0\ll\phi_0$.  Accordingly, we can consider two
temperature regions where the unitary Fermi gas exhibits different
thermodynamics.

One is the low temperature region where $T\sim\eps\phi_0$.  In this
region, the energy gap of the fermion quasiparticle $\Delta\sim\phi_0$ 
is still large compared to the temperature. 
Therefore, thermal excitations of the fermion quasiparticle are
exponentially suppressed by a factor $e^{-\Delta/T}\sim e^{-1/\eps}$.
The thermodynamics in this region is dominated by the bosonic phonon
excitations. 
The other temperature region is the high temperature region where
$T\sim\phi_0$.  ($\phi_0$ represents the condensate at zero
temperature.)  In this region, the condensate decreases and eventually
vanishes at the critical temperature $\Tc$.  Fermions and bosons are
equally excited here.  We defer our discussion on the high temperature
region to Chap.~\ref{sec:above-Tc} and concentrate on the thermodynamics
at the low temperature region $T\ll\Tc$ in this Chapter.

\section{Finite temperature formalism}
The extension to finite temperature $T$ follows from the prescription of
the imaginary time formalism.  The system under consideration is
described by the following Lagrangian density with the imaginary time
$0\leq\tau\leq1/T$:
\begin{align}
  \mathcal{L}_0 & = \Psi^\+\left(\d_\tau - \frac{\sigma_3\grad^2}{2m}
  - \sigma_+\phi_0 - \sigma_-\phi_0\right)\Psi 
  + \varphi^*\left(\d_\tau - \frac{\grad^2}{4m}\right)\varphi
  + \frac{\phi_0^{\,2}}{c_0}\,, \\
  \mathcal{L}_1 & = - g\Psi^\+\sigma_+\Psi\varphi 
  - g\Psi^\+\sigma_-\Psi\varphi^* - \mu\Psi^\+\sigma_3\Psi
  - \left(2\mu-\frac{g^2}{c_0}\right)\varphi^*\varphi 
  + \frac{g\phi_0}{c_0}\varphi + \frac{g\phi_0}{c_0}\varphi^*\,,
 \phantom{\frac{\frac\int\int}{\frac\int\int}}\hspace{-4.5mm} \\
 \mathcal{L}_2 & = -\varphi^*\left(\d_\tau
 - \frac{\grad^2}{4m}\right)\varphi + 2\mu\varphi^*\varphi\,.
\end{align}
The propagators of fermion and boson are generated by $\mathcal{L}_0$.
The fermion propagator is a $2\times2$ matrix, 
\begin{equation}
 G(i\omega_n,\p) = \frac1{(i\omega_n)^2-E_\p^{\,2}}
  \begin{pmatrix}
   i\omega_n + \ep & -\phi_0 \\
   -\phi_0 & i\omega_n-\ep
  \end{pmatrix},
\end{equation}
where $\ep=\p^2/2m$, $E_\p=\sqrt{\ep^{\,2}+\phi_0^{\,2}}$ and $\phi_0$
is the condensate in the superfluid ground state as before. 
The boson propagator $D$ is 
\begin{equation}
  D(i\nu_n,\p) = \left(i\nu_n - \frac{\ep}2\right)^{-1}.
\end{equation}
$\omega_n=2\pi T(n+\frac12)$ and $\nu_n=2\pi Tn$ are discrete Matsubara
frequencies for fermion and boson with an integer
$n=0,\pm1,\pm2,\cdots$.  The unitary Fermi gas around four spatial
dimensions is described by the weakly-interacting system of fermionic
and bosonic quasiparticles, whose coupling $g\sim\eps^{1/2}$ in
$\mathcal{L}_1$ was introduced in Eq.~(\ref{eq:coupling}) as
\begin{equation}
 g = \frac{(8\pi^2\epsilon)^{1/2}}m
  \left(\frac{m\phi_0}{2\pi}\right)^{\epsilon/4}.
\end{equation}
We define the counter vertex in $\mathcal{L}_2$ for the boson propagator
by $\Pi_0(p_0,\p)=p_0-\ep/2$ and the boson chemical potential by
$\muB=2\mu-g^2/c_0$.  When $c_0$ is negative, $-g^2/c_0\simeq\eb$ gives
the binding energy of boson to the leading order in $\eps$.  We consider
the vicinity of the unitary point where $\eb\sim\eps\phi_0$. 

The power counting rule of $\eps$ developed at zero temperature in
Sec.~\ref{sec:counting} holds in the low temperature region where the
condensate is still large compared to the chemical potential
$\mu/\phi_0\sim\eps$, while it breaks down near the critical temperature
because $\phi_0\to0$ at $T\to\Tc$.  In the high temperature region
$T\sim\Tc$, a minor modification of the power counting rule is necessary 
as we discuss in Chap.~\ref{sec:above-Tc}.

\section{Phonon spectrum}

\begin{figure}[tp]
 \begin{center}
  \includegraphics[width=0.9\textwidth,clip]{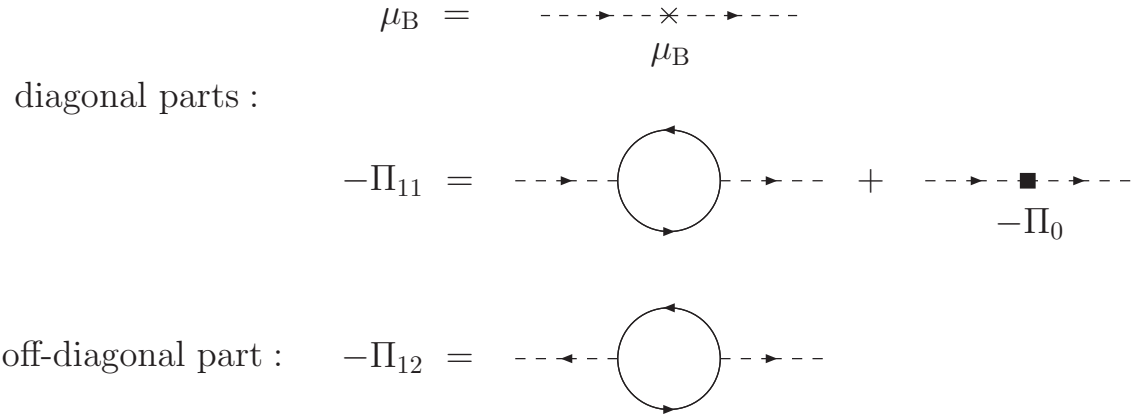}
  \caption{Boson's self-energies contributing to the order $O(\eps)$. 
  Solid (dotted) lines represent the fermion (boson) propagator $-G$
  $(-D)$, while the cross in the first diagram represents the $\muB$
  insertion to the boson propagator.
  The vertex $\Pi_0$ from $\mathcal{L}_2$ needs to be added to the
  second diagram. The last diagram gives the off-diagonal part of the
  self-energy. \label{fig:boson}}  
 \end{center}
\end{figure}

The thermodynamics at the low temperature region $T\ll\phi_0$ is
dominated by the phonon excitations.  In order to determine the phonon
spectrum, we first study the boson self-energy at zero temperature.  To
the order of $O(\eps)$, there are three types of contributions to the
boson self-energy as depicted in Fig.~\ref{fig:boson}.  In addition to
the chemical potential insertion $\muB=2\mu+\eb$, the one-loop diagrams 
contribute to the diagonal part $\Pi_{11}$ and off-diagonal part
$\Pi_{12}$ of the boson self-energy.  The vertex $\Pi_0$ from
$\mathcal{L}_2$ is necessary for $\Pi_{11}$ according to the power
counting rule described in Sec.~\ref{sec:counting}.  Then the diagonal
part of the boson self-energy is given by 
\begin{equation}
 \Pi_{11}(p) = \Pi_0(p) + \Pi_\mathrm{a}(p),
\end{equation}
where $\Pi_0$ is defined in Eq.~(\ref{eq:Pi_0}) and
$\Pi_\mathrm{a}$ is given by
\begin{align}
 -\Pi_\mathrm{a}(p) 
 &= g^2\int\!\frac{idk_{0\,}dk}{(2\pi)^{d+1}}\, 
 G_{11}\!\left(k+\frac p2\right)G_{22}\!\left(k-\frac p2\right) \notag\\ 
 &= g^2
 \int\!\frac{d\k}{(2\pi)^d}\,\frac1{4E_{\k-\frac\p2}E_{\k+\frac\p2}} \\ 
 &\qquad\quad\times\left[\frac{(E_{\k-\frac\p2}+\varepsilon_{\k-\frac\p2})
 (E_{\k+\frac\p2}+\varepsilon_{\k+\frac\p2})}
 {E_{\k-\frac\p2}+E_{\k+\frac\p2}-p_0}
 +\frac{(E_{\k-\frac\p2}-\varepsilon_{\k-\frac\p2})
 (E_{\k+\frac\p2}-\varepsilon_{\k+\frac\p2})}
 {E_{\k-\frac\p2}+E_{\k+\frac\p2}+p_0}\right]. \notag
\end{align}
Since we are interested in physics at the scale of temperature
$T\ll\phi_0$, it is sufficient to evaluate the self-energy when the
external momentum is small $p\sim T\ll\phi_0$.  Expanding $\Pi_{11}(p)$
in terms of $p/\phi_0$ and performing the $\k$ integration with the use
of the formula 
\begin{equation}
 \int_0^\infty\!dz\,\frac{z^{\alpha-1}}{(z+1)^\beta}
  =\frac{\Gamma(\alpha)\Gamma(\beta-\alpha)}{\Gamma(\beta)},
\end{equation}
we obtain
\begin{equation}
 \Pi_{11}(p) \simeq -g^2 \int\!\frac{d\k}{(2\pi)^d}
  \frac{E_\k^{\,2}+\ek^{2}}{4E_\k^{\,3}}
  =\frac32\eps\phi_0+O(\eps^2).
\end{equation}

Similarly, the off-diagonal part of the boson-self energy is given by
\begin{equation}
 \begin{split}
  -\Pi_{12}(p)&=g^2\int\!\frac{idk_{0\,}dk}{(2\pi)^{d+1}}\,
  G_{12}\left(k+\frac p2\right)G_{12}\left(k-\frac p2\right) \\
  &=-g^2\int\!\frac{d\k}{(2\pi)^d}\,
  \frac{\phi_0^{\,2}}{4E_{\k-\frac\p2}E_{\k+\frac\p2}}
  \left[\frac{1}{E_{\k-\frac\p2}+E_{\k+\frac\p2}-p_0}
  +\frac{1}{E_{\k-\frac\p2}+E_{\k+\frac\p2}+p_0}\right].
 \end{split}
\end{equation}
Expanding $\Pi_{12}(p)$ in terms of $p/\phi_0$ and performing the 
integration over $\k$, we obtain
\begin{equation}
  \Pi_{12}(p) \simeq g^2 \int\!\frac{d\k}{(2\pi)^d}
  \frac{\phi^{\,2}}{4E_\k^{\,3}}=\frac12\eps\phi_0+O(\eps^2).
\end{equation}

As a result of the resummation of these self-energies, the resumed 
boson propagator $\mathcal D$ is expressed by the following $2\times2$
matrix:  
\begin{equation}\label{eq:resum}
 \mathcal D(p_0,\p) =
  \begin{pmatrix}
   D(p)^{-1}+\muB-\Pi_{11} & -\Pi_{12} \\
   -\Pi_{21} & D(-p)^{-1}+\muB-\Pi_{22}
  \end{pmatrix}^{-1},
\end{equation}
where $\Pi_{22}=\Pi_{11}$ and $\Pi_{12}=\Pi_{21}$.  The dispersion
relation of the boson $\wph(\p)$ can be obtained by solving the equation 
$\det[D^{-1}(\omega,\p)]=0$ in terms of $\omega$ as 
\begin{equation}\label{eq:boson}
 \wph(\p)=\sqrt{\left(\frac\ep2-\muB+\eps\phi_0\right)
  \left(\frac\ep2-\muB+2\eps\phi_0\right)}.
\end{equation}
Note that this expression is valid as long as $\ep\ll\phi_0$ because of
the expansions made to evaluate the boson self-energies.  Substituting
the leading order solution of the gap equation at zero temperature in 
Eq.~(\ref{eq:phi_0}), $\muB=\eps\phi_0$, the phonon spectrum is
determined to be
\begin{equation}\label{eq:phonon}
 \wph(\p)=\sqrt{\frac\ep2\left(\frac\ep2+\eps\phi_0\right)}.
\end{equation}
For the small momentum $\ep\ll\eps\phi_0$, the dispersion relation 
becomes linear in the momentum as $\wph\simeq c_\mathrm{s}|\p|$, 
remaining gapless in accordance with the Nambu--Goldstone theorem. 
The sound velocity of phonon $c_\mathrm{s}$ is given by 
\begin{equation}
 c_\mathrm{s}=\sqrt{\frac{\eps\phi_0}{4m}}\sim\eps^{1/2}.
\end{equation}
For the large momentum $\ep\gg\eps\phi_0$, the dispersion relation
approaches that of the free boson as $\wph\simeq\ep/2$.

\section{Effective potential and condensate}

\begin{figure}[tp]
 \begin{center}
  \includegraphics[width=0.2\textwidth,clip]{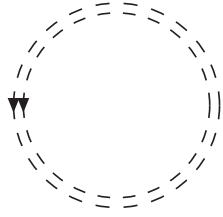}
 \caption{One-loop diagram of boson contributing the effective potential
 at finite temperature. The dotted double line represents the resumed
 boson propagator $\mathcal D$ in Eq.~(\ref{eq:resum}).
 \label{fig:phonon}}
 \end{center}
\end{figure}

At finite temperature, the phonon excitations contribute to the
effective potential, and consequently, the magnitude of the condensate 
decreases.  The temperature dependent part of the effective potential
$V_T(\phi_0)$ to the lowest order in $\eps$ is given by the one-loop
diagram of the boson with the resumed propagator in Eq.~(\ref{eq:resum})
[Fig.~\ref{fig:phonon}]: 
\begin{equation}\label{eq:V_T}
 \begin{split}
  V_T(\phi_0)
  &=\frac T2\sum_n\int\!\frac{d\p}{(2\pi)^4}
  \Tr\ln\left[\mathcal D(i\nu_n,\p)^{-1}\right]\\
  &=\int\!\frac{d\p}{(2\pi)^4}\,T 
  \ln\left[1-e^{-\wph(\p)/T}\right].
 \end{split}
\end{equation}
The $n$-th interaction vertex among phonons $\varphi^n$ is of the order
$\eps^{n/2}$ and appears in the effective potential only at higher
orders.  Then the contribution of $V_T(\phi_0)$ to the gap equation is
\begin{equation}\label{eq:gap_eq}
 \frac{\d V_T(\phi_0)}{\d\phi_0}=\int\!\frac{d\p}{(2\pi)^4}\,
  \bose(\wph)\,\frac{\d\wph}{\d\phi_0},
\end{equation}
where $\bose(x)=1/(e^{x/T}-1)$ is the Bose distribution function and 
$\d\wph/\d\phi_0$ is given from Eq.~(\ref{eq:boson}) by
\begin{equation}
 \frac{\d\wph}{\d\phi_0}=\eps\frac{3\ep+2\eps\phi_0}{4\wph}.
\end{equation}

There are two limiting cases where the integration over $\p$ in
Eq.~(\ref{eq:gap_eq}) can be analytically performed.  Since the integral 
is dominated by the integration region where $\ep\sim T$, we can
approximate the phonon spectrum by its linear branch 
$\wph(\p)\simeq c_\mathrm{s}|\p|$ when the temperature is very low
$T\ll\eps\phi_0$.  In this case, the integration over $\p$ in
Eq.~(\ref{eq:gap_eq}) leads to 
\begin{equation}
 \frac{\d V_T(\phi_0)}{\d\phi_0}\simeq
  \frac{8\zeta(3)T}{\phi_0}\left(\frac{mT}{2\pi}\right)^2.
\end{equation}
On the other hand, when the temperature is located in the intermediate
region $\eps\phi_0\ll T\ll\phi_0$, the phonon spectrum can be
approximated by its quadratic branch $\wph\simeq\ep/2$.  In this case,
the integration over $\p$ in Eq.~(\ref{eq:gap_eq}) results in
\begin{equation}
 \frac{\d V_T(\phi_0)}{\d\phi_0}\simeq\eps\frac{(mT)^2}4.
\end{equation}

Now, from the gap equation $\d(\Veff(\phi)+V_T(\phi))/\d\phi=0$ with
$\phi\equiv\phi_0+\phi_T$, one finds the temperature dependent
correction of the condensate $\phi_T$ satisfies
\begin{equation}
 \frac{\d V_\mathrm{eff}(\phi_0)^2}{\d\phi_0^{\,2}}\phi_T
  +\frac{\d V_T(\phi_0)}{\d\phi_0}=0,
\end{equation}
where $\Veff$ is the effective potential at zero temperature in
Eq.~(\ref{eq:Veff}).  To the leading order in $\eps$, $\phi_T$ at
$T\ll\eps\phi_0$ is given by
\begin{equation}
 \phi_T=-\frac{8\zeta(3)T^3}{\phi_0^{\,2}},
\end{equation}
while at $\eps\phi_0\ll T\ll\phi_0$,
\begin{equation}
 \phi_T=-\eps\frac{\pi^2T^2}{\phi_0}. 
\end{equation}
The condensate in total is $\phi=\phi_0+\phi_T$, which decreases as
the temperature increases.  Note that since $\phi_T\ll\eps\phi_0$, the
leading part of the condensate does not change in the temperature region
considered here $T\ll\phi_0$.  The effective potential is given by the
sum of the zero temperature and finite temperature parts; 
$\Veff(\phi_0+\phi_T)+V_T(\phi_0+\phi_T)\simeq\Veff(\phi_0)+V_T(\phi_0)$.

\section{Thermodynamic functions at low temperature}
The temperature dependent part of the pressure $P_\mathrm{ph}$ at the
low temperature region $T\ll\phi_0$ is given from the effective potential
in Eq.~(\ref{eq:V_T}) by
\begin{equation}\label{eq:P_ph}
 P_\mathrm{ph} = -V_T(\phi_0) = -\int\!\frac{d\p}{(2\pi)^4}\,T 
  \ln\left[1-e^{-\wph(\p)/T}\right].
\end{equation}
The phonon contributions to the fermion number density, the
entropy density, and the energy density are computed from the
thermodynamic relations, 
$N_\mathrm{ph}=\d P_\mathrm{ph}/\d\mu$, 
$S_\mathrm{ph}=\d P_\mathrm{ph}/\d T$, and 
$E_\mathrm{ph}=\mu N_\mathrm{ph}+TS_\mathrm{ph}-P_\mathrm{ph}$,
respectively.  Here we show analytic expressions for these thermodynamic
functions in the two cases where the analytic evaluation of the $\p$
integration in Eq.~(\ref{eq:P_ph}) is available. 

When the temperature is very low $T\ll\eps\phi_0$, only the linear
branch of the phonon spectrum 
$\omega_\mathrm{ph}(\p)\simeq c_\mathrm{s}|\p|$ is important to the
thermodynamic functions.  In this case, the integration over $\p$ can
be performed analytically to lead to
\begin{equation}
 P_\mathrm{ph}\simeq
  \frac{12\pi^2\zeta(5)}{(2\pi)^4}\frac{T^5}{c_\mathrm{s}^{\,4}}
  =\frac{12\zeta(5)}{\pi^2}\frac{m^2T^5}{(\eps\phi_0)^2}.
\end{equation}
Accordingly, we obtain the phonon contributions to the fermion number
density, the entropy density, and the energy density;
\begin{align}
 N_\mathrm{ph}&=-\frac{48\zeta(5)}{\pi^2}\frac{m^2T^5}{(\eps\phi_0)^3},
 \label{eq:N_ph} \\
 S_\mathrm{ph}&=\frac{60\zeta(5)}{\pi^2}\frac{m^2T^4}{(\eps\phi_0)^2},
 \phantom{\frac{\frac\int\int}{\frac\int\int}}\hspace{-4.5mm} \\
 E_\mathrm{ph}&=\frac{24\zeta(5)}{\pi^2}\frac{m^2T^5}{(\eps\phi_0)^2}
 \left(1+\frac\eb{\eps\phi_0}\right).
\end{align}

Since actual experiments or simulations are performed with the fixed
fermion density, it is useful to show the thermodynamic functions at
fixed $N$ instead of fixed $\mu$.  From Eqs.~(\ref{eq:phi_0}),
(\ref{eq:N}), and (\ref{eq:N_ph}), we find the chemical
potential for the fixed fermion density increases as a function of the
temperature as
\begin{equation}
 \mu=\mu_0+48\zeta(5)\frac{T^5}{\eps\phi_0^{\,4}}, 
\end{equation}
where $\mu_0$ represents the chemical potential at zero temperature in
Eq.~(\ref{eq:mu}).  Normalizing $\mu$ by the Fermi energy in 
Eq.~(\ref{eq:eF}), we have 
\begin{equation}\label{eq:mu-N1}
 \frac\mu\eF =\frac{\mu_0}\eF
 +\frac{3\zeta(5)}{2\eps^3}\left(\frac{2T}\eF\right)^5.
\end{equation}
The other thermodynamic functions for the fixed fermion number density 
are given by 
\begin{align}
 \frac{P}{\eF N} &=\frac{P_0}{\eF N}
 +\frac{3\zeta(5)}{\eps^3}\left(\frac{2T}\eF\right)^5,\\
 \frac{E}{\eF N} &=\frac{E_0}{\eF N}
 +\frac{6\zeta(5)}{\eps^3}\left(\frac{2T}\eF\right)^5,
 \phantom{\frac{\frac\int\int}{\frac\int\int}}\hspace{-4.5mm} \\
 \frac{S}{N} &=\frac{15\zeta(5)}{\eps^3}\left(\frac{2T}\eF\right)^4,
 \label{eq:S-N1}
\end{align}
where $P_0$ and $E_0$ represent the pressure and energy density at zero
temperature in Eqs.~(\ref{eq:P}) and (\ref{eq:E}), respectively.  
These expressions are valid in the low temperature region where
$T\ll\eps\phi_0$. 

On the other hand, when the temperature is located in the intermediate
region $\eps\phi_0\ll T\ll\phi_0$, we can expand the phonon spectrum
$\omega_\mathrm{ph}(\p)$ in terms of $\eps\phi_0/\ep$ up to its first
order
\begin{equation}
 \wph(\p)\simeq\frac\ep2+\frac{\eps\phi_0}2. 
\end{equation}
In this case, the integration over $\p$ can be performed analytically
again to result in
\begin{equation}
 P_\mathrm{ph}\simeq\frac{\zeta(3)}{\pi^2}m^2T^3
  -\frac{m^2T^2}{12}\eps\phi_0.
\end{equation}
Accordingly, we obtain the temperature dependent parts of the fermion
number density, the entropy density, and the energy density;
\begin{align}
 N_\mathrm{ph}&=-\frac{m^2T^2}{6},\label{eq:N_ph2}\\
 S_\mathrm{ph}&=\frac{3\zeta(3)}{\pi^2}m^2T^2-\frac{m^2T}{6}\eps\phi_0,
 \phantom{\frac{\frac\int\int}{\frac\int\int}}\hspace{-4.5mm} \\
 E_\mathrm{ph}&=\frac{2\zeta(3)}{\pi^2}m^2T^3
 -\frac{m^2T^2}{6}\eps\phi_0\left(1-\frac\eb{2\eps\phi_0}\right).
\end{align}

From Eqs.~(\ref{eq:phi_0}), (\ref{eq:N}), and (\ref{eq:N_ph2}), we find
the chemical potential for the fixed fermion density increases as a
function of the temperature as
\begin{equation}
  \mu=\mu_0+\eps^{2}\frac{\pi^2}6\frac{T^2}{\phi_0}.
\end{equation}
Normalizing $\mu$ by the Fermi energy in Eq.~(\ref{eq:eF}), we
have
\begin{equation}\label{eq:mu-N2}
 \frac\mu\eF=\frac{\mu_0}\eF
 +\eps^{3/2}\frac{\pi^2}6\left(\frac T\eF\right)^2. 
\end{equation}
The other thermodynamic functions for the fixed fermion number density 
are given by 
\begin{align}
 \frac{P}{\eF N} &=\frac{P_0}{\eF N}+4\zeta(3)\left(\frac T\eF\right)^3
 -\eps^{3/2}\frac{\pi^2}6\left(\frac T\eF\right)^2,\\
 \frac{E}{\eF N} &=\frac{E_0}{\eF N}+8\zeta(3)\left(\frac T\eF\right)^3
 -\eps^{3/2}\frac{\pi^2}3\left(\frac T\eF\right)^2,
 \phantom{\frac{\frac\int\int}{\frac\int\int}}\hspace{-4.5mm} \\
 \frac{S}{N} &=12\zeta(3)\left(\frac T\eF\right)^2
 -\eps^{3/2}\frac{2\pi^2}3\frac T\eF.  \label{eq:S-N2}
\end{align}
These expressions are valid in the intermediate temperature region where
$\eps\phi_0\ll T\ll\phi_0$.

\section{Effective potential near $\Tc$}
At the end of this Chapter, we study the behavior of the condensate
$\phi$ as a function of the temperature for a given $\mu$.  The
effective potential to the leading order in $\eps$ is given by one-loop
diagrams of fermion with and without one $\mu$ insertion, which is
equivalent to the mean-field approximation.  Since the critical
dimension of the superfluid-normal phase transition is four, the
mean-field approximation remains as a leading part at any temperature in
the limit $d\to 4$.
Then the leading contribution to the effective potential at finite
temperature is given by
\begin{equation}
 \Veff(\phi) = -\frac{\eb}{g^2}\phi^{2}-\int\!\frac{d\p}{(2\pi)^d}
 \left[E_\p-\frac{\ep}{\Ep}\mu+2T\ln\left(1+e^{-\Ep/T}\right)
 +\fermi(\Ep)\frac{2\ep}{\Ep}\mu\right], \notag
\end{equation}
where $\fermi(x)=1/(e^{x/T}+1)$ is the Fermi distribution function.

\begin{figure}[tp]
 \begin{center}
  \includegraphics[width=0.5\textwidth,clip]{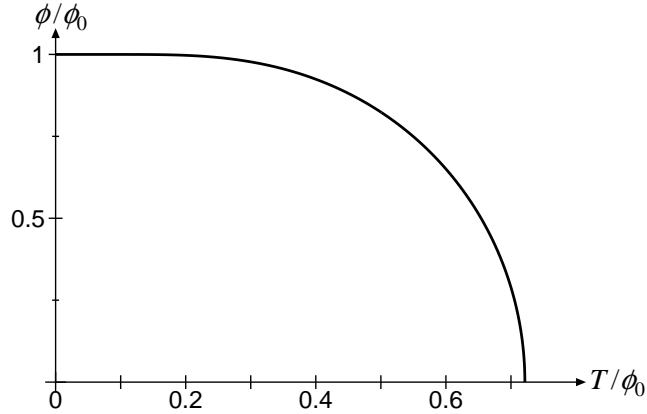}
 \caption{The behavior of the condensate $\phi$ as a function of the 
 temperature $T$ to the lowest order in $\eps$.  $\phi_0=\muB/\eps$ is 
 the value of the condensate at $T=0$ and the critical temperature is 
 located at $\Tc/\phi_0=1/(2\ln2)\approx0.721348$.
 \label{fig:mean-field}}  
 \end{center}
\end{figure}

For the low temperature $T\ll\phi$, we can neglect the exponentially
small factor $e^{-\Ep/T}\ll1$ and the integration over $\p$ reproduces 
the leading order effective potential at zero temperature in
Eq.~(\ref{eq:Veff}).  In the opposite limit where $\phi\ll T$,
we can expand $\Veff$ in terms of $\phi/T$ to lead to  
\begin{equation}
 \Veff(\phi) = \Veff(0)+
 \left[T\ln2-\frac\muB{2\eps}\right]\left(\frac{m\phi}{2\pi}\right)^2
 +\frac{\phi^{2}}{16T}\left(\frac{m\phi}{2\pi}\right)^2+\cdots,
\end{equation}
where $\muB=2\mu+\eb$. 
From the coefficient of the quadratic term in $\phi$, we can read the
critical temperature $\Tc$ to the leading order in $\eps$ as
\begin{equation}\label{eq:Tc-muB}
 \Tc=\frac\muB{\eps\,2\ln2}+O(\eps),
\end{equation}
and the value of the condensate $\phi$ just below $\Tc$ as
\begin{equation}
 \phi^{2}=8T\left(\Tc-T\right)\ln2+O(\eps).
\end{equation}
The critical exponent of the condensate $\phi\sim(\Tc-T)^{1/2}$ will
be shifted if we include higher order corrections to the effective
potential.  The condensate $\phi$ in the intermediate range of the
temperature is obtained by solving the gap equation $\d\Veff/\d\phi=0$;
\begin{equation}
 \phi-\frac\muB\eps+\int\!d\ep\frac{2\ep}\Ep\fermi(\Ep)=0.
\end{equation}
The numerical solution of the gap equation as a function of $T$ is shown
in Fig.~\ref{fig:mean-field}.

\chapter{Thermodynamics above $\Tc$ \label{sec:above-Tc}}

\section{Power counting rule of $\eps$ near $\Tc$} 
In the $\eps$ expansion at the zero or low temperature region, the
chemical potential is small compared to the condensate
$\mu\sim\eps\phi_0$ and we made expansions in terms of
$\mu/\phi_0\sim\eps$ as well as the small coupling $g\sim\eps^{1/2}$.
Near the critical temperature, the ratio $\mu/\phi$ is no longer small
because $\phi$ vanishes at $T=\Tc$, but $\mu/\Tc$ is $O(\eps)$ as it is
clear from $\Tc=\muB/(\eps\,2\ln2)$.  Therefore, we can still treat the 
chemical potential as a small perturbation near $\Tc$ and the same power
counting rule of $\eps$ described in Sec.~\ref{sec:counting} holds
even above $\Tc$ just by replacing $\phi_0$ with $T$.  Hereafter we
consider $T\sim\Tc$ to be $O(1)$.

\section{Boson's thermal mass}
First we study the self-energy of boson at $T\geq\Tc$.  The leading 
contribution to the self-energy is the chemical potential insertion
$\muB$ as well as the one-loop diagram $\Pi_{11}$ shown in
Fig.~\ref{fig:boson}: 
\begin{equation}
 \begin{split}
  \Pi_{11}(i\nu,\p)-\Pi_0(i\nu,\p)
  &=g^2\,T\sum_n\int\!\frac{d\k}{(2\pi)^d}
  G_{11}(i\omega_n+i\nu,\k+\p)G_{22}(i\omega_n,\k) \\
  &=-g^2\int\!\frac{d\k}{(2\pi)^d}
  \frac{1-\fermi(\varepsilon_{\k-\p/2})
  -\fermi(\varepsilon_{\k+\p/2})}{2\ek-i\nu+\ep/2}.
 \end{split}
\end{equation}
For the zero Matsubara frequency mode $\nu_n=0$ at the small momentum 
$\ep\ll T$, we have 
\begin{equation}\label{eq:thermal}
 \Pi_{11}(0,\bm{0})\simeq g^2\int\!\frac{d\k}{(2\pi)^4}
  \frac{\fermi(\ek)}{\ek} =\eps\,T\,2\ln2.
\end{equation}
Therefore, the zero Matsubara frequency mode has the non-negative
thermal mass $\Pi_T=\eps\,T\,2\ln2-\muB\sim\eps$ at $T\geq\Tc$.  The
condition of the vanishing thermal mass $\Pi_T=0$ gives the critical
temperature $\Tc=\muB/(\eps\,2\ln2)$ equivalent to
Eq.~(\ref{eq:Tc-muB}).  As we will see below, at a sufficiently high
order in the perturbation theory near $\Tc$ [$\eps^2$ ($\eps$) compared 
to the leading term in the pressure (fermion density)], the resummation 
of the boson self-energy is needed to avoid infrared singularities
appearing in the zero Matsubara frequency mode.  

\begin{figure}[tp]
 \begin{center}
  \includegraphics[width=0.6\textwidth,clip]{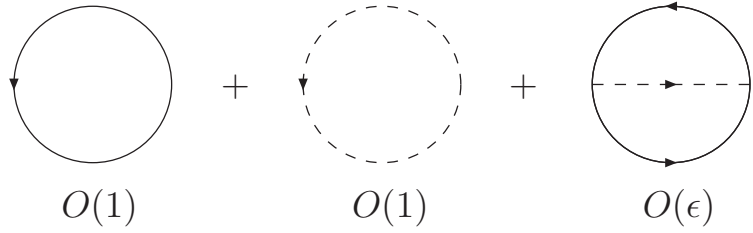}
 \caption{Three types of diagrams contributing to the pressure up to the 
 next-to-leading order in $\eps$.  Each $\mu$ $(\muB)$ insertion to the 
 fermion (boson) line reduces the power of $\eps$ by one. 
 \label{fig:potential_finite-T}} 
 \end{center}
\end{figure}

\section{Pressure}
Now we calculate the thermodynamic functions at $T\geq\Tc$ to the
leading and next-to-leading orders in $\eps$.  There are three types of
diagrams contributing to the pressure up to the next-to-leading order in 
$\eps$ as depicted in Fig.~\ref{fig:potential_finite-T}; one-loop
diagrams with and without one $\mu$ $(\muB)$ insertion and two-loop
diagram with a boson exchange.  Note that at $T\geq\Tc$, the boson's
one-loop diagram contributes as $O(1)$ as well as the fermion's one-loop
diagram.  Then the pressure from the one-loop diagrams is given by 
\begin{equation}\label{eq:P1}
 \begin{split}
  P_1&=\int\!\frac{d\p}{(2\pi)^d}
  \left[2T\ln\left(1+e^{-\ep/T}\right)-T\ln\left(1-e^{-\ep/2T}\right)
  +2\mu\fermi(\ep)+\muB\bose(\ep/2)\right]\\
  &= T\left[\frac{11}2\zeta(3)-\frac{9\ln2\,\zeta(3)+11\zeta'(3)}4\eps 
  +\frac{\pi^2}6\frac\mu T+\frac{2\pi^2}3\frac\muB T\right]
  \left(\frac{mT}{2\pi}\right)^{d/2}. 
 \end{split}
\end{equation}
The contribution from the two-loop diagram to the pressure, which is
$O(\eps)$, is given by
\begin{equation}
 \begin{split}
  P_2 & = g^2\,T^2\sum_{n,m}\int\!\frac{d\p\,d\q}{(2\pi)^{2d}}\,
  G_{11}(i\omega_n,\p)G_{22}(i\omega_m,\q)D(i\omega_n-i\omega_m,\p-\q)\\ 
  &=-g^2\int\!\frac{d\p\,d\q}{(2\pi)^{2d}}\,
  \frac{\fermi(\ep)\fermi(\eq)+\left[\fermi(\ep)+\fermi(\eq)\right]
  \bose(\varepsilon_{\p-\q}/2)}{\ep+\eq-\varepsilon_{\p-\q}/2}.
 \end{split}
\end{equation}
The numerical integrations over $\p$ and $\q$ result in
\begin{equation}\label{eq:P2}
  P_2 = -C_P\eps\left(\frac{mT}{2\pi}\right)^{d/2}T,
\end{equation}
where $C_P\approx8.4144$.  From Eqs.~(\ref{eq:P1}) and (\ref{eq:P2}), we
obtain the pressure up to the next-to-leading order in $\eps$ as
\begin{equation}\label{eq:P-Tc}
 \begin{split}
  P &= P_1+P_2 \\
  &= T\left[\frac{11}2\zeta(3)-\frac{9\ln2\,\zeta(3) 
  +11\zeta'(3)}4\eps -\eps\,C_P +\frac{\pi^2}6\frac\mu T
  +\frac{2\pi^2}3\frac\muB T\right]\left(\frac{mT}{2\pi}\right)^{d/2}. 
 \end{split}
\end{equation}
The entropy density $S$ and the energy density $E$ to the same order can
be computed from the thermodynamic relations $S=\d P/\d T$ and 
$E=\mu N+TS-P$.

\section{Fermion number density}
The fermion number density to the next-to-leading order in $\eps$ can
not be obtained simply by differentiating the pressure in
Eq.~(\ref{eq:P-Tc}) with respect to the chemical potential 
$N=\d P/\d\mu$.  Since the pressure to the leading order in $\eps$ does
not depend on $\mu$ and the $\mu$ derivative $\d/\d\mu\sim1/\eps$
enhances the power of $\eps$ by one, we need to compute the one-loop 
diagrams with two $\mu$ $(\muB)$ insertions and two-loop diagrams with
one $\mu$ $(\muB)$ insertion.  Then the fermion density from the
fermion's one-loop diagrams is given by 
\begin{align}\label{eq:NF}
 \begin{split}
  N_{\mathrm{F}}&=2\int\!\frac{d\p}{(2\pi)^d}
  \left[\fermi(\ep)+\frac{\mu}T\fermi(\ep)\fermi(-\ep)\right]\\
  &=\left[\frac{\pi^2}6-\frac{\pi^2\ln2+6\zeta'(2)}{12}\eps
  +\frac{2\ln2}T\mu\right]\left(\frac{mT}{2\pi}\right)^{d/2}. 
 \end{split}
\end{align}
On the other hand, the boson's one-loop diagrams contribute to the
fermion density as
\begin{equation}
 N_{\mathrm{B}}=2\int\!\frac{d\p}{(2\pi)^d}
 \left[\bose(\ep/2)-\frac{\muB}T\bose(\ep/2)\bose(-\ep/2)\right]. 
\end{equation}
Apparently, the last term has an infrared singularity because the Bose
distribution function behaves as $\bose(\ep/2)\simeq2T/\ep$ at the small
momentum $\ep\ll T$.  
In order to resolve this infrared singularity, the resummation of the
boson self-energy $\Pi_{11}=\eps\,T\,2\ln2$ is needed at the small
momentum region $\ep\sim\mu$.  Within the accuracy we are working, we
can rewrite $N_{\mathrm{B}}$ as
\begin{align}\label{eq:NB}
 N_{\mathrm{B}}=2\int\!\frac{d\p}{(2\pi)^d} \left[\bose(\ep/2+\Pi_T)
 -\eps\,2\ln2\,\bose(\ep/2)\bose(-\ep/2)\right],
\end{align}
where $\Pi_T=\eps\,T\,2\ln2-\muB$.  Now the first term is infrared
finite, where the boson's thermal mass $\Pi_T/T=(2\ln2)\epsilon-\mu_B/T$
plays a role of an infrared cutoff.  Integrating over $\p$ and expanding
up to the next-to-leading order in $\epsilon$, we have  
\begin{equation}\label{eq:NB_regular}
 \begin{split}
  &2\int\!\frac{d\p}{(2\pi)^d}\bose(\ep/2+\Pi_T) \\
  &\qquad\quad=\left[\frac{4\pi^2}3-\frac{2\pi^2\ln2+12\zeta'(2)}3\eps
  -8\left(1-\ln\frac{\Pi_T}T\right)\frac{\Pi_T}T+O(\eps^2)\right]
  \left(\frac{mT}{2\pi}\right)^{d/2}. 
 \end{split}
\end{equation}
The logarithmic term $\sim\ln\Pi_T/T$ appears as a consequence of the
resummation.  The second term in Eq.~(\ref{eq:NB}), which is still
infrared divergent, will cancel with the infrared singularity existing
in the two-loop diagram.

The contribution from the two-loop diagrams to the fermion density is 
given by
\begin{align}
 N_2 &= -g^2\int\!\frac{d\k\,d\p}{(2\pi)^{2d}} \\
 &\qquad\times
 \frac{\fermi(\varepsilon_{\k+\p/2})\fermi(-\varepsilon_{\k+\p/2})
 \left[\fermi(\varepsilon_{\k-\p/2})+\bose(\ep/2)\right]
 -2\fermi(\varepsilon_{\k+\p/2})\bose(\ep/2)\bose(-\ep/2)}{\ek T}. \notag
\end{align}
The second term in the numerator contains the infrared singularity at
small $\ep$.  Extracting the divergent part, we can rewrite $N_2$ as
\begin{align}
 N_2 &= -g^2\int\!\frac{d\k\,d\p}{(2\pi)^{2d}}\,
 \frac{\fermi(\varepsilon_{\k+\p/2})\fermi(-\varepsilon_{\k+\p/2})
 \left[\fermi(\varepsilon_{\k-\p/2})+\bose(\ep/2)\right]}{\ek T} \notag\\
 &\quad+g^2\int\!\frac{d\k\,d\p}{(2\pi)^{2d}}\,
 \frac{2\left[\fermi(\varepsilon_{\k+\p/2})-\fermi(\ek)\right]
 \bose(\ep/2)\bose(-\ep/2)}{\ek T} \notag\\
 &\quad +2g^2\int\!\frac{d\k\,d\p}{(2\pi)^{2d}}\,
 \frac{\fermi(\ek)}{\ek T} \bose(\ep/2)\bose(-\ep/2).
\end{align}
One finds the $\k$ integration in the last term can be performed to
lead to $\Pi_{11}=\eps\,T\,2\ln2$ in Eq.~(\ref{eq:thermal}), which 
exactly cancels out the infrared divergent part in Eq.~(\ref{eq:NB}).
The numerical integrations over $\k$ and $\p$ in the first two terms 
result in 
\begin{equation}\label{eq:N2}
 N_2=-C_N\eps\left(\frac{mT}{2\pi}\right)^{d/2}
  +\eps\,4\ln2\int\!\frac{d\p}{(2\pi)^d}\bose(\ep/2)\bose(-\ep/2),
\end{equation}
where $C_N\approx1.92181$.  Gathering up all contributions,
Eqs.~(\ref{eq:NF}), (\ref{eq:NB}), and (\ref{eq:N2}), the fermion number
density to the leading and next-to-leading orders is given by
\begin{align}\label{eq:N-Tc}
 \begin{split}
  N &= N_{\mathrm{F}}+N_{\mathrm{B}}+N_2\\
  &=\left[\frac{3\pi^2}2-\frac{3\pi^2\ln2+18\zeta'(2)}4\eps-\eps\,C_N
  +\frac{2\ln2}T\mu-8\left(1-\ln\frac{\Pi_T}T\right)\frac{\Pi_T}T
  \right]\left(\frac{mT}{2\pi}\right)^{d/2}. 
 \end{split}
\end{align}
We define the Fermi energy $\eF$ through the relationship in
Eq.~(\ref{eq:eF}) as 
\begin{align}\label{eq:eF-Tc}
 \frac\eF{T} &= \frac{2\pi}m
 \left[ \frac12\Gamma\left(\frac d2+1\right) N \right]^{2/d} \notag\\
 &=\sqrt{\frac{3\pi^2}2}
 \left[1+\left\{\frac{2\gamma-3-2\ln2}8-\frac{3\zeta'(2)}{2\pi^2}
 -\frac{C_N}{3\pi^2}+\frac18\ln\!\left(\frac{3\pi^2}2\right)\right\}\eps
 \right.\\ &\qquad\qquad\quad\left.
 +\frac{2\ln2}{3\pi^2}\frac\mu{T}-\frac8{3\pi^2}
 \left(1-\ln\frac{\Pi_T}T\right)\frac{\Pi_T}T\right]. \notag
\end{align}
The logarithmic correction $(\Pi_T/T)\ln\Pi_T/T\sim\eps\ln\eps$ is a
consequence of the resummation to avoid the infrared singularities,
while it vanishes at the critical temperature
$\Pi_\Tc=\eps\,\Tc\,2\ln2-\muB=0$.

\section{Critical temperature}
The critical temperature in units of the Fermi energy directly follows
from Eq.~(\ref{eq:eF-Tc}) with the use of the relationship
$2\mu+\eb=\eps\,\Tc\,2\ln2$:
\begin{align}\label{eq:Tc}
 \frac\Tc\eF &= \sqrt{\frac2{3\pi^2}} \notag
 \left[1-\left\{\frac{2\gamma-3-2\ln2}8-\frac{3\zeta'(2)}{2\pi^2}
 -\frac{C_N-2(\ln2)^2}{3\pi^2}+\frac18\ln\!\left(\frac{3\pi^2}2\right)
 \right\}\eps\right]+\frac{\ln2}{3\pi^2}\frac\eb\eF \\
 &=0.260-0.0112\,\eps+0.0234\,\frac\eb\eF+O(\eps^2),
 \phantom{\frac{\frac\int{}}{}}\hspace{-4.5mm}
\end{align}
where the numerical value $C_N\approx1.92181$ is substituted.  We find
the critical temperature $\Tc$ is an increasing function of the binding 
energy $\eb$ near the unitarity limit.  The next-to-leading order
correction is reasonably small compared to the leading term even at
$\eps=1$.  The naive extrapolation of the critical temperature to the
physical case of $d=3$ gives $\Tc/\eF\approx0.249$ in the unitarity
limit $\eb=0$. 
This value is surprisingly close to results from two Monte Carlo
simulations, $\Tc/\eF=0.23(2)$~\cite{bulgac-Tc} and
$\Tc/\eF\approx0.25$~\cite{Akkineni-Tc}, while other two simulations
provide smaller values, $\Tc/\eF<0.14$~\cite{Lee-Schafer} and
$\Tc/\eF=0.152(7)$~\cite{burovski-Tc}. 

It is also interesting to compare the critical temperature in the
unitarity limit with that in the BEC limit $T_\mathrm{BEC}$.  In the BEC
limit, all fermion pairs are confined into tightly bound molecules and
the system becomes a non-interacting Bose gas where the boson mass is
$2m$ and the boson density is $N/2$.  The critical temperature for the
Bose-Einstein condensation of such an ideal Bose gas at $d>2$ spatial
dimensions becomes
\begin{equation}
  \frac{T_\mathrm{BEC}}\eF
  =\frac12\left[\zeta\!\left(\frac d2\right)
  \Gamma\!\left(1+\frac d2\right)\right]^{-2/d}. 
\end{equation}
To the leading and next-to-leading orders in $\eps=4-d$, the ratio of
the critical temperatures in the unitarity limit $\Tc$ and in the BEC
limit $T_\mathrm{BEC}$ at the same fermion density is given by
\begin{equation}
  \frac{\Tc}{T_\mathrm{BEC}}
  =\sqrt{\frac89}\left[1+0.0177\,\eps+O(\eps^2)\right]  
  =0.943 + 0.0167\,\eps + O(\eps^2).
  \phantom{\frac{\int}{}}  
\end{equation}
The ratio is slightly below unity, indicating the lower critical
temperature in the unitarity limit $\Tc<T_\mathrm{BEC}$.  The leading
order term of the above ratio, $\Tc/T_\mathrm{BEC}=\sqrt{8/9}$, has the
following clear physical interpretation: The critical temperature for
the Bose-Einstein condensation at $d=4$ is proportional to a square root
of the boson's density.  In the BEC limit, all fermion pairs form the
bound bosons, while only 8 of 9 fermion pairs form the bosons and 1 of 9
fermion pairs is dissociated in the unitarity limit [see the leading
order terms in Eqs.~(\ref{eq:NF}) and (\ref{eq:NB_regular})].  Thus,
their ratio in the critical temperature should be
$\Tc/T_\mathrm{BEC}=\sqrt{8/9}<1$ at $d=4$. 

The more appropriate estimate of $\Tc$ at $d=3$ will be obtained by
matching the $\eps$ expansion with the exact result around $d=2$.  The
critical temperature at unitarity in the expansion over $\bar\eps=d-2$
is given by  $\Tc=\left(e^\gamma/\pi\right)\Delta$, where
$\Delta/\eF=\left(2/e\right)e^{-1/\bar\eps}$ is the energy gap of the
fermion quasiparticle at zero temperature~\cite{Nishida-Son2}: 
\begin{equation}\label{eq:Tc-2d}
 \frac\Tc\eF = \frac{2e^{\gamma-1}}{\pi}
  \,e^{-1/\bar\eps}\left[1+O(\bar\eps)\right].
\end{equation}
We shall write the power series of $\bar\eps$ in the form of the Borel
transformation, 
\begin{equation}\label{eq:borel-Tc}
 \frac{\Tc(\bar\eps)}\eF = \frac{2e^{\gamma-1}}{\pi}\,e^{-1/\bar\eps} 
  \int_0^\infty\!dt\, e^{-t}B_\Tc(\bar\eps t).
\end{equation}
$B_\Tc(t)$ is the Borel transform of the power series in
$\Tc(\bar\eps)$, whose Taylor coefficients at origin is given by
$B_\Tc(t)=1+\cdots$.  In order to perform the integration over $t$ in
Eq.~(\ref{eq:borel-Tc}), the analytic continuation of the Borel
transform $B_\Tc(t)$ to the real positive axis of $t$ is necessary.
Here we employ the Pad\'e approximant, where $B_\Tc(t)$ is
replaced by the following rational functions 
\begin{equation}
 B_\Tc(t) = \frac{1+p_1 t+\cdots+p_M t^M}{1+q_1 t+\cdots+q_N t^N}\,.
\end{equation}
Then we incorporate the results around four spatial dimensions in
Eq.~(\ref{eq:Tc}) by imposing 
\begin{equation}\label{eq:boundary}
 \frac{\Tc(2-\eps)}\eF=0.260-0.0112\,\eps+\cdots
\end{equation}
on the Pad\'e approximants as a boundary condition. 
Since we have two known coefficients from the $\eps$ expansion, the
Pad\'e approximants $[M/N]$ satisfying $M+N=2$ are possible. 
Since we could not find a solution satisfying the boundary condition 
in Eq.~(\ref{eq:boundary}) for $[M/N]=[1/1]$, we adopt other two Pad\'e
approximants with $[M/N]=[2/0],\,[0/2]$, whose coefficients $p_m$ and
$q_n$ are determined uniquely by the above conditions.

\begin{figure}[tp]
 \begin{center}
  \includegraphics[width=0.6\textwidth,clip]{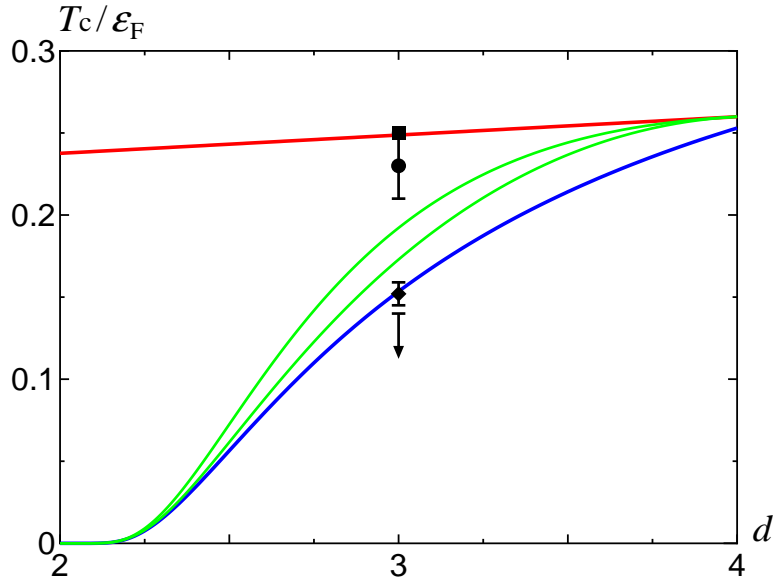}
 \caption{The critical temperature $\Tc$ at unitarity as a function of
 the spatial dimensions $d$.  The upper solid line is from the expansion
 around $d=4$ in Eq.~(\ref{eq:Tc}), while the lower solid line is from the
 expansion around $d=2$ in Eq.~(\ref{eq:Tc-2d}).  The middle two
 curves show the different Borel--Pad\'e approximants connecting the
 two expansions.  The symbols at $d=3$ indicate the results from the
 Monte Carlo simulations; $\Tc/\eF=0.23(2)$~\cite{bulgac-Tc} (circle), 
 $\Tc/\eF<0.14$~\cite{Lee-Schafer} (down arrow),
 $\Tc/\eF=0.152(7)$~\cite{burovski-Tc} (diamond), and
 $\Tc/\eF\approx0.25$~\cite{Akkineni-Tc} (square).  \label{fig:Tc}}
 \end{center}
\end{figure}

Fig.~\ref{fig:Tc} shows the critical temperature $\Tc$ in units of the
Fermi energy $\eF$ as a function of the spatial dimensions $d$.  The 
middle two curves show $\Tc/\eF$ in the different Pad\'e approximants
connecting the two expansions around $d=4$ and $d=2$.  These
Borel--Pad\'e approximants give $\Tc/\eF=0.173$ and $0.192$ at $d=3$,
which are located between the naive extrapolations to $d=3$ from the 
$\eps=4-d$ expansion $(\Tc/\eF\to0.249)$ and the $\bar\eps=d-2$
expansion $(\Tc/\eF\to0.153)$. 
It is also interesting to compare our results with those from the recent
Monte Carlo simulations, where $\Tc/\eF=0.23(2)$~\cite{bulgac-Tc},
$\Tc/\eF<0.14$~\cite{Lee-Schafer}, $\Tc/\eF=0.152(7)$~\cite{burovski-Tc},
and $\Tc/\eF\approx0.25$~\cite{Akkineni-Tc}.  Although these results
from the Monte Carlo simulations seem not to be settled, the
interpolation of the two expansions provides the moderate value 
$\Tc/\eF=0.183\pm0.014$ not too far from the Monte Carlo simulations.

\section{Thermodynamic functions at $\Tc$}
Finally we show the thermodynamic functions at $\Tc$ in the unitarity
limit $\eb=0$ to the leading and next-to-leading orders in $\eps$.  The
pressure $P$ normalized by the fermion density $\eF N$ follows from
Eqs.~(\ref{eq:P-Tc}), (\ref{eq:N-Tc}), and (\ref{eq:eF-Tc}). Introducing
the numerical values $C_P\approx8.4144$ and $C_N\approx1.92181$, we
obtain the pressure up to the next-to-leading order in $\eps$ as
\begin{equation}
 \left.\frac P{\eF N}\right|_{\Tc} = 0.116+0.0188\,\eps. 
\end{equation}
From the universal relationship in the unitarity limit $E=(d/2)P$, the 
energy density is given by
\begin{equation}
 \left.\frac E{\eF N}\right|_{\Tc} = 0.232-0.0205\,\eps.
\end{equation}
The chemical potential at the critical temperature $\mu=\eps\,\Tc\ln2$
is $O(\eps)$.  Normalizing $\mu$ by the Fermi energy in
Eq.~(\ref{eq:Tc}), we have
\begin{equation}
 \left.\frac\mu\eF\right|_{\Tc} = \epsilon\ln2\sqrt{\frac2{3\pi^2}} 
  = 0.180\,\eps. 
\end{equation}
Then the entropy density $\Tc S=(d/2+1)P-\mu N$ is given by
\begin{equation}
 \left.\frac S{N}\right|_{\Tc} = 1.340 - 0.642\,\eps.
\end{equation}
The next-to-leading order corrections to the pressure and energy density
are reasonably small compared to the leading order terms, while that is
large for the entropy density.

We match the thermodynamic functions at $\Tc$ in the expansions over
$\eps=4-d$ with those around $d=2$ as we demonstrated for $\Tc/\eF$. 
The critical temperature around $d=2$ is $\Tc/\eF\sim e^{-1/\bar\eps}$, 
which is exponentially small and negligible compared to any power series
of $\bar\eps$.  Therefore, the pressure, energy density, and chemical
potential at $\Tc$ in the expansions over $\bar\eps=d-2$ is simply given
by those at zero temperature~\cite{Nishida-Son2};
\begin{align}
 \left.\frac{P}{\eF N}\right|_{\Tc} 
 &= \frac{2}{d+2}\,\xi = \frac12-\frac58\bar\eps,\\
 \left.\frac{E}{\eF N}\right|_{\Tc}
 &= \frac{d}{d+2}\,\xi = \frac12-\frac38\bar\eps,
 \phantom{\frac{\frac\int\int}{\frac\int\int}}\hspace{-4.5mm} \\ 
 \left.\frac{\mu}{\eF}\right|_{\Tc} &= \xi = 1-\bar\eps.
\end{align}
A straightforward calculation shows that the entropy per particle at
$\Tc$ to the leading order in $\bar\eps$ is given by 
\begin{equation}
 \left.\frac S{N}\right|_{\Tc} = \frac{\pi^2}3\frac\Tc\eF
  = \frac{2\pi e^{\gamma-1}}{3}\,e^{-1/\bar\eps}.
\end{equation}

Using the Borel--Pad\'e approximants connecting the two expansions
above, thermodynamic functions at $d=3$ are found to be 
$\left.P/(\eF N)\right|_\Tc=0.172\pm0.022$, 
$\left.E/(\eF N)\right|_\Tc=0.270\pm0.004$, 
$\left.\mu/\eF\right|_\Tc=0.294\pm0.013$, 
and $\left.S/N\right|_\Tc=0.642$. 
The errors here indicates only the uncertainty due to the choice of
different Pad\'e approximants.  In Table~\ref{tab:comparison_Tc}, we
summarize our results on the critical temperature $\Tc$ and the
thermodynamic functions at $\Tc$ as well as results from other analytic
and numerical calculations.  In the recent Monte Carlo simulation, the
thermodynamic functions at the critical temperature is 
$\left.P/(\eF N)\right|_\Tc=0.207(7)$,
$\left.E/(\eF N)\right|_\Tc=0.31(1)$,
$\left.\mu/\eF\right|_\Tc=0.493(14)$, and
$\left.S/N\right|_\Tc=0.16(2)$~\cite{burovski-Tc}. 
We see that the interpolations of the two expansions indeed improve the 
series summations compared to the naive extrapolations from $d=4$ or 
$d=2$, while there still exist deviations between our results and the
Monte Carlo simulation.  We can understand these deviations partially 
due to the difference in the determined critical temperature.  The large
deviations existing in $\mu/\eF$ and $S/N$ may be because we know only
the leading term for $\mu$ and the next-to-leading order correction for
$S$ is sizable.  

\begin{table}[tp]
 \begin{center}
  \begin{tabular}{l|l|l|l|l|l}
   & $\Tc/\eF$ & $P/(\eF N)$ & $E/(\eF N)$ & $\mu/\eF$ & $S/N$ \\\hline
   $\eps$ expansion (NLO) & $0.249$ & $0.135$ & $0.212$ & $0.180$ 
   & $0.698$ \\
   Borel--Pad\'e approximant & $0.183$ & $0.172$ & $0.270$ & $0.294$
   & $0.642$ \\
   self-consistent approach~\cite{Zwerger} & $0.160$ & $0.204$ & $0.304$
   & $0.394$ & $0.71$ \\
   Monte Carlo simulation~\cite{bulgac-Tc} & $0.23(2)$ & $0.27$ & $0.41$
   & $0.45$ & $0.99$ \\
   Monte Carlo simulation~\cite{burovski-Tc} & $0.152(7)$ & $0.207(7)$ 
   & $0.31(1)$ & $0.493(14)$ & $0.16(2)$
  \end{tabular}
  \caption{Comparison of the results by the $\eps$ expansion with other
  analytic and numerical calculations in the unitarity limit.
  \label{tab:comparison_Tc}}
 \end{center}
\end{table}

\chapter{Summary and concluding remarks \label{sec:summary}}
We have developed the systematic expansion for the Fermi gas near the 
unitarity limit treating the dimensionality of space $d$ as close to 
four.  To the leading and next-to-leading orders in the expansion over 
$\eps=4-d$, the thermodynamic functions and the fermion quasiparticle
spectrum were calculated as functions of the binding energy $\eb$ of the
two-body state.  Results for the physical case of three spatial
dimensions were obtained by extrapolating the series expansions to
$\eps=1$.  We found the universal parameter of the unitary Fermi gas to
be
\begin{equation}
 \xi\equiv\frac\mu\eF =\frac12\eps^{3/2}+\frac1{16}\eps^{5/2}\ln\eps
  -0.0246\,\eps^{5/2}+O(\eps^{7/2})\approx0.475.
\end{equation}
The fermion quasiparticle spectrum in the unitarity limit was given by
the form
$\omega_\mathrm{F}(\p)\simeq\sqrt{(\ep-\varepsilon_0)^2+\Delta^2}$ with 
the energy gap 
\begin{equation}
 \frac\Delta\mu=\frac2\eps-0.691+O(\eps)\approx1.31
\end{equation}
and the location of the minimum of the dispersion curve
\begin{equation}
 \frac{\varepsilon_0}\mu=2+O(\epsilon)\approx2.
\end{equation}
We also found the condensate fraction in the fermion density to be
\begin{equation}
 \frac{N_0}N=1-0.0966\,\eps-0.2423\,\eps^2+O(\eps^3)\approx0.661.
\end{equation}
Although we have only the first two terms in the $\eps$ expansion, these
extrapolated values give reasonable results consistent with the Monte
Carlo simulations or the experimental measurements.  Furthermore, the
corrections are not too large even when extrapolated to $\eps=1$, which
suggests that the picture of the unitary Fermi gas as a collection of
weakly interacting fermionic and bosonic quasiparticles may be a useful
starting point even in three spatial dimensions. 

We have also formulated the systematic expansion for the unitary Fermi
gas around two spatial dimensions.  We used the results around $d=2$ as
boundary conditions which should be satisfied by the series summations
of the expansion over $\eps=4-d$.  The simple Borel--Pad\'e
approximants connecting the two expansions yielded $\xi=0.378\pm0.013$
at $d=3$, which is small compared to the naive extrapolation of the
$\eps$ expansion [Fig.~\ref{fig:xi}].  In order for the accurate
determination of $\xi$ at $d=3$, the precise knowledge on the large
order behavior of the expansion around four spatial dimensions as well
as the calculation of higher order corrections would be desirable.  Once
these information become available, a conformal mapping technique, if
applicable, will further improve the series summations~\cite{justin}. 

In Chapters \ref{sec:polarization} and \ref{sec:unequal}, the phase
structure of the polarized Fermi gas in the unitary regime with equal
and unequal masses has been studied based on the $\eps$ expansion.  The
gapless superfluid phase and the superfluid phase with spatially varying
condensate were found to exist between the gapped superfluid phase and
the polarized normal phase in a certain range of the binding energy and
the mass difference.  In the equal mass limit, our study gives a
microscopic foundation to the phase structure around the splitting point
which has been proposed using the effective field theory~\cite{son05}.
Moreover, we found the gapless phase with spatially varying condensate
around $d=4$ exists only at $0.494\lesssim(a\kF)^{-1}\lesssim1$ and
terminates near the unitarity limit.  Our result suggests that the phase
with spatially varying condensate existing in the unitary regime may be
separated from the FFLO phase in the BCS regime~\cite{FF,LO}.  
We also found the splitting point and the same phase structure
around it in the unitarity limit but with finite mass difference when
the majority is the lighter fermions.  The range of the superfluid phase
with spatially varying condensate can be estimated to be
$1.37\lesssim m_\mathrm{minority}/m_\mathrm{majority}\lesssim3$.  The
splitting points form a smooth line in the three-dimensional phase
diagram of $1/(a\kF)$, the mass difference $\kappa$, and polarization
chemical potential $H$.  It gets away from the unitarity with increasing
the mass of major fermions.  Eventually the splitting point becomes
unstable due to the competition with the polarized normal state at the
point given by $(a\kF)^{-1}\approx1.81$ and
$m_\mathrm{majority}/m_\mathrm{minority}\approx5.54$.  Further
investigation will be worthwhile to confirm these possibility. 

In Chapters \ref{sec:below-Tc} and \ref{sec:above-Tc}, the
thermodynamics of the Fermi gas near the unitarity limit at finite
temperature has been investigated using the systematic expansion over
$\eps=4-d$.  We discussed that the thermodynamics in the low temperature
region $T\ll\Tc$ is dominated by the bosonic phonon excitations.  The
analytic formulas for the thermodynamic functions at the fixed fermion
density are derived in the two limiting cases; $T\ll\eps\Tc$ in
Eqs.~(\ref{eq:mu-N1})--(\ref{eq:S-N1}) and $\eps\Tc\ll T\ll\Tc$ in
Eqs.~(\ref{eq:mu-N2})--(\ref{eq:S-N2}).
In the high temperature region $T\sim\Tc$, the fermionic quasiparticles
are excited as well as the bosonic quasiparticles.  We showed that the
similar power counting rule of $\eps$ to that developed at zero
temperature works even above $\Tc$.  The critical temperature $\Tc$ and
the thermodynamic functions around $\Tc$ were calculated to the leading
and next-to-leading orders in $\eps$.  We found the critical temperature
is an increasing function of the binding energy $\eb$ near the unitarity
limit: 
\begin{equation}
 \frac\Tc\eF=0.260-0.0112\,\eps+0.0234\,\frac\eb\eF+O(\eps^2).
\end{equation}
The next-to-leading order correction is reasonably small compared to the
leading term even at $\eps=1$.  In the unitarity limit $\eb=0$, the
naive extrapolation of the critical temperature to the physical case of
$d=3$ gives $\Tc/\eF\approx0.249$. 

We also discussed the matching of the $\eps$ expansion with the
expansion around $d=2$.  The critical temperature at unitarity in the
expansion over $\bar\eps=d-2$ is given by
$\Tc/\eF=\left(2e^{\gamma-1}/\pi\right)e^{-1/\bar\eps}$, whose naive 
extrapolation to $\bar\eps=1$ gives $\Tc/\eF\approx0.153$. 
The Borel--Pad\'e approximants connecting the two expansions yielded
$\Tc/\eF=0.183\pm0.014$ at $d=3$, which is a moderate value located
between the two naive extrapolations [Fig.~\ref{fig:Tc}].  These values
are not too far from the results obtained by the recent Monte Carlo
simulations where
$\Tc/\eF=0.15\sim0.25$~\cite{bulgac-Tc,burovski-Tc,Akkineni-Tc}. 
We also applied the Borel--Pad\'e approximants to the thermodynamic
functions at $\Tc$, which yielded 
$\left.P/(\eF N)\right|_\Tc\approx0.172$, 
$\left.E/(\eF N)\right|_\Tc\approx0.270$,
$\left.\mu/\eF\right|_\Tc\approx0.294$,  
and $\left.S/N\right|_\Tc\approx0.642$ at $d=3$. 

\begin{figure}[tp]
 \begin{center}
  \includegraphics[scale=1.5,clip]{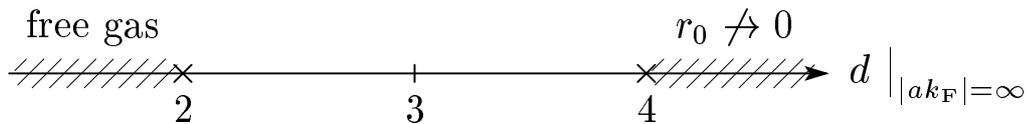}
  \caption{The Fermi gas at infinity scattering length $a\kF\to\infty$
  as a function of the spatial dimension $d$.  At $d<2$, such a system
  becomes a free Fermi gas, while the limit of zero-range interaction
  $r_0\kF\to0$ can not be taken at $d>4$.   Therefore, the non-trivial
  universal limit exists only at $2<d<4$.\label{fig:dimension}}
 \end{center}
\end{figure}

We conclude that the $\eps$ expansion is not only theoretically 
interesting but also provides us an useful analytical tool to
investigate the properties of the unitary Fermi gas.  Although our
results are consistent with those from Monte Carlo simulation, there are
several problems which should be clarified in future.  For example, the
Borel--Pad\'e approximants employed here to match the two expansions
around four and two spatial dimensions do not take into account the
large order behavior of the expansions around $d=4$ and $d=2$.  Most
probably, the expansions over $\eps=4-d$ and $\bar\eps=d-2$ are not 
convergent because $d=4$ and $d=2$ corresponds to the quantum phase
transition points as depicted in Fig.~\ref{fig:dimension}.  Thus the
Borel transform $B(t)$ of such series expansions will have singularities
somewhere in the complex $t$-plane.  Furthermore, an understanding of
the analytic structure of high-order terms in the perturbation theory
around $d=4$ is currently lacking.  In order for the accurate
determination of the universal quantities, e.g., $\xi$, $\Tc/\eF$, at
$d=3$, it will be important to appropriately take into account the
knowledge on the large order behaviors of the expansions both around
four and two spatial dimensions.  The calculation of higher order
corrections to our results is also important for this purpose.  In
Appendix, we show a type of diagrams which grows as a factorial of $n$
at order $\eps^n$ and may be related with the large order behavior of
the $\eps$ expansion.  These problems should be further studied in
future works.  Also the application of the $\eps$ expansion to
investigate other problems of the unitary Fermi gas, including its
dynamical properties (dynamical response functions and kinetic
coefficients) and the structure of superfluid vortices, will be
extremely interesting.

\chapter*{Acknowledgments \label{sec:acknowledge}}
\addcontentsline{toc}{section}{Acknowledgments}

I am grateful to all members of the theoretical nuclear physics group in
the University of Tokyo for many useful discussions during I was a
graduate student there.  My supervisor Prof.\ Tetsuo Hatsuda drew my
attention to the interesting field of quantum chromodynamics (QCD) at
finite temperature and density, which is an intersection of condensed
matter physics and high energy physics, when I was an undergraduate
student.  I would like to thank him for continuous and stimulating
discussions on a large variety of subjects in physics.  He also
encouraged me to study abroad, without which I could not finish my
thesis as its present form.

All works in this thesis were completed during I was visiting the
Institute for Nuclear Theory in Seattle to collaborate with Prof.\ Dam
Thanh Son.  I am grateful to him for accepting me as a visiting scholar
and for continuous and stimulating discussions on the physics of QCD and
cold atomic gases.  It was very fruitful to discuss and work with him
together and I could learn many things as a researcher.  I would really
like to thank all people in the Institute and Robert Ingalls for their
hospitality.  Especially the discussions with Dr.\ Michael McNeil Forbes
and Dr.\ Gautam Rupak were very helpful when I could not figure out how
to organize the $\epsilon$ expansion correctly. 

Finally I would like to express my sincere gratitude to my parents and
family, who have kept me studying what I like to through moral and
financial support.  My father and grandfather raised my interest in
physics, astronomy, and science during I was a child.  Minako always
gave me much encouragement and pleasure whenever I was in hard times and
good times.  I was supported by the predoctoral research fellowship of
the Japan Society for the Promotion of Science for the last three years
of my PhD course.

\appendix
\chapter{Large orders in the $\eps=4-d$ expansion \label{sec:renormalon}}

\begin{figure}[tp]
 \begin{center}
  \includegraphics[width=0.65\textwidth,clip]{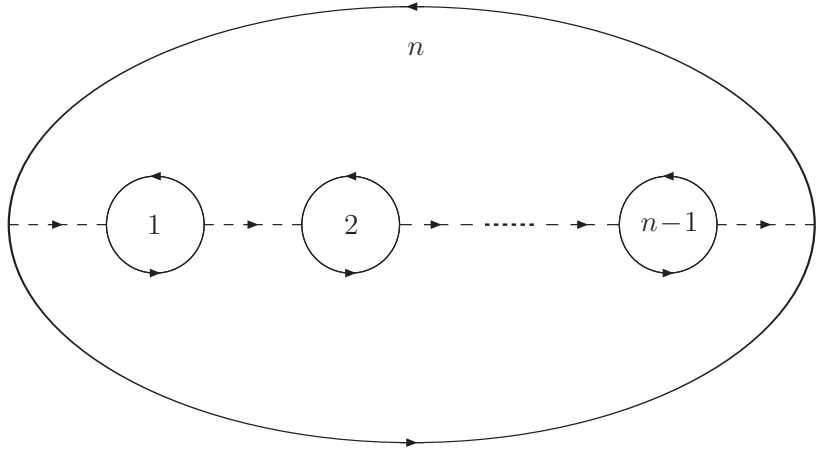}
  \caption{A $n$-th order diagram at $d=4$ which contributes to the
  effective potential as $n!$ by itself.  The counter vertex $-i\Pi_0$
  for each bubble diagram is understood implicitly.
  \label{fig:renormalon}}
 \end{center}
\end{figure}

In this Appendix, we show that there exists a type of diagrams which
grows as $n!$ by itself at order $\eps^n$ of the $\eps=4-d$ expansion. 
Such a factorial contribution originates from the large momentum region
of the loop integrals which resembles the \textit{ultraviolet
renormalon} in relativistic field
theories~\cite{Gross-Neveu,Lautrup,tHooft}.  An example of the
$n+1$-loop diagram contributing to the effective potential as $n!$ at
$O(\eps^n)$ is depicted in Fig.~\ref{fig:renormalon}, which can be
written as
\begin{equation}
 V_n=\frac in\int\!\frac{dk}{(2\pi)^{d+1}}
  \left[\left(\Pi_0(k)+\Pi_\mathrm{a}(k)\right)D(k)\right]^n,
\end{equation}
where $\Pi_0(k)$, $\Pi_\mathrm{a}(k)$, and $D(k)$ are respectively
defined in Eqs.~(\ref{eq:Pi_0}), (\ref{eq:Pi_a}), and (\ref{eq:D}). 
Introducing these definitions and integrating over $k_0$, we obtain the
following expression for $V_n$,
\begin{equation}\label{eq:V_n}
 \begin{split}
  V_n &=\frac in\int\!\frac{dk}{(2\pi)^{d+1}}
  \left[1+\frac{-g^2}{k_0-\frac\ek2+i\delta}
  \int\!\frac{d\p}{(2\pi)^d}\,\frac1{4E_{\p-\frac\k2}E_{\p+\frac\k2}}
  \right.\\ &\qquad\qquad\qquad\qquad\left.\times
  \left\{\frac{(E_{\p-\frac\k2}+\varepsilon_{\p-\frac\k2})
  (E_{\p+\frac\k2}+\varepsilon_{\p+\frac\k2})}
  {E_{\p-\frac\k2}+E_{\p+\frac\k2}-k_0-i\delta}
  +\frac{(E_{\p-\frac\k2}-\varepsilon_{\p-\frac\k2})
  (E_{\p+\frac\k2}-\varepsilon_{\p+\frac\k2})}
  {E_{\p-\frac\k2}+E_{\p+\frac\k2}+k_0-i\delta}\right\}\right]^n \\ 
  &= -\frac{g^2}4\int\!\frac{d\k d\p}{(2\pi)^{2d}}\,
  \frac{(E_{\p-\frac\k2}-\varepsilon_{\p-\frac\k2})
  (E_{\p+\frac\k2}-\varepsilon_{\p+\frac\k2})}
  {E_{\p-\frac\k2}E_{\p+\frac\k2}
  \left(E_{\p-\frac\k2}+E_{\p+\frac\k2}+\frac\ek2\right)}\\
  &\qquad\qquad\times
  \left[1+\frac{g^2}{E_{\p-\frac\k2}+E_{\p+\frac\k2}+\frac\ek2}
  \int\!\frac{d\q}{(2\pi)^d}\,\frac1{4E_{\q-\frac\k2}E_{\q+\frac\k2}}
  \right.\\ &\qquad\qquad\qquad\quad\left.\times
  \left\{\frac{(E_{\q-\frac\k2}+\varepsilon_{\q-\frac\k2})
  (E_{\q+\frac\k2}+\varepsilon_{\q+\frac\k2})}
  {E_{\q-\frac\k2}+E_{\q+\frac\k2}+E_{\p-\frac\k2}+E_{\p+\frac\k2}}
  +\frac{(E_{\q-\frac\k2}-\varepsilon_{\q-\frac\k2})
  (E_{\q+\frac\k2}-\varepsilon_{\q+\frac\k2})}
  {E_{\q-\frac\k2}+E_{\q+\frac\k2}-E_{\p-\frac\k2}-E_{\p+\frac\k2}}
  \right\}\right]^{n-1}.
 \end{split}
\end{equation}
Then we consider the $\q$ integration in the bracket.  Since the $\q$
integration contains a logarithmic divergence at $d=4$, we subtract and
add its divergent piece as
\begin{equation}
 \begin{split}
  &1+\frac{g^2}{E_{\p-\frac\k2}+E_{\p+\frac\k2}+\frac\ek2}
  \int\!\frac{d\q}{(2\pi)^d}\,\frac1{4E_{\q-\frac\k2}E_{\q+\frac\k2}}
  \biggl\{\ \cdots\ \biggr\}\\
  &\qquad =1+\frac{g^2}{E_{\p-\frac\k2}+E_{\p+\frac\k2}+\frac\ek2}
  \int\!\frac{d\q}{(2\pi)^d}\left[
  \frac1{4E_{\q-\frac\k2}E_{\q+\frac\k2}}\biggl\{\ \cdots\ \biggr\} 
  -\frac{1}{2\eq+E_{\p-\frac\k2}+E_{\p+\frac\k2}+\frac\ek2}\right]\\
  &\qquad\qquad+\frac{g^2}{E_{\p-\frac\k2}+E_{\p+\frac\k2}+\frac\ek2}
  \int\!\frac{d\q}{(2\pi)^d}\,
  \frac{1}{2\eq+E_{\p-\frac\k2}+E_{\p+\frac\k2}+\frac\ek2}.
 \end{split}
\end{equation}
The $\q$ integration in the second line becomes finite at $d=4$ and does
not produce singular logarithmic terms.  So we concentrate on the $\q$
integration in the last line, which can be evaluated in the dimensional
regularization as
\begin{equation}
 \begin{split}
  &1+\frac{g^2}{E_{\p-\frac\k2}+E_{\p+\frac\k2}+\frac\ek2}
  \int\!\frac{d\q}{(2\pi)^d}\,
  \frac{1}{2\eq+E_{\p-\frac\k2}+E_{\p+\frac\k2}+\frac\ek2} \\
  &\qquad = 1+\frac\eps2\,\Gamma\!\left(1-\frac d2\right)
  \left(\frac{E_{\p-\frac\k2}+E_{\p+\frac\k2}
  +\frac\ek2}{2\phi_0}\right)^{-\frac\eps2} \\
  &\qquad = \frac\eps2\,\ln\left(\frac{E_{\p-\frac\k2}+E_{\p+\frac\k2}
  +\frac\ek2}{2\phi_0}\right)+O(\eps^2).
 \end{split}
\end{equation}
Then $V_n$ in Eq.~(\ref{eq:V_n}) to the order $\eps^n$ becomes
\begin{align}
 V_n &= -\frac{g^2}4\int\!\frac{d\k d\p}{(2\pi)^{2d}}\,
 \frac{(E_{\p-\frac\k2}-\varepsilon_{\p-\frac\k2})
 (E_{\p+\frac\k2}-\varepsilon_{\p+\frac\k2})}
 {E_{\p-\frac\k2}E_{\p+\frac\k2}
 \left(E_{\p-\frac\k2}+E_{\p+\frac\k2}+\frac\ek2\right)}
 \left[\frac\eps2\,\ln\left(\frac{E_{\p-\frac\k2}
 +E_{\p+\frac\k2}+\frac\ek2}{2\phi_0}\right)
 +\mathrm{\,regular\ terms\,}\right]^{n-1} \notag\\
  &= -\frac{g^2}4\int\!\frac{d\p d\q}{(2\pi)^{2d}}\,
 \frac{(\Ep-\ep)(\Eq-\eq)}
 {\Ep\Eq\left(\Ep+\Eq+\frac{\varepsilon_{\p-\q}}2\right)}
 \left[\frac\eps2\,\ln\left(\frac{\Ep+\Eq+
 \frac{\varepsilon_{\p-\q}}2}{2\phi_0}\right)
 +\mathrm{\,regular\ terms\,}\right]^{n-1}.
\end{align}
By picking up the logarithmic terms, we find the $\p$ integration from
its large momentum region $\ep\gg\eq$ and $\ep\gg\phi_0$ gives the
following contribution to the effective potential, 
\begin{equation}
 \begin{split}
  V_n &\sim \eps\int\!\frac{d\p}{(2\pi)^{d}}\,\left(\frac1\ep\right)^3
  \left(\frac\eps2\ln\ep\right)^{n-1} \\
  &\sim \left(\frac\eps2\right)^n \int\!d\ep\,\left(\frac1\ep\right)^2
  \left(\ln\ep\right)^{n-1} \\
  &\sim \left(\frac\eps2\right)^n \Gamma(n),
 \end{split}
\end{equation}
which grows as a factorial of $n$.  Note that this $n!$ contribution
comes from the single diagram depicted in Fig.~\ref{fig:renormalon}.  At
the moment, it is not clear whether such a $n!$ contribution survives at
$O(\eps^n)$ of the effective potential and dominates the large order
behavior of the $\eps$ expansion.  Possibly there might be cancellations
with other $n$-th order diagrams and/or with subleading contributions
from lower-order diagrams.


\begin{thebibliography}{99}
\addcontentsline{toc}{chapter}{Bibliography}


\bibitem{Cooper}
 L.~N.~Cooper, 
 {\em Bound Electron Pairs in a Degenerate Fermi Gas\/},
 Phys.\ Rev.\ \journal{104,1189,1956}.

\bibitem{BCS}
 J.~Bardeen, L.~N.~Cooper, and J.~R.~Schrieffer, 
 {\em Microscopic Theory of Superconductivity\/}, 
 Phys.\ Rev.\ \journal{106,162,1957};
 {\em Theory of Superconductivity\/},
 Phys.\ Rev.\ \journal{108,1175,1957}.

\bibitem{Onnes}
 H.~Kamerlingh~Onnes, 
 {\em On the sudden change in the rate at which the resistance of
 mercury disappears\/}, 
 Akad.\ van~Wetenschappen \journal{14(113),818,1911}.

\bibitem{Osheroff}
 D.~D.~Osheroff, R.~C.~Richardson, and D.~M.~Lee, 
 {\em Evidence for a New Phase of Solid He\/$^3$\/}, 
 \PRL{28,885,1972}.

\bibitem{Bednorz}
 J.~G.~Bednorz and K.~M\"{u}ller,
 {\em Possible high-$\Tc$ superconductivity in the Ba-La-Cu-O system\/},
 Z.\ Physik B \journal{64,189,1986}.

\bibitem{Regal04}
 C.~A.~Regal, M.~Greiner, and D.~S.~Jin,
 {\em Observation of Resonance Condensation of Fermionic Atom Pairs\/}, 
 \PRL{92,040403,2004} [arXiv:cond-mat/0401554].

\bibitem{Zwierlein04}
 M.~W.~Zwierlein, C.~A.~Stan, C.~H.~Schunck, S.~M.~F.~Raupach,
 A.~J.~Kerman, and W.~Ketterle,
 {\em Condensation of Pairs of Fermionic Atoms near a Feshbach
 Resonance\/}, 
 \PRL{92,120403,2004} [arXiv:cond-mat/0403049].


\bibitem{Bailin}
  D.~Bailin and A.~Love,
 {\em Superfluidity And Superconductivity In Relativistic Fermion
 Systems\/}, 
 \PR{107,325,1984}.

\bibitem{Dean}
 D.~J.~Dean and M.~H.-Jensen,
 {\em Pairing in nuclear systems: from neutron stars to finite
 nuclei\/}, 
 \RMP{75,607,2003} [arXiv:nucl-th/0210033].

\bibitem{Rajagopal}
 K.~Rajagopal and F.~Wilczek,
 {\em The condensed matter physics of QCD\/},
 in {\em At the frontier of particle physics: Handbook of QCD\/},
 edited by B.~I.~Festschrift and M.~Shifman 
 (World Scientific, Singapore, 2001) [arXiv:hep-ph/0011333]. 

\bibitem{Alford}
 M.~G.~Alford,
 {\em Color superconducting quark matter\/},
 Ann.\ Rev.\ Nucl.\ Part.\ Sci.\ \journal{51,131,2001}
 [arXiv:hep-ph/0102047].

\bibitem{Kapusta}
 J.~I.~Kapusta,
 {\em Neutrino Superfluidity\/},
 \PRL{93,251801,2004} [arXiv:hep-th/0407164].


 \bibitem{Stwalley}
 W.~C.~Stwalley,
 {\em Stability of Spin-Aligned Hydrogen at Low Temperatures and High
 Magnetic Fields: New Field-Dependent Scattering Resonances and
 Predissociations\/},
 \PRL{37,1628,1976}.

\bibitem{Tiesinga}
 E.~Tiesinga, B.~J.~Verhaar, and H.~T.~C.~Stoof,
 {\em Threshold and resonance phenomena in ultracold ground-state
 collisions\/}, 
 \PRA{47,4114,1993}.

\bibitem{s_length1}
 K.~M.~O'Hara, S.~L.~Hemmer, S.~R.~Granade, M.~E.~Gehm, J.~E.~Thomas, 
 V.~Venturi, E.~Tiesinga, and C.~J.~Williams, 
 {\em Measurement of the zero crossing in a Feshbach resonance of
 fermionic $^6$Li\/}, 
 \PRA{66,041401,2002} [arXiv:cond-mat/0207717].

\bibitem{s_length2}
 C.~A.~Regal and D.~S.~Jin,
 {\em Measurement of Positive and Negative Scattering Lengths in a Fermi
 Gas of Atoms\/},
 \PRL{90,230404,2003} [arXiv:cond-mat/0302246].


\bibitem{Eagles}
 D.~M.~Eagles, 
 {\em Possible pairing without superconductivity at low carrier
 concentration in bulk and thin-film superconducting semiconductors\/},
 Phys.\ Rev.\ \journal{186,456,1969}.

\bibitem{Leggett}
 A.~J.~Leggett, in {\em Modern Trends in the Theory of Condensed
 Matter\/} edited by A. Pekalski and R. Przystawa (Springer-Verlag,
 Berlin, 1980); J.\ Phys.\ (Paris), Colloq.\ \journal{41,C7-19,1980}. 

\bibitem{Nozieres}
 P.~Nozi\`eres and S.~Schmitt--Rink,
 {\em Bose condensation in an attractive fermion gas: from weak to
 strong coupling superconductivity\/},
 \JLTP{59,195,1985}.

\bibitem{Randeria-review}
 M.~Randeria, 
 {\em Crossover from BCS Theory to Bose-Einstein Condensation\/},
 in {\em Bose--Einstein Condensation\/}, edited by A.~Griffin 
 {\em et al.\/} (Cambridge University Press, New York, 1995).

\bibitem{Fetter-Walecka}
 See, e.g., A.~L.~Fetter and J.~D.~Walecka, 
 {\em Quantum Theory of Many-Particle Systems\/}
 (McGraw Hill, New York, 1971);

 A.~A.~Abrikosov, L.~P.~Gorkov, and I.~E.~Dzialoshinskii,
 {\em Methods of Quantum Field Theory in Statistical Physics\/}
 (Dover, New York, 1975).


\bibitem{OHara02}
 K.~M.~O'Hara, S.~L.~Hemmer, M.~E.~Gehm, S.~R.~Granade, and
 J.~E.~Thomas, 
 {\em Observation of a Strongly Interacting Degenerate Fermi Gas of
 Atoms\/}, 
 Science \journal{298,2179,2002} [arXiv:cond-mat/0212463].

\bibitem{Dieckmann02}
 K.~Dieckmann, C.~A.~Stan, S.~Gupta, Z.~Hadzibabic, C.~H.~Schunck,
 and W. Ketterle, 
 {\em Decay of an Ultracold Fermionic Lithium Gas near a Feshbach
 Resonance\/}, 
 \PRL{89,203201,2002} [arXiv:cond-mat/0207046].

\bibitem{Bourdel03}
 T.~Bourdel, J.~Cubizolles, L.~Khaykovich, K.~M.~F.~Magalh\~{a}es,
 S.~J.~J.~M.~F.~Kokkelmans, G.~V.~Shlyapnikov, and C.~Salomon, 
 {\em Measurement of the Interaction Energy near a Feshbach Resonance in
 a $^6$Li Fermi Gas\/}, 
 \PRL{91,020402,2003} [arXiv:cond-mat/0303079].

\bibitem{Regal03}
 C.~A.~Regal, C.~Ticknor, J.~L.~Bohn, and D.~S.~Jin,
 {\em Creation of ultracold molecules from a Fermi gas of atoms\/},
 Nature \journal{424,47,2003} [arXiv:cond-mat/0305028].

\bibitem{Strecker03}
 K.~E.~Strecker, G.~B.~Partridge, and R.~G.~Hulet, 
 {\em Conversion of an Atomic Fermi Gas to a Long-Lived Molecular Bose Gas\/},
 \PRL{91,080406,2003} [arXiv:cond-mat/0308318].

\bibitem{Cubizolles03}
 J.~Cubizolles, T.~Bourdel, S.~J.~J.~M.~F.~Kokkelmans,
 G.~V.~Shlyapnikov, and C.~Salomon, 
 {\em Production of Long-Lived Ultracold Li$_2$ Molecules from a Fermi
 Gas\/}, 
 \PRL{91,240401,2003} [arXiv:cond-mat/0308018].

\bibitem{Jochim03-PRL}
 S.~Jochim, M.~Bartenstein, A.~Altmeyer, G.~Hendl, C.~Chin, 
 J.~Hecker~Denschlag, and R.~Grimm, 
 {\em Pure Gas of Optically Trapped Molecules Created from Fermionic
 Atoms\/}, 
 \PRL{91,240402,2003} [arXiv:cond-mat/0308095].

\bibitem{Greiner03}
 M.~Greiner, C.~A.~Regal, and D.~S.~Jin,
 {\em Emergence of a molecular Bose--Einstein condensate from a Fermi
 gas\/}, 
 Nature \journal{426,537,2003} [arXiv:cond-mat/0311172].

\bibitem{Zwierlein03}
 M.~W.~Zwierlein, C.~A.~Stan, C.~H.~Schunck, S.~M.~F.~Raupach, S.~Gupta,
 Z.~Hadzibabic, and W.~Ketterle 
 {\em Observation of Bose--Einstein Condensation of Molecules\/}, 
 \PRL{91,250401,2003} [arXiv:cond-mat/0311617].

\bibitem{Jochim03}
 S.~Jochim, M.~Bartenstein, A.~Altmeyer, G.~Hendl, S.~Riedl, C.~Chin,
 J.~Hecker~Denschlag, and R.~Grimm, 
 {\em Bose--Einstein Condensation of Molecules\/}, 
 Science \journal{302,2101,2003}.


\bibitem{Regal04-lifetime}
 C.~A.~Regal, M.~Greiner, and D.~S.~Jin,
 {\em Lifetime of Molecule-Atom Mixtures near a Feshbach Resonance in
 $^{40}$K\/}, 
 \PRL{92,083201,2004} [arXiv:cond-mat/0308606].

\bibitem{Bartenstein04}
 M.~Bartenstein, A.~Almeyer, S.~Riedl, S.~Jochim, C.~Chin, 
 J.~Hecker~Denschlag, and R.~Grimm,
 {\em Crossover from a Molecular Bose-Einstein Condensate to a
 Degenerate Fermi Gas\/},
 \PRL{92,120401,2004} [arXiv:cond-mat/0401109].


\bibitem{Kinast04}
 J.~Kinast, S.~L.~Hemmer, M.~E.~Gehm, A.~Turlapov, and J.~E.~Thomas, 
 {\em Evidence for Superfluidity in a Resonantly Interacting Fermi Gas\/},
 \PRL{92,150402,2004} [arXiv:cond-mat/0403540].

\bibitem{Bartenstein04-collective}
 M.~Bartenstein, A.~Altmeyer, S.~Riedl, S.~Jochim, C.~Chin, 
 J.~Hecker~Denschlag, and R.~Grimm, 
 {\em Collective Excitations of a Degenerate Gas at the BEC-BCS
 Crossover\/}, 
 \PRL{92,203201,2004} [arXiv:cond-mat/0403716].

\bibitem{Bourdel04}
 T.~Bourdel, L.~Khayakovich, J.~Cubizolles, J.~Zhang, F.~Chevy,
 M.~Teichmann, L.~Tarruell, S.~J.~J.~M.~F.~Kokkelmans, and C.~Salomon, 
 {\em Experimental Study of the BEC-BCS Crossover Region in Lithium 6\/},
 \PRL{93,050401,2004} [arXiv:cond-mat/0403091].

\bibitem{Chin04}
 C.~Chin, M.~Bartenstein, A.~Altmeyer, S.~Riedl, S.~Jochim,
 J.~Hecker~Denschlag, and R.~Grimm, 
 {\em Observation of the Pairing Gap in a Strongly Interacting Fermi
 Gas\/}, 
 Science \journal{305,1128,2004} [arXiv:cond-mat/0405632]. 

\bibitem{Kinast04-breakdown}
 J.~Kinast, A.~Turlapov, and J.~E.~Thomas, 
 {\em Breakdown of hydrodynamics in the radial breathing mode of a
 strongly interacting Fermi gas\/}, 
 \PRA{70,051401(R),2004} [arXiv:cond-mat/0408634]. 

\bibitem{Kinast05} 
 J.~Kinast, A.~Turlapov, J.~E.~Thomas, Q.~Chen, J.~Stajic, and K.~Levin, 
 {\em Heat Capacity of a Strongly Interacting Fermi Gas\/},
 Science \journal{307,1296,2005} [arXiv:cond-mat/0502087].

\bibitem{Partridge05}
 G.~B.~Partridge, K.~E.~Strecker, R.~I.~Kamar, M.~W.~Jack,
 and R.~G.~Hulet,
 {\em Molecular probe of pairing in the BEC-BCS crossover\/}, 
 \PRL{95,020404,2005} [arXiv:cond-mat/0505353].

\bibitem{Zwierlein05-vortices}
 M.~W.~Zwierlein, J.~R.~Abo-Shaeer, A.~Schirotzek, C.~H.~Schunck and
 W.~Ketterle, 
 {\em Vortices and superfluidity in a strongly interacting Fermi gas\/},
 Nature \journal{435,1047,2005} [arXiv:cond-mat/0505635].


\bibitem{Zwierlein05}
 M.~W.~Zwierlein, A.~Schirotzek, C.~H.~Schunck, and W.~Ketterle,
 {\em Fermionic Superfluidity with Imbalanced Spin Populations\/},
 Science \journal{311,492,2006} [arXiv:cond-mat/0511197].

\bibitem{Partridge06}
 G.~B.~Partridge, W.~Li, R.~I.~Kamar, Y.~Liao, and R.~G.~Hulet,
 {\em Pairing and Phase Separation in a Polarized Fermi Gas\/},
 Science \journal{311,503,2006} [arXiv:cond-mat/0511752].

\bibitem{Zwierlein06}
 M.~W.~Zwierlein, C.~H.~Schunck, A.~Schirotzek, and W.~Ketterle,
 {\em Direct Observation of the Superfluid Phase Transition in
 Ultracold Fermi Gases\/},
 Nature \journal{442,54,2006} [arXiv:cond-mat/0605258].

\bibitem{Shin06}
 Y.~Shin, M.~W.~Zwierlein, C.~H.~Schunck, A.~Schirotzek, and 
 W.~Ketterle, 
 {\em Observation of Phase Separation in a Strongly-Interacting
 Imbalanced Fermi Gas\/}, 
 \PRL{97,030401,2006} [arXiv:cond-mat/0606432].

\bibitem{Partridge06-2}
 G.~B.~Partridge, W.~Li, Y.~A.~Liao, and R.~G.~Hulet
 M.~Haque, and H.~T.~C.~Stoof, 
 {\em Deformation of a Trapped Fermi Gas with Unequal Spin
 Populations\/}, 
 \PRL{97,190407,2006} [arXiv:cond-mat/0608455]. 


\bibitem{Bertsch}
 G.~Bertsch, {\em Many-Body X Challenge\/},
 in: Proc. X Conference on Recent Progress in Many-Body Theories, 
 eds. R.~F.~Bishop {\em et al.\/} (World Scientific, Singapore, 2000).

\bibitem{Gardestig}
 A.~Gardestig and D.~R.~Phillips,
 {\em Using chiral perturbation theory to extract the neutron-neutron
 scattering length from $\pi^- d \to n n \gamma$\/},
 \PRC{73,014002,2006} [arXiv:nucl-th/0501049].

\bibitem{highTc}
 See, e.g., Q.~Chen, J.~Stajic, S.~Tan, and K.~Levin,
 {\em BCS-BEC Crossover: From High Temperature Superconductors to 
 Ultracold Superfluids\/},
 \PR{412,1,2005} [arXiv:cond-mat/0404274], and references therein.


\bibitem{Phillips}
 W.~D.~Phillips,
 {\em Laser cooling and trapping of neutral atoms\/},
 \RMP{70,721,1998}.

 \bibitem{DeMarco}
 B.~DeMarco and D.~S.~Jin,
 {\em Onset of Fermi degeneracy in a trapped atomic gas\/}, 
 Science \journal{285,1703,1999}.



\bibitem{BRAHMS}
 I.~Arsene {\itshape et al.}  [BRAHMS Collaboration],
 {\em Quark gluon plasma and color glass condensate at RHIC? The
 perspective from the BRAHMS experiment\/},
 \NPA{757,1,2005} [arXiv:nucl-ex/0410020].

\bibitem{PHOBOS}
 B.~B.~Back {\itshape et al.}  [PHOBOS Collaboration],
 {\em The PHOBOS perspective on discoveries at RHIC\/},
 \NPA{757,28,2005} [arXiv:nucl-ex/0410022].

\bibitem{STAR}
 J.~Adams {\itshape et al.}  [STAR Collaboration],
 {\em Experimental and theoretical challenges in the search for the
 quark gluon plasma: The STAR collaboration's critical assessment of the
 evidence from RHIC collisions\/},
 \NPA{757,102,2005} [arXiv:nucl-ex/0501009].

\bibitem{PHENIX}
 K.~Adcox {\itshape et al.}  [PHENIX Collaboration],
 {\em Formation of dense partonic matter in relativistic nucleus nucleus
 collisions at RHIC: Experimental evaluation by the PHENIX
 collaboration\/}, 
 \NPA{757,184,2005} [arXiv:nucl-ex/0410003].

\bibitem{BNL}
 Brookhaven National Laboratory, 
 {\em RHIC Scientists Serve Up ``Perfect'' Liquid\/},
 in Laboratory News (April 2005).\\
 \verb#http://www.bnl.gov/bnlweb/pubaf/pr/PR_display.asp?prID=05-38#

\bibitem{Hatsuda}
 See, e.g., K.~Yagi, T.~Hatsuda, and Y.~Miake, 
 {\em Quark-Gluon Plasma\/}
 (Cambridge University Press, Cambridge, 2005).


\bibitem{Anderson95}
 M.~H.~Anderson, J.~R.~Ensher, M.~R.~Matthews, C.~E.~Wieman, 
 and E.~A.~Cornell, 
 {\em Observation of Bose--Einstein condensation in a dilute atomic
 vapor\/}, 
 Science \journal{269,198,1995}.

\bibitem{Davis95}
 K.~B.~Davis, M.--O.~Mewes, M.~R.~Andrews, N.~J.~van~Druten,
 D.~S.~Durfee, D.~M.~Kurn, and W.~Ketterle, 
 {\em Bose--Einstein condensation in a gas of sodium atoms\/}, 
 \PRL{75,3969,1995}.


\bibitem{Stan04}
 C.~A.~Stan, M.~W.~Zwierlein, C.~H.~Schunck, S.~M.~F.~Raupach,
 and W.~Ketterle,
 {\em Observation of Feshbach Resonances between Two Different Atomic
 Species\/}, 
 \PRL{93,143001,2004} [arXiv:cond-mat/0406129].

\bibitem{Inoue04}
 S.~Inouye, J.~Goldwin, M.~L.~Olsen, C.~Ticknor, J.~L.~Bohn, 
 and D.~S.~Jin,
 {\em Observation of Heteronuclear Feshbach Resonances in a Mixture of
 Bosons and Fermions\/},
 \PRL{93,183201,2004} [arXiv:cond-mat/0406208].

\bibitem{Alford99}
 M.~G.~Alford, J.~Berges, and K.~Rajagopal,
 {\em Gapless color superconductivity\/},
 \PRL{84,598,2000} [arXiv:hep-ph/9908235].

\bibitem{Shovkovy}
 I.~Shovkovy and M.~Huang,
 {\em Gapless two-flavor color superconductor\/},
 \PLB{564,205,2003} [arXiv:hep-ph/0302142].

\bibitem{Huang}
 M.~Huang and I.~Shovkovy,
 {\em Gapless color superconductivity at zero and at finite temperature\/},
 \NPA{729,835,2003} [arXiv:hep-ph/0307273].

\bibitem{Gubankova03}
 E.~Gubankova, W.~V.~Liu, and F.~Wilczek,
 {\em Breached pairing superfluidity: Possible realization in QCD\/},
 \PRL{91,032001,2003} [arXiv:hep-ph/0304016].

\bibitem{Alford03}
 M.~Alford, C.~Kouvaris, and K.~Rajagopal,
 {\em Gapless color-flavor-locked quark matter\/},
 \PRL{92,222001,2004} [arXiv:hep-ph/0311286].


\bibitem{Melo}
 C.~A.~R.~S\'a~de~Melo, M.~Randeria, and J.~R.~Engelbrecht,
 {\em Crossover from BCS to Bose Superconductivity: Transition
 Temperature and Time-Dependent Ginzburg-Landau Theory\/},
 \PRL{71,3202,1993}.

\bibitem{Engelbrecht}
 J.~R.~Engelbrecht, M.~Randeria and C.~A.~R.~S\'a~de~Melo, 
 {\em BCS to Bose crossover: Broken-symmetry state\/},
 \PRB{55,15153,1997}. 

\bibitem{Haussmann}
 R.~Haussmann, 
 {\em Crossover from BCS superconductivity to Bose--Einstein
 condensation: A self-consistent theory\/}, 
 Z.\ Phys.\ B \journal{91,291,1993};
 {\em Properties of a Fermi liquid at the superfluid transition in the
 crossover region between BCS superconductivity and Bose--Einstein
 condensation\/}, \PRB{49,12975,1994}.

\bibitem{Holland}
 M.~Holland, S.~J.~J.~M.~F.~Kokkelmans, M.~L.~Chiofalo, and R.~Walser, 
 {\em Resonance Superfluidity in a Quantum Degenerate Fermi Gas\/},
 \PRL{87,120406,2001} [arXiv:cond-mat/0103479].

\bibitem{Timmermans}
 E.~Timmermans, K.~Furuyab, P.~W.~Milonnia, and A.~K.~Kermanc, 
 {\em Prospect of creating a composite Fermi--Bose superfluid\/},
 \PLA{285,228,2001} [arXiv:cond-mat/0103327].

\bibitem{Ohashi}
 Y.~Ohashi and A.~Griffin, 
 {\em BCS-BEC Crossover in a Gas of Fermi Atoms with a Feshbach
 Resonance\/}, 
 \PRL{89,130402,2002} [arXiv:cond-mat/0201262]; 
 {\em Superfluid transition temperature in a trapped gas of Fermi atoms
 with a Feshbach resonance\/}, 
 \PRA{67,033603,2003} [arXiv:cond-mat/0210185].

\bibitem{Milstein}
 J.~N.~Milstein, S.~J.~J.~M.~F.~Kokkelmans, and M.~J.~Holland, 
 {\em Resonance theory of the crossover from Bardeen-Cooper-Schrieffer
 superfluidity to Bose--Einstein condensation in a dilute Fermi gas\/}, 
 \PRA{66,043604,2002} [arXiv:cond-mat/0204334].

\bibitem{Perali}
 A.~Perali, P.~Pieri, L.~Pisani, and G.~C.~Strinati, 
 {\em BCS-BEC Crossover at Finite Temperature for Superfluid
 Trapped Fermi Atoms\/}, 
 \PRL{92,220404,2004} [arXiv:cond-mat/0311309].

\bibitem{Liu}
 X.-J.~Liu and H.~Hu,
 {\em Self-consistent theory of atomic Fermi gases with a Feshbach
 resonance at the superfluid transition\/}, 
 \PRA{72,063613,2005} [arXiv:cond-mat/0505572].

\bibitem{Nishida-Abuki}
 Y.~Nishida and H.~Abuki,
 {\em BCS-BEC crossover in a relativistic superfluid and its
 significance to quark matter\/},
 \PRD{72,096004,2005} [arXiv:hep-ph/0504083].

\bibitem{Abuki}
 H.~Abuki, 
 {\em BCS/BEC crossover in Quark Matter and Evolution of its Static and
 Dynamic properties\/}, 
 arXiv:hep-ph/0605081.

\bibitem{Zwerger}
 R. Haussmann, W. Rantner, S. Cerrito, and W. Zwerger,
 {\em Thermodynamics of the BCS-BEC crossover\/},
 arXiv:cond-mat/0608282.


\bibitem{Ho}
 T.-L.~Ho and E.~J.~Mueller,
 {\em High Temperature Expansion Applied to Fermions near Feshbach
 Resonance\/},
 \PRL{92,160404,2004} [arXiv:cond-mat/0306187].

\bibitem{Horowitz}
 C.~J.~Horowitz and A.~Schwenk,
 {\em The Virial Equation of State of Low-Density Neutron Matter\/}, 
 \PLB{638,153,2006} [arXiv:nucl-th/0507064].

\bibitem{Rupak_finite-T}
 G.~Rupak, 
 {\em Universality in a 2-component Fermi System at Finite
 Temperature\/}, 
 arXiv:nucl-th/0604053.


\bibitem{Carlson2003}
 J.~Carlson, S.--Y.~Chang, V.~R.~Pandharipande, and K.~E.~Schmidt,
 {\em Superfluid Fermi Gases with Large Scattering Length\/},
 \PRL{91,050401,2003} [arXiv:physics/0303094].

\bibitem{Chang2004}
 S.~Y.~Chang, V.~R.~Pandharipande, J.~Carlson, and K.~E.~Schmidt,
 {\em Quantum Monte Carlo studies of superfluid Fermi gases\/},
 \PRA{70,043602,2004} [arXiv:physics/0404115].

\bibitem{Chen:2003vy}
 J.~W.~Chen and D.~B.~Kaplan,
 {\em A lattice theory for low energy fermions at finite chemical
 potential\/}, 
 \PRL{92,257002,2004} [arXiv:hep-lat/0308016].

\bibitem{Astrakharchik2004}
 G.~E.~Astrakharchik, J.~Boronat, J.~Casulleras, and S.~Giorgini,
 {\em Equation of State of a Fermi Gas in the BEC-BCS Crossover: A Quantum 
 Monte Carlo Study\/},
 \PRL{93,200404,2004} [arXiv:cond-mat/0406113].

\bibitem{Carlson:2005kg}
 J.~Carlson and S.~Reddy,
 {\em Asymmetric Two-component Fermion Systems in Strong Coupling\/},
 \PRL{95,060401,2005} [arXiv:cond-mat/0503256].

\bibitem{Lee}
 D.~Lee, 
 {\em Ground state energy of spin-1/2 fermions in the unitary limit\/},
 \PRB{73,115112,2006} [arXiv:cond-mat/0511332].


\bibitem{Wingate-Tc}
 M.~Wingate,
 {\em Critical temperature for fermion pairing using lattice field
 theory\/}, 
 arXiv:cond-mat/0502372.

\bibitem{bulgac-Tc}
 A.~Bulgac, J.~E.~Drut, and P.~Magierski, 
 {\em Spin 1/2 Fermions in the Unitary Regime: A Superfluid of a
 New Type\/},
 \PRL{96,090404,2006} [arXiv:cond-mat/0505374].

\bibitem{Lee-Schafer}
 D.~Lee and T.~Schafer,
 {\em Cold dilute neutron matter on the lattice I: Lattice virial
 coefficients and large scattering lengths\/},
 \PRC{73,015201,2006} [arXiv:nucl-th/0509017];
 {\em Cold dilute neutron matter on the lattice II: Results in the
 unitary limit\/},
 \PRC{73,015202,2006} [arXiv:nucl-th/0509018].

\bibitem{burovski-Tc}
 E.~Burovski, N.~Prokofev, B.~Svistunov, and M.~Troyer,
 {\em Critical Temperature and Thermodynamics of Attractive
 Fermions at Unitarity\/},
 \PRL{96,160402,2006} [arXiv:cond-mat/0602224]; 
 {\em The Fermi-Hubbard model at unitarity\/},
 \NJP{8,153,2006} [arXiv:cond-mat/0605350].

\bibitem{Akkineni-Tc}
 V.~K.~Akkineni, D.~M.~Ceperley, and N.~Trivedi,
 {\em Pairing and Superfluid Properties of Dilute Fermion Gases at 
 Unitarity\/},
 arXiv:cond-mat/0608154. 


\bibitem{Nishida-Son1}
 Y.~Nishida and D.~T.~Son,
 {\em $\epsilon$ expansion for a Fermi gas at infinite scattering
 length\/}, 
 \PRL{97,050403,2006} [arXiv:cond-mat/0604500].

\bibitem{Nishida-Son2}
 Y.~Nishida and D.~T.~Son, 
 {\em Fermi gas near unitarity around four and two spatial dimensions\/},
 arXiv:cond-mat/0607835.  

\bibitem{Nishida_finite-T}
 Y.~Nishida,
 {\em Unitary Fermi gas at finite temperature in the $\epsilon$
 expansion\/}, 
 arXiv:cond-mat/0608321.  

\bibitem{nussinov04}
 Z.~Nussinov and S.~Nussinov,
 {\em The BCS-BEC Crossover In Arbitrary Dimensions\/},
 arXiv:cond-mat/0410597;
 {\em Triviality of the BCS-BEC crossover in extended dimensions:
 Implications for the ground state energy\/},
 \PRA{74,053622,2006}.

\bibitem{Sauli}
 F.~Sauli and P.~Kopietz,
 {\em Low-density expansion for the two-dimensional electron gas\/}, 
 arXiv:cond-mat/0608423.

\bibitem{Sachdev}
 P.~Nikolic and S.~Sachdev, 
 {\em Renormalization group fixed points, universal phase diagram, and
 1/N expansion for quantum liquids with interactions near the unitarity
 limit\/},  
 arXiv:cond-mat/0609106.

\bibitem{Rupak-dimer}
 G.~Rupak,
 {\em Dimer scattering in the $\varepsilon$ expansion\/},
 arXiv:nucl-th/0605074.

\bibitem{WilsonKogut} 
 K.~G.~Wilson and J.~Kogut,
 {\em The renormalization group and the $\epsilon$ expansion\/},
 \PR{12,75,1974}.

\bibitem{Rupak-polarized}
 G.~Rupak, T.~Schaefer, and A.~Kryjevski,
 {\em Polarized fermions in the unitarity limit\/},
 arXiv:cond-mat/0607834.

\bibitem{Arnold}
 P.~Arnold, J.~E.~Drut, and D.~T.~Son,
 {\em NNLO $\epsilon$ expansion for a Fermi gas at infinite scattering
 length\/}, 
 arXiv:cond-mat/0608477.

\bibitem{Chen}
 J.~W.~Chen and E.~Nakano,
 {\em BEC-BCS Crossover in the $\epsilon$ Expansion\/}, 
 arXiv:cond-mat/0610011.



\bibitem{Veillette}
 M.~Y.~Veillette, D.~E.~Sheehy, and L.~Radzihovsky, 
 {\em Large-N expansion for unitary superfluid Fermi gases\/},
 arXiv:cond-mat/0610798. 

\bibitem{NS_unequal-mass}
 Y.~Nishida and D.~T.~Son, 
 {\em Phase structure of unitary Fermi gas with unequal densities
 and masses\/},  
 in preparation.  


\bibitem{randeria-2d}
 M.~Randeria, J.--M.~Duan, and L.--Y.~Shieh, 
 {\em Bound states, Cooper pairing, and Bose condensation in two
 dimensions\/}, 
 \PRL{62,981,1989}; 
 {\em Superconductivity in a two-dimensional Fermi gas: Evolution from
 Cooper pairing to Bose condensation\/}, 
 \PRB{41,327,1990}.


\bibitem{Stratonovich}
 R.~L.~Stratonovich, 
 {\em On a method of calculating quantum distribution functions\/}, 
 Dokl.\ Akad.\ Nauk SSSR \journal{115,1097,1957}; 
 [Soviet Phys.\ Doklady \journal{2,416,1958}].

\bibitem{Hubbard}
 J.~Hubbard, 
 {\em Calculation of Partition Functions\/},
 \PRL{3,77,1959}.

\bibitem{Gorkov}
 L.~P.~Gorkov, 
 {\em On the energy spectrum of superconductors\/},
 Sov.\ Phys.\ JETP \journal{7,505,1958}. 

\bibitem{Nambu}
 Y.~Nambu, 
 {\em Quasi-Particles and Gauge Invariance in the Theory of
 Superconductivity\/},
 Phys.\ Rev.\ \journal{117,648,1960}. 

\bibitem{Peskin:1995ev}
 See, e.g., M.~E.~Peskin and D.~V.~Schroeder,
 {\em An Introduction to Quantum Field Theory\/} 
 (Addison-Wesley, Reading MA, 1995).

\bibitem{son05}
 D.~T.~Son and M.~A.~Stephanov,
 {\em Phase Diagram of Cold Polarized Fermi Gas\/},
 \PRA{74,013614,2006} [arXiv:cond-mat/0507586].

\bibitem{Virerit}
 L.~Viverit, S.~Giorgini, L.~P.~Pitaevskii, and S.~Stringari, 
 {\em Momentum distribution of a trapped Fermi gas with large scattering
 length\/}, 
 \PRA{69,013607,2004} [arXiv:cond-mat/0307538].

\bibitem{Tan}
 S.~Tan,
 {\em Energetics of the Fermi gas which has BEC-BCS crossover\/},
 arXiv:cond-mat/0505200;
 {\em Large momentum part of fermions with large scattering length\/}, 
 arXiv:cond-mat/0508320.

\bibitem{Regal-momentum}
 C.~A.~Regal, M.~Greiner, S.~Giorgini, M.~Holland, and D.~S.~Jin,
 {\em Momentum Distribution of a Fermi Gas of Atoms in the BCS-BEC
 Crossover\/}, 
 \PRL{95,250404,2005} [arXiv:cond-mat/0507316]. 

\bibitem{Astrakharchik}
 G.~E.~Astrakharchik, J.~Boronat, J.~Casulleras, and S.~Giorgini, 
 {\em Momentum distribution and condensate fraction of a Fermi gas in the
 BCS-BEC crossover\/}, 
 arXiv:cond-mat/0507483. 

\bibitem{Ortiz}
 G.~Ortiz and J.~Dukelsky,
 {\em BCS-to-BEC crossover from the exact BCS solution\/}, 
 \PRA{72,043611,2005} [arXiv:cond-mat/0503664]. 

\bibitem{Salasnich}
 L.~Salasnich and N.~Manini, 
 {\em Condensate fraction of a Fermi gas in the BCS-BEC crossover\/}, 
 \PRA{72,023621,2005} [arXiv:cond-mat/0506074]. 

\bibitem{Yang}
 C.~N.~Yang, 
 {\em Concept of Off-Diagonal Long-Range Order and the Quantum Phases of
 Liquid He and of Superconductors\/}, 
 \RMP{34,694,1962}.

\bibitem{Luttinger-Ward}
 J.~M.~Luttinger and J.~C.~Ward, 
 {\em Ground-State Energy of a Many-Fermion System. II\/},
 Phys.\ Rev.\ \journal{118,1417,1960}.

\bibitem{DeDominicis-Martin}
 C.~De~Dominicis and P.~C.~Martin, 
 {\em Stationary Entropy Principle and Renormalization in Normal and
 Superfluid Systems. I. Algebraic Formulation\/},
 J.\ Math.\ Phys.\ \journal{5,14,1964};
 {\em Stationary Entropy Principle and Renormalization in Normal and
 Superfluid Systems. II. Diagrammatic Formulation\/}, 
 J.\ Math.\ Phys.\ \journal{5,31,1964}.

\bibitem{Baym-Kadanoff}
 G.~Baym and L.~P.~Kadanoff, 
 {\em Conservation Laws and Correlation Functions\/},
 Phys.\ Rev.\ \journal{124,287,1961}.

\bibitem{Baym}
 G.~Baym, 
 {\em Self-Consistent Approximations in Many-Body Systems\/}, 
 Phys.\ Rev.\ \journal{127,1391,1962}. 



\bibitem{FF}
 P.~Fulde and R.~A.~Ferrell, 
 {\em Superconductivity in a Strong Spin-Exchange Field\/},
 Phys.\ Rev.\ \journal{135,A550,1964}.

\bibitem{LO}
 A.~I.~Larkin and Y.~N.~Ovchinnikov, 
 {\em Superconductivity in a Strong Spin-Exchange Field\/},
 Zh.\ Eksp.\ Teor.\ Fiz.\ \journal{47,1136,1964} 
 [Sov.\ Phys.\ JETP \journal{20,762,1965}].


\bibitem{iskin06}
 M.~Iskin and C.~A.~R.~S\`{a}~de~Melo,
 {\em Two-species fermion mixtures with population imbalance\/}, 
 arXiv:cond-mat/0604184;
 {\em Asymmetric two-component Fermi gas with unequal masses\/},
 arXiv:cond-mat/0606624.

\bibitem{yip06}
 S.--T.~Wu, C.--H.~Pao, and S.--K.~Yip,
 {\em Resonant pairing between Fermions with unequal masses\/}, 
 arXiv:cond-mat/0604572.

\bibitem{lin06}
 G.--D.~Lin, W.~Yi, and L.--M.~Duan,
 {\em Superfluid shells for trapped fermions with mass and population
 imbalance\/}, 
 \PRA{74,031604(R),2006} [arXiv:cond-mat/0607664].

\bibitem{parish06}
 M.~M.~Parish, F.~M.~Marchetti, A.~Lamacraft, and B.~D.~Simons, 
 {\em Polarized Fermi condensates with unequal masses: Tuning the
 tricritical point\/}, 
 arXiv:cond-mat/0608651.


\bibitem{gorkov}
 L.~P.~Gor'kov and T.~K.~Melik--Barkhudarov, 
 {\em Contribution to the Theory of Superfluidity in an Imperfect Fermi
 Gas\/}, 
 Sov.\ Phys.\ JETP \journal{13,1018,1961}. 

\bibitem{Heiselberg}
 H.~Heiselberg, C.~J.~Pethick, H.~Smith, and L.~Viverit,
 {\em Influence of Induced Interactions on the Superfluid Transition in 
 Dilute Fermi Gases\/}, 
 \PRL{85,2418,2000} [arXiv:cond-mat/0004360].

\bibitem{luttinger}
 J.~M.~Luttinger, 
 {\em Fermi Surface and Some Simple Equilibrium Properties of a System
 of Interacting Fermions\/}, 
 Phys.\ Rev.\ \journal{119,1153,1960}. 


\bibitem{justin}
 See, e.g., J.~Z.--Justin, 
 {\em Quantum Field Theory and Critical Phenomena\/} 
 (Clarendon Press, Oxford, 2002).


\bibitem{Efimov}
 V.~Efimov, 
 {\em Energy levels arising from resonant two-body forces in a
 three-body system\/},
 \PLB{33,563,1970};
 {\em Weakly-bound states of three resonantly-interacting particles\/},
 Sov.\ J.\ Nucl.\ Phys.\ \journal{12,589,1971}; 
 {\em Energy levels of three resonantly interacting particles\/},
 \NPA{210,157,1973}.

\bibitem{Braaten}
 E.~Braaten and H.--W.~Hammer,
 {\em Universality in few-body systems with large scattering length\/}, 
 \PR{428,259,2006} [arXiv:cond-mat/0410417]. 

\bibitem{Nielsen}
 E.~Nielsen, D.~V.~Fedorov, A.~S.~Jensen, and E.~Garrido,
 {\em The three-body problem with short-range interactions\/},
 \PR{347,373,2001}.


\bibitem{Gross-Neveu}
 D.~J.~Gross and A.~Neveu,
 {\em Dynamical symmetry breaking in asymptotically free field theories\/},
 \PRD{10,3235,1974}.

\bibitem{Lautrup}
 B.~Lautrup,
 {\em On high order estimates in QED\/},
 \PLB{69,109,1977}.

\bibitem{tHooft}
 G.~'t Hooft, in {\em The Whys of subnuclear physics\/},
 Proc.\ Int.\ School, Erice, Italy, 1977, 
 edited by A.~Zichichi (Plenum, New York, 1978).

\end{thebibliography}
\end{document}